\documentclass[acmsmall,screen]{acmart}

\AtBeginDocument{%
  }

\setcopyright{acmlicensed}
\copyrightyear{2025}
\acmYear{2025}
\acmDOI{XXXXXXX.XXXXXXX}

\usepackage{graphicx}
\usepackage{diagbox}
\usepackage{multirow}
\usepackage{booktabs}
\usepackage{threeparttable, array}
\usepackage{colortbl}
\usepackage{tkz-graph}
\usepackage{tikz}
\usepackage{pgfplots}
\usetikzlibrary{calc}

\usepackage{textcomp}

\usepackage{xcolor}

\usepackage{hyperref}
\definecolor{mydarkblue}{rgb}{0,0.08,0.45}
\hypersetup{ %
    pdftitle={},
    pdfsubject={},
    pdfkeywords={},
    pdfborder=0 0 0,
    pdfpagemode=UseNone,
    colorlinks=true,
    linkcolor=mydarkblue,
    citecolor=mydarkblue,
    filecolor=mydarkblue,
    urlcolor=mydarkblue,
}

\usepackage{amsmath,amsfonts}
\usepackage{bm, bbm}
\usepackage{mathtools}
\usepackage{nicefrac}
\usepackage{physics}

\mathtoolsset{showonlyrefs,showmanualtags}

\makeatletter
\newcommand\addstarred[1]{%
    \expandafter\let\csname\string#1@nostar\endcsname#1%
    \edef#1{\noexpand\@ifstar\expandafter\noexpand\csname\string#1@star\endcsname\expandafter\noexpand\csname\string#1@nostar\endcsname}%
    \expandafter\newcommand\csname\string#1@star\endcsname%
}
\makeatother
\newcommand{\1}[1]{\mathbbm{1}{\{#1\}}}
\addstarred\1[1]{\mathbbm{1}{\left\{#1\right\}}}

\renewcommand{\ge}{\geqslant}

\newcommand{\E}{\mathbb{E}}

\newcommand{\type}[1]{Type-\uppercase\expandafter{\romannumeral#1}}

\usepackage{amsthm}
\newtheorem{theorem}{Theorem}
\newtheorem{lemma}[theorem]{Lemma}

\newtheorem{remark}[theorem]{Remark}

\newtheorem{definition}[theorem]{Definition}

\usepackage{algorithm}
\usepackage{algpseudocode}

\let\oldComment=\Comment
\renewcommand{\Comment}[1]{\oldComment{\texttt{#1}}}
\algnewcommand{\LeftComment}[1]{\Statex $\triangleright$ \texttt{#1}}
\algnewcommand{\RightComment}[1]{\Statex \leavevmode\hfill$\triangleright$ \texttt{#1}}

\algnewcommand\algorithmicinput{\textbf{Input:}}
\algnewcommand\Input{\item[\algorithmicinput]}%

\algnewcommand\algorithmicoutput{\textbf{Output:}}
\algnewcommand\Output{\item[\algorithmicoutput]}%

\algnewcommand\algorithmicinitial{\textbf{Initialize:}}
\algnewcommand\Initial{\item[\algorithmicinitial]}%

\usepackage{xspace}
\usepackage{comment}

\usepackage{fontawesome5}

\usepackage[colorinlistoftodos,prependcaption,textsize=tiny,textwidth=1.5cm,color=green,disable]{todonotes}
\setuptodonotes{inline}

\usepackage{subcaption}

\usetikzlibrary{shapes, shapes.multipart, positioning}
\usetikzlibrary{quantikz2}

\usepackage{wrapfig}
\usepackage{pdflscape}

\def\cG{\mathcal{G}}
\def\cV{\mathcal{V}}

\def\cE{\mathcal{E}}

\newcommand{\nodeAlice}{\texttt{A}\xspace}
\newcommand{\nodeBob}{\texttt{B}\xspace}
\newcommand{\nodeCharlie}{\texttt{C}\xspace}

\newcommand{\mergecast}{\texttt{Mergecast}\xspace}
\newcommand{\bypassunicast}{\texttt{BypassUnicast}\xspace}

\DeclarePairedDelimiter{\bbra}{\langle\!\langle}{\rvert}
\DeclarePairedDelimiter{\kett}{\lvert}{\rangle\!\rangle}
\newcommand{\diag}[1]{\operatorname{diag}\left(#1\right)}

\begin{document}

\title[Quantum Network Tomography for General Topology with SPAM Errors]{Quantum Network Tomography \\
  for General Topology with SPAM Errors
}


\author{Xuchuang Wang}
\affiliation{%
  \institution{University of Massachusetts Amherst}
  \city{Amherst}
  \state{MA}
  \country{USA}}
\email{xuchuangw@gmail.com}

\author{Matheus Guedes de Andrade}
\affiliation{%
  \institution{University of Massachusetts Amherst}
  \city{Amherst}
  \state{MA}
  \country{USA}}
\email{mguedesdeand@cs.umass.edu}

\author{Guus Avis}
\affiliation{%
  \institution{University of Massachusetts Amherst}
  \city{Amherst}
  \state{MA}
  \country{USA}}
\email{gavis@cs.umass.edu}

\author{Yu-Zhen Janice Chen}
\affiliation{%
  \institution{University of Massachusetts Amherst}
  \city{Amherst}
  \state{MA}
  \country{USA}}
\email{yuzhenchen@cs.umass.edu}

\author{Mohammad Hajiesmaili}
\affiliation{%
  \institution{University of Massachusetts Amherst}
  \city{Amherst}
  \state{MA}
  \country{USA}}
\email{hajiesmaili@cs.umass.edu}

\author{Don Towsley}
\affiliation{%
  \institution{University of Massachusetts Amherst}
  \city{Amherst}
  \state{MA}
  \country{USA}}
\email{towsley@cs.umass.edu}









\renewcommand{\shortauthors}{}

\begin{abstract}
  The goal of quantum network tomography (QNT) is the characterization of internal quantum channels in a quantum network from external peripheral operations.
  Prior research has primarily focused on star networks featuring bit-flip and depolarizing channels, leaving the broader problem---such as QNT for networks with arbitrary topologies and general Pauli channels---largely unexplored. Moreover, establishing channel identifiability remains a significant challenge even in simplified quantum star networks.

  In the first part of this paper, we introduce a novel network tomography method, termed \mergecast, in quantum networks.
  We demonstrate that \mergecast, together with a progressive etching procedure, enables the unique identification of all internal quantum channels in networks characterized by arbitrary topologies and Pauli channels.
  As a side contribution, we introduce a subclass of Pauli channels, termed \emph{bypassable} Pauli channels, and propose a more efficient unicast-based tomography method, called \bypassunicast, for networks exclusively comprising these channels.
  In the second part, we extend our investigation to a more realistic QNT scenario that incorporates state preparation and measurement (SPAM) errors. We rigorously formulate SPAM errors in QNT, propose estimation protocols for such errors within QNT, and subsequently adapt our \mergecast approaches to handle networks affected by SPAM errors.
  Lastly, we conduct \texttt{NetSquid}-based simulations to corroborate the effectiveness of our proposed protocols in identifying internal quantum channels and estimating SPAM errors in quantum networks.
  In particular, we demonstrate that \mergecast maintains a good performance under realistic conditions, such as photon loss and quantum memory decoherence.
\end{abstract}



\maketitle

\section{Introduction}

Quantum networks (QNs) are pivotal to the realization of critical quantum technologies, including quantum key distribution~\citep{bennett1984update}, distributed quantum computing~\citep{wehner2018quantum}, and advanced quantum sensing applications~\citep{degen2017quantum}.
A QN can be regarded as a collection of quantum links connecting devices such as quantum repeaters~\citep{munro2015inside,azuma2023quantum}, switches~\citep{davidovich1993quantum,vardoyan2021exact,avis2023analysis}, and other network nodes.
While the quantum links have various implementations, e.g., transmitting quantum states directly~\citep{munro2012quantum}, or relying on pre-established entanglement and teleportation,
they are typically abstracted as \emph{quantum channels}---parametric formulas describing how quantum information evolves during transmission~\citep{wilde2013quantum}.

To fully harness the potential of QNs, it is essential to characterize the quantum channels that constitute the network infrastructure.
For individual quantum channels, extensive work has been done on estimating parameters when both ends of the channels are accessible~\citep{bennett1992communication,chen2022quantum,chen2024tight,nielsen2021gate,chuang1997prescription, magesan2011scalable,harper2020efficient}.
However, in real-world QNs, internal quantum channels may not be directly accessible, rendering standard approaches inadequate and highlighting the challenge of performing reliable characterization in complex, networked settings.
To address this, one needs to \emph{characterize internal quantum channels of a network solely through end-to-end measurements}---i.e., without direct access to internal nodes, and only through external nodes, called \emph{monitors} (see Figure~\ref{fig:qnt} as an example).

This task of estimating quantum channel parameters in QNs is known as \emph{Quantum Network Tomography} (QNT).
This concept draws a parallel to its classical analog, ``network tomography''~\citep{he2021network}.
It is vital for numerous networking applications. For instance, network path selection usually operates without access
to internal quantum devices (internal nodes), yet it depends on accurate assessments of internal channel properties to identify optimal paths~\citep{van2013path,wang2025learn, abane2025entanglement}.
The QNT task mirrors classical network tomography, which has been
instrumental in monitoring the performance of the internet by enabling the evaluation of internal network performance through external observations. Given the anticipated complexity and heterogeneity of future QNs, similar end-to-end characterization techniques are expected to play a crucial role in their advancement.

\noindent\textbf{Related Works.} To date, QNT is first studied by~\citet{de2022quantum,de2023characterization,de2024quantum},
where they have focused exclusively on simple star network topologies, with internal channels modeled only as bit-flip or depolarizing channels, leaving the more general setting of arbitrary network topologies and general quantum channels largely unexplored. In particular, previous research has addressed channel identifiability---the ability to uniquely determine network channel parameters, as defined in Definition 2.1 of~\citet{he2021network}---\emph{only for bit-flip channels (and, in certain special cases, depolarizing channels) within the star topology}. Finally, prior work~\cite{de2024quantum} emphasizes that establishing the QNT protocol with identifiability for general quantum channels across arbitrary network topologies remains a challenging \emph{open problem}.

\usetikzlibrary{shapes, positioning, decorations.pathreplacing}

\begin{figure}[tbp]
    \centering
    \resizebox{0.4\linewidth}{!}{%
        \begin{tikzpicture}
            \node[draw, circle, minimum size=6pt, inner sep=1pt, fill] (A1) at (0,1) {};
            \node[draw, circle, minimum size=6pt, inner sep=1pt, fill] (A2) at (0,-1) {};

            \node[draw, circle, minimum size=6pt, inner sep=1pt, fill] (B1) at (-1.732,2) {};
            \node[draw, circle, minimum size=6pt, inner sep=1pt, fill] (B2) at (-3.464,1) {};
            \node[draw, circle, minimum size=6pt, inner sep=1pt, fill] (B3) at (-3.464,-1) {};
            \node[draw, circle, minimum size=6pt, inner sep=1pt, fill] (B4) at (-1.732,-2) {};

            \node[draw, circle, minimum size=6pt, inner sep=1pt, fill] (C1) at (1.732,2) {};
            \node[draw, circle, minimum size=6pt, inner sep=1pt, fill] (C2) at (3.464,1) {};
            \node[draw, circle, minimum size=6pt, inner sep=1pt, fill] (C3) at (3.464,-1) {};
            \node[draw, circle, minimum size=6pt, inner sep=1pt, fill] (C4) at (1.732,-2) {};

            \node[draw, rectangle, minimum size=7pt, inner sep=1pt, fill=blue] (D1) at (-1.732,4) {};
            \node[draw, rectangle, minimum size=7pt, inner sep=1pt, fill=blue] (D2) at (-5.196,2) {};
            \node[draw, rectangle, minimum size=7pt, inner sep=1pt, fill=blue] (D3) at (-5.196,-2) {};
            \node[draw, rectangle, minimum size=7pt, inner sep=1pt, fill=blue] (D4) at (-1.732,-4) {};

            \node[draw, rectangle, minimum size=7pt, inner sep=1pt, fill=blue] (E1) at (1.732,4) {};
            \node[draw, rectangle, minimum size=7pt, inner sep=1pt, fill=blue] (E2) at (5.196,2) {};
            \node[draw, rectangle, minimum size=7pt, inner sep=1pt, fill=blue] (E3) at (5.196,-2) {};
            \node[draw, rectangle, minimum size=7pt, inner sep=1pt, fill=blue] (E4) at (1.732,-4) {};

            \draw[line width=2pt] (B1) -- node[draw, rectangle, midway, fill=white] {\(\mathcal{P}_{12}\)} (D1);
            \draw[line width=2pt] (B2) -- node[draw, rectangle, midway, fill=white] {\(\mathcal{P}_{13}\)} (D2);
            \draw[line width=2pt] (B3) -- node[draw, rectangle, midway, fill=white] {\(\mathcal{P}_{14}\)} (D3);
            \draw[line width=2pt] (B4) -- node[draw, rectangle, midway, fill=white] {\(\mathcal{P}_{15}\)} (D4);

            \draw[line width=2pt] (C1) -- node[draw, rectangle, midway, fill=white] {\(\mathcal{P}_{19}\)} (E1);
            \draw[line width=2pt] (C2) -- node[draw, rectangle, midway, fill=white] {\(\mathcal{P}_{18}\)} (E2);
            \draw[line width=2pt] (C3) -- node[draw, rectangle, midway, fill=white] {\(\mathcal{P}_{17}\)} (E3);
            \draw[line width=2pt] (C4) -- node[draw, rectangle, midway, fill=white] {\(\mathcal{P}_{16}\)} (E4);

            \draw[line width=2pt] (A1) -- node[draw, rectangle, midway, fill=white] {\(\mathcal{P}_2\)} (B1);
            \draw[line width=2pt] (B1) -- node[draw, rectangle, midway, fill=white] {\(\mathcal{P}_3\)} (B2);
            \draw[line width=2pt] (B2) -- node[draw, rectangle, midway, fill=white] {\(\mathcal{P}_4\)} (B3);
            \draw[line width=2pt] (B3) -- node[draw, rectangle, midway, fill=white] {\(\mathcal{P}_5\)} (B4);
            \draw[line width=2pt] (B4) -- node[draw, rectangle, midway, fill=white] {\(\mathcal{P}_6\)} (A2);

            \draw[line width=2pt] (A1) -- node[draw, rectangle, midway, fill=white] {\(\mathcal{P}_{11}\)} (C1);
            \draw[line width=2pt] (C1) -- node[draw, rectangle, midway, fill=white] {\(\mathcal{P}_{10}\)} (C2);
            \draw[line width=2pt] (C2) -- node[draw, rectangle, midway, fill=white] {\(\mathcal{P}_9\)} (C3);
            \draw[line width=2pt] (C3) -- node[draw, rectangle, midway, fill=white] {\(\mathcal{P}_{8}\)} (C4);
            \draw[line width=2pt] (C4) -- node[draw, rectangle, midway, fill=white] {\(\mathcal{P}_{7}\)} (A2);

            \draw[line width=2pt] (A1) -- node[draw, rectangle, midway, fill=white] {\(\mathcal{P}_1\)} (A2);

            \node[
                draw,
                cloud, cloud puffs=10,
                cloud ignores aspect,
                fill=cyan,
                fill opacity=0.4,
                draw=cyan,
                opacity=0.3,
                line width=0pt,
                inner xsep=90pt,
                inner ysep=70pt,
                midway
            ] at (0,0) {};

            \draw[->, purple, line width=2.5pt]
            ($(D1)+(0.4,0)$) -- ($(B1)+(0.4,0)$) ;

            \draw[->, purple, line width=2.5pt]
            ($(B1)+(0.4,0)$) -- ($(A1)+(0.4,0)$);

            \draw[->, purple, line width=2.5pt]
            ($(A1)+(0.4,0)$) -- ($(A2)+(0.4,0)$);
        \end{tikzpicture}
    }

    \caption{Quantum network tomography: We study the problem of inferring the parameters of the internal Pauli channels (inside the cloud) from the state preparation and measurement operations of peripheral nodes, called \emph{monitors} (square \textcolor{blue}{blue} nodes outside the cloud).
        The \textcolor{purple}{purple} arrow sequence across the channels \(\mathcal{P}_{12}\), \(\mathcal{P}_2\), and \(\mathcal{P}_1\) indicates the \emph{``progressive etching''} protocol detailed in Section~\ref{subsec:progressive-etching}.
    }
    \label{fig:qnt}
\end{figure}
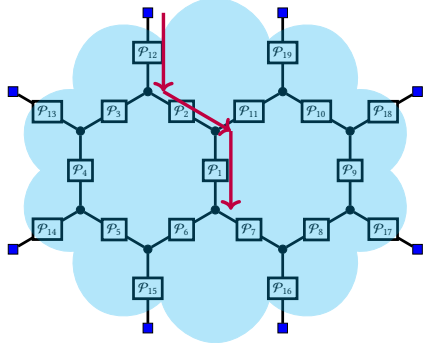

\noindent\textbf{Contributions.} In this paper, we study the QNT problem, depicted in Figure~\ref{fig:qnt}, in a general setting,
where the quantum network can have an arbitrary topology, internal channels are arbitrary Pauli channels (a typical and general class of quantum channels)~\citep{wilde2013quantum}, and state preparation and measurements are erroneous, all of which have not been considered in prior works~\citep{de2024quantum,de2023characterization,de2022quantum}.
We note that focusing only Pauli channels is not restrictive, as other noisy quantum channels can be transformed to Pauli channels via the Pauli twirling technique~\citep{bennett1996mixed,dur2005standard}.
Specifically, this paper proposes novel QNT approaches, addressing open problems raised in prior work~\citep{de2024quantum}
and tackling three key challenges in QNT:
(C1) identifiability of Pauli channels,
(C2) general topologies,
and (C3) the existence of state preparation and measurement (SPAM) errors, which are detailed as follows.


    {\em  (1)}
We propose a novel QNT procedure that \emph{uniquely} identifies all Pauli channels in a \emph{general network topology} (e.g., Figure~\ref{fig:qnt}) in Section~\ref{sec:qnt-identifiability}, addressing the challenges of both (C1) and (C2).
The procedure consists of a \mergecast protocol and a progressive etching process. \mergecast ``merges'' two quantum states generated by two end nodes via a controlled-NOT (CNOT) gate at an intermediate node and then sends the merged state to a third end node for measurement.
\mergecast can identify the parameters of the \emph{peripheral} channels---those connecting to \emph{monitors} (peripheral nodes, e.g., blue squares in Figure~\ref{fig:qnt}).
Our QNT procedure then performs a \emph{progressive etching} procedure to infer the \emph{internal} channels---whose end nodes are \emph{not} monitors---inside the network in a peripheral-to-internal manner  (e.g., along the purple arrow sequence in Figure~\ref{fig:qnt}).

This procedure pushes forward the frontier of QNT in three aspects: (1) it attains identifiability for general Pauli channels, (2) it works for arbitrary network topologies, and (3) it eliminates a previous assumption that all intermediate nodes must be monitors (i.e., with the ability to perform state preparations), while in prior works~\citep{de2024quantum,de2023characterization,de2022quantum}, only star networks consisting of bit-flip or depolarizing channels were considered under the assumption that states can be prepared at intermediate nodes.

    {\em (2)}
We propose
protocols
to account for state preparation and measurement (SPAM) errors in Section~\ref{sec:qnt-spam}, addressing challenge (C3).
These protocols are applicable in settings where monitors are not capable of \emph{both} state preparation and measurement and hence cannot estimate SPAM errors locally.
Specifically, we first devise two protocols to estimate the error parameters of state preparation and measurement, respectively, \emph{without knowledge of channels in the network}.
We then adapt \mergecast to identify the network channels once the SPAM error parameters are known.

    {\em (3)}
As a side contribution, we discover a subclass of Pauli channels, termed \emph{bypassable} Pauli channels (Definition~\ref{def:bypassable-channel}), which can be ``bypassed'' by certain input states.
When a quantum network consists solely of bypassable Pauli channels, one can utilize this property to ``single out'' each channel in a network and estimate its parameters directly (detailed in Section~\ref{subsec:bypassable-channel-protocol}), reducing QNT to standard quantum channel tomography.
We propose a unicast-based protocol, called \bypassunicast, utilizing this reduction to simplify the overall process for more efficient parameter estimation, and addresses all three challenges.

The paper is organized as follows.
In Section~\ref{sec:preliminary}, we provide preliminaries, including a formal definition of the quantum network tomography problem and introducing the Pauli-Liouville representation of quantum states and channels, which is convenient for later derivations.
In Section~\ref{sec:qnt-identifiability}, we study the identifiability of internal Pauli channels in a quantum network, and propose \mergecast and progressive etching
to identify quantum channels in a general quantum network.
In Section~\ref{sec:qnt-spam}, we account for state preparation and measurement (SPAM) errors and propose protocols to estimate SPAM parameters without knowledge of the network channel parameters.
Later in Section~\ref{sec:qnt-experiment}, we present experiments that demonstrate the effectiveness of our proposed protocols using the \texttt{NetSquid} simulator~\citep{coopmans2021netsquid}.
Finally, we conclude this paper in Section~\ref{sec:conclusion}.

\section{Preliminaries}\label{sec:preliminary}

This section provides background for the main content of this manuscript. We first formulate the quantum network tomography problem and then introduce the
Pauli-Liouville representation of quantum states and channels for convenience of later derivations.

\subsection{Quantum Network Tomography}


We represent a quantum network as a graph \(\cG = (\cE, \cV)\), where \(\cV\) represents a set of quantum devices (e.g., quantum repeaters) and \(\cE\) corresponds to the set of quantum channels (edges) connecting these devices (vertices) as illustrated in Figure~\ref{fig:qnt}.
The quantum channel is a model~\citep{wilde2013quantum} that describes how a quantum state changes when it is transmitted between two connected quantum devices through a noisy environment.
This paper focuses on Pauli channels (detailed in Section~\ref{subsubsec:Pauli-channel}), a class of quantum channels of common and practical interest in various quantum systems~\citep{wilde2013quantum}.

The quantum devices at the \emph{periphery} of the graph (square blue nodes in Figure~\ref{fig:qnt}) are assumed to be able to prepare and measure quantum states, called \emph{monitors}.
Such capabilities, if present at internal devices, are assumed not to be available for QNT, possibly because they are managed by a different party than that managing the monitors in the periphery.
This is the standard setting in classical network tomography~\citep{he2021network} as well.
While internal quantum devices cannot conduct SPAM, we assume that they support basic quantum gate operations, such as Hadamard and controlled-NOT gates.
The goal of QNT is to estimate (infer) parameters of the quantum channels (edges) through state preparation and measurement at the monitors in the network.



\emph{Unicast and Multicast.} Two classes of protocols have been proposed in classical network tomography~\citep{he2021network}: unicast and multicast.
A unicast protocol sends a signal from one monitor, through a designated path, to another monitor (which analyzes the received signal to infer the network parameters), whereas a multicast protocol sends a signal from one monitor to \emph{multiple} monitors. This is accomplished by duplicating the signal at intermediate nodes.
While unicast QNT protocols have not been explored,
\citet{de2022quantum,de2023characterization,de2024quantum} proposed multicast protocols for star quantum network tomography, which relies on intermediate nodes performing state preparation (i.e., preparing ancilla states) for ``duplicating'' the quantum state to mimic classical multicast.

In this paper, we propose a QNT protocol, {\mergecast}, that removes the need for state preparation at intermediate nodes as in prior works.
This protocol is based on the idea of \emph{``merging''} quantum states at intermediate nodes, relying on operations specific to quantum systems and thus is not an extension of known classical network tomography protocols.\footnote{The terminology \mergecast was used in the classical network, e.g.,~\citet{ghosh2008software}, referring to merge two data streams into one to optimize data transmission efficiency. This is different from our quantum \mergecast, which merges two quantum states into one state (discarding the other) via unitary operations.
}
Coupled with a unicast protocol, \mergecast can estimate channel parameters in an arbitrary network composed of Pauli channels.

\subsection{Pauli-Liouville (Superoperator) Representation}

The Pauli-Liouville representation~\citep{emerson2005scalable,lin2021independent} is an alternative to the standard Dirac (bra-ket) and unitary representations of quantum states and processes in the Pauli basis \(\mathfrak{P} \coloneqq \{I, X, Y, Z\}\), where \(X, Y, Z \in \mathbb{C}^{2\times 2}\) are Pauli matrices, and \(I\) is the \(2\)-by-\(2\) identity matrix.
We find it convenient for our later development of the QNT protocols and SPAM error formulation.

\subsubsection{Basic Quantum States}
Any single-qubit~\footnote{
    For simplicity, we mainly focus on single qubit states and processes in this paper, with some two-qubit operations when necessary.
    Both this representation and our results can be generalized to quantum states with any number of qubits.
}
quantum state (density matrix) \(\rho\) can be represented as a vector in the Pauli-Liouville representation via decomposition in the Pauli basis as follows,
\begin{align}
    \rho = \frac{1}{2} \sum_{P \in \mathfrak{P}} \Tr[P \rho] P
    = \frac{1}{2} \sum_{P \in \mathfrak{P}} x_P P,
\end{align}
where \(x_P \coloneqq \Tr[P \rho]\) is the Pauli coefficient of state \(\rho\) in the Pauli basis \(P \in \{I, X, Y, Z\}\).
Hence, the Pauli-Liouville representation of \(\rho\) is the following \(4\times 1\) vector,
\begin{align}
    \kett{\rho} \coloneqq [x_I, x_X, x_Y, x_Z]^T,
\end{align}
where \(\kett{\cdot}\) denotes the Pauli-Liouville representation (also known as \emph{superoperator}) of a quantum state, and \([\dots]^T\) denotes the transpose of a vector.
For example,
we have
\begin{align}
    \ketbra{0}{0}
    = \begin{bmatrix}
          1 & 0 \\
          0 & 0
      \end{bmatrix}
    = \frac{1}{2} \left( I + Z \right)
     & \Longrightarrow
    \kett{0} = [1, 0, 0, 1]^T,
    \\
    \ketbra{1}{1}
    = \begin{bmatrix}
          0 & 0 \\
          0 & 1
      \end{bmatrix}
    = \frac{1}{2} \left( I - Z \right)
     & \Longrightarrow
    \kett{1} = [1, 0, 0, -1]^T,
    \\
    \ketbra{+}{+}
    = \frac{1}{2}
    \begin{bmatrix}
        1 & 1 \\
        1 & 1
    \end{bmatrix}
    = \frac{1}{2} \left( I + X \right)
     & \Longrightarrow
    \kett{+} = [1, 1, 0, 0]^T.
\end{align}
Denote \(\bbra{\rho} \coloneqq \kett{\rho}^\dagger\) as the conjugate transpose. The probability of measuring a state \(\kett{\rho}\)  is
\begin{align}
    \mathbb{P}(\kett{0}) {=} \frac{1}{2}\bbra{\rho}\kett{0} {=} \frac{1 + x_Z}{2},\quad
    \mathbb{P}(\kett{1}) {=} \frac{1}{2}\bbra{\rho}\kett{1} {=} \frac{1 - x_Z}{2}.
\end{align}
\subsubsection{Pauli Transfer Matrix (PTM)}

A quantum process (linear map), e.g., quantum gates (unitary) and quantum channels (non-unitary), maps a quantum state (density matrix) to a quantum state.
This process can also be represented in the Pauli-Liouville space as a Pauli transfer matrix (PTM).
A single-qubit quantum process \(G: \mathbb{C}^{2\times 2} \to \mathbb{C}^{2\times 2}\) is represented by a \(4\times 4\) matrix $\cG$ as follows,
\begin{align}\label{eq:Pauli-transfer-matrix}
    \cG \coloneqq \frac{1}{2} \left[ \Tr[P {G}(Q)] \right]_{P, Q \in \mathfrak{P}}.
\end{align}
The entries of the PTM correspond to the transformation of the Pauli basis by quantum process \(G\).
For example, the Hadamard gate \(H\) is represented by the following PTM,
\begin{align}
    \mathcal{H}
    =\begin{bmatrix}
         1 & 0 & 0  & 0
         \\
         0 & 0 & 0  & 1
         \\
         0 & 0 & -1 & 0
         \\
         0 & 1 & 0  & 0
     \end{bmatrix},
    \text{ since }
    \begin{cases}
        HIH^\dagger & = I
        \\
        HXH^\dagger & = Z
        \\
        HYH^\dagger & = -Y
        \\
        HZH^\dagger & = X
    \end{cases},
\end{align}
where \(H^\dagger\) is the conjugate transpose of \(H\).
That is, the Hadamard gate swaps Pauli bases \(X\) and \(Z\) and flips the sign of \(Y\).

\subsubsection{Pauli Channels: Bypassable and Non-bypassable} \label{subsubsec:Pauli-channel}

In this paper, we focus on Pauli channels. Since any noise channel can be converted to a Pauli channel via the Pauli twirling technique~\citep{wallman2016noise}, this is a very general class of quantum channels.
A Pauli channel is defined on the Pauli basis as follows, \(\mathcal{P}(\rho) = (1-p_X - p_Y - p_Z)\rho + p_X X\rho X + p_Y Y\rho Y + p_Z Z\rho Z\), where \(p_X, p_Y, p_Z \in [0,1]\) are the probabilities of applying  Pauli matrices $X,Y,Z$ respectively to the input state \(\rho\).
Here, we slightly abuse the notation \(\mathcal{P}\) to denote both the Pauli channel and its Pauli transfer matrix.

It follows from~\eqref{eq:Pauli-transfer-matrix}, that the Pauli transfer matrix of a single-qubit Pauli channel is the following \(4\times 4\) diagonal matrix
\begin{align}
    \mathcal P   = \diag{1, q_X, q_Y, q_Z},
    \label{eq:Pauli-channel-parameters}
\end{align}
where \(q_X \coloneqq 1 - 2(p_Y + p_Z)\), \(q_Y \coloneqq 1 - 2(p_X + p_Z)\), and \(q_Z \coloneqq 1 - 2(p_X + p_Y)\).
Then, we have \(q_X, q_Y, q_Z \in [-1, 1]\).
Later on, we will see that the challenge of identifiability in QNT stems from the symmetric \([-1,1]\) range of \(q_X, q_Y, q_Z\).
For simplicity of presentation and without loss of generality, we assume all three parameters are non-zero in QNT.
\footnote{Because (1) if all three entries are zero, the channel becomes the depolarization with mixing probability \(1\), that is, across this channel, any state \(\rho\) would be mapped to the maximally mixed state \(\frac{1}{2}I\)---no information about the input state \(\rho\) after passing through the channel.
    This corresponds to the link failure in network tomography. One may adapt the link failure identification techniques~\citep{nguyen2004active,ahuja2008srlg} to solve the problem, which is not the focus of this paper.
    (2) If at least one of the three entries is non-zero, one can use the channel dressing technique later proposed in Figure~\ref{fig:channel-dressing} to locate the non-zero entries and then estimate them with the proposed protocols in this paper---reduce back to the all non-zero case.}

Next, we introduce the concepts of bypassable and non-bypassable Pauli channels.
\begin{definition}[Bypassable Pauli channel in Pauli basis]\label{def:bypassable-channel}
    A single-qubit Pauli channel \(\mathcal{P}\) is called \emph{bypassable} if there exists at least one non-identity Pauli operator \(P\in\mathfrak{P}\) such that \(\mathcal{P}(P) = P\);
    otherwise, it is \emph{non-bypassable}.
\end{definition}
Definition~\ref{def:bypassable-channel} immediately implies the following lemma, which we find more convenient for verifying whether or not a channel is bypassable (proof deferred to Appendix~\ref{app:bypassable-lemma-proof}).
\begin{lemma}\label{lem:bypassable-channel}
    A single-qubit Pauli channel is bypassable if and only if
    the diagonal of its Pauli transfer matrix has at least two entries equal to one.
\end{lemma}



For example, the bit-flip channel \(\mathcal{B}(\rho) \coloneqq (1-p)\rho + pX\rho X\), with flip probability \(p\in (0,1)\), is bypassable, since its PTM is \(\diag{1, 1, 1-2p, 1-2p}\) with three entries equal to one.
Note that sending the \(\kett{+}\) state through the bit-flip channel \(\mathcal{B}\), produces $\kett{+}$,
\begin{align}\label{eq:bypass-bit-flip-channel}
    \mathcal{B}\kett{+}
    =
    \begin{bmatrix}
        1 & 0 & 0    & 0
        \\
        0 & 1 & 0    & 0
        \\
        0 & 0 & 1-2p & 0
        \\
        0 & 0 & 0    & 1-2p
    \end{bmatrix}
    \cdot
    \begin{bmatrix}
        1 \\
        1 \\
        0 \\
        0
    \end{bmatrix}
    =
    \begin{bmatrix}
        1 \\
        1 \\
        0 \\
        0
    \end{bmatrix}
    = \kett{+}.
\end{align}
The input state \(\kett{+}\) is unchanged after passing through the bit-flip channel \(\mathcal{B}\); hence, the name ``bypassable.''
Other common bypassable channels include the dephasing (phase-flip) channel \(\mathcal{F}(\rho) \coloneqq (1-p)\rho + p Z \rho Z\), and
the bit-phase-flip channel \(\mathcal{E}(\rho) \coloneqq (1-p)\rho + p Y \rho Y\), whose PTMs are \(\diag{1, 1 -2p, 1-2p, 1}\) and \(\diag{1, 1 -2p, 1, 1-2p}\), respectively.
One typical non-bypassable channel is the depolarization channel, expressed as \(
\mathcal{D}(\rho) \coloneqq (1-p)\rho + \frac{p}{4}\left(I\rho I + X\rho X + Y\rho Y + Z\rho Z \right)
\), with mixing probability \(p\in (0,1)\), and its PTM is \(\diag{1, 1-p, 1-p, 1-p}\).
In the next section, we propose QNT protocols to identify networks composed of arbitrary Pauli channels.
Especially, for networks composed only of bypassable channels, we present more efficient protocols (Section~\ref{subsec:bypassable-channel-protocol}).


\section{QNT Protocols for General Topologies}\label{sec:qnt-identifiability}

In this section, we propose QNT protocols that identify all channel parameters in an arbitrary network.
We begin with a discussion of the challenges that identifiability introduces to quantum network tomography (QNT) and explain why existing protocols from classical network tomography cannot be directly applied to QNT.
Then, we introduce the \mergecast protocol for general Pauli channels in Section~\ref{subsec:mergecast}. In Section~\ref{subsec:progressive-etching} we extend \mergecast to a general network topology
by introducing a progressive etching protocol to identify all channels in the network.
Lastly, we propose the \bypassunicast protocol for a special class of quantum networks composed exclusively by bypassable channels (Definition~\ref{def:bypassable-channel}) that is more efficient than \mergecast.

\begin{figure}
    \centering
    \begin{subfigure}[b]{0.24\textwidth}
        \centering
        \resizebox{\textwidth}{!}{%
            \begin{tikzpicture}
                \node[draw, circle, minimum size=6pt, inner sep=1pt, fill] (A) at (0,0) {};
                \node[draw, rectangle, minimum size=7pt, inner sep=1pt, fill=blue]  (B) at (0,2) {};
                \node[draw, rectangle, minimum size=7pt, inner sep=1pt, fill=blue]  (C) at (1.732,-1) {};
                \node[draw, rectangle, minimum size=7pt, inner sep=1pt, fill=blue]  (D) at (-1.732,-1) {};

                \draw[line width=1.5pt] (A) -- node[draw, rectangle, midway, fill=white] {\(\mathcal{P}_{1}\)} (B);
                \draw[line width=1.5pt] (A) -- node[draw, rectangle, midway, fill=white] {\(\mathcal{P}_{3}\)} (C);
                \draw[line width=1.5pt] (A) -- node[draw, rectangle, midway, fill=white] {\(\mathcal{P}_{2}\)} (D);

                \draw[dashed, line width=1pt] (-0.3, 1.7) rectangle (0.3, 2.3);
                \node at (0.5, 1.5) {\nodeAlice};

                \draw[dashed, line width=1pt] (-0.3, -0.3) rectangle (0.3, 0.3);
                \node at (0, -0.6) {\nodeCharlie};


                \draw[dashed, line width=1pt] (-2.032, -1.3) rectangle (-1.432, -0.7);
                \node at (-1.732, -1.5) {\(\nodeBob\)};
            \end{tikzpicture}
        }
        \caption{\(3\)-link star network}
        \label{subfig:star}
    \end{subfigure}
    \hspace{0.1\textwidth}
    \begin{subfigure}[b]{0.4\textwidth}
        \centering
        \resizebox{\textwidth}{!}{%
            \begin{quantikz}
                \inputD{\text{SP}}\slice{\(\begin{bmatrix}
                        1 \\0\\0\\1
                    \end{bmatrix}\)}\gategroup[1,steps=1,style={dashed,rounded corners,fill=blue!20, inner xsep=2pt},background,label style={label position=below,anchor=north,yshift=-0.2cm}]{\nodeAlice}
                & \gate{\textcolor{blue}{\mathcal{P}_1}}
                \slice{\(\begin{bmatrix}
                        1 \\0\\0\\q_{Z,1}
                    \end{bmatrix}\)}
                & \ctrl{0}\gategroup[1,steps=1,style={dashed,rounded corners,fill=blue!20, inner xsep=2pt},background,label style={label position=below,anchor=north,yshift=-0.2cm}]{\nodeCharlie}
                & \gate{\textcolor{blue}{\mathcal{P}_2}}
                \slice{\(\begin{bmatrix}
                        1 \\0\\0\\q_{Z,1}q_{Z,2}
                    \end{bmatrix}\)}
                & \meter{}\gategroup[1,steps=1,style={dashed,rounded corners,fill=blue!20, inner xsep=2pt},background,label style={label position=below,anchor=north,yshift=-0.2cm}]{\nodeBob}
                & \setwiretype{c}\rstick[1]{
                } \end{quantikz}
        }
        \caption{Unicast for \(\mathcal{P}_1\) and \(\mathcal{P}_2\)}
        \label{fig:simple-unicast}
    \end{subfigure}
    \caption{Unicast protocol in a \(3\)-link star network. Figure~\ref{fig:simple-unicast} illustrates the unicast protocol for two channels \(\mathcal{P}_1\) and \(\mathcal{P}_2\) in the star network in Figure~\ref{subfig:star}.
        Symmetrically, one can apply the same unicast protocol to the other two pairs of channels \(\{\mathcal{P}_2, \mathcal{P}_3\}\) and \(\{\mathcal{P}_1, \mathcal{P}_3\}\).
        However, this approach leads to a sign ambiguity (i.e., not identifiable) in estimating the channel parameters, as discussed in Section~\ref{sec:qnt-identifiability}.
    }
    \label{fig:unicast-for-star}
\end{figure}
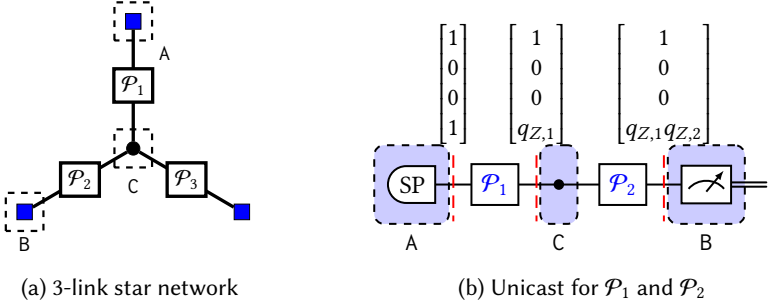


\noindent\textbf{Challenge of Identifiability.}
Below, we take unicast as an example to illustrate the identifiability challenge of QNT in a star network.\footnote{This unicast example is mainly for demonstrating the challenge of identifiability in QNT, and one may devise other variants of unicast that can address the challenge, e.g., the variant presented in Section~\ref{subsec:bypassable-channel-protocol}.}
Figure~\ref{fig:simple-unicast} presents how unicast in this example estimates the parameters of
two channels of a 3-link star:
Prepare quantum state \(\kett{0}=[1,0,0,1]^T\) at node \nodeAlice (SP for State Preparation),
send the state across the first Pauli channel \(\mathcal{P}_1\), node \nodeCharlie, and the second Pauli channel \(\mathcal{P}_2\),
receive and measure the state in the computational (\(Z\)) basis at node \nodeBob.\footnote{This paper follows the convention of labeling nodes as \nodeAlice, \nodeBob, \nodeCharlie in quantum communications, where \nodeAlice for ``Alice'' represents the sender, \nodeBob for ``Bob'' represents the receiver, and \nodeCharlie for ``Charlie'' represents the intermediate node.}~\footnote{Whenever a protocol is picked, all nodes (including the monitors at the peripheral) involved are notified by classical communications and follow the prescribed steps accordingly.}

As illustrated in Figure~\ref{fig:simple-unicast}, the final received quantum state is \(\kett{\rho} = [1, 0, 0, q_{Z,1} q_{Z,2}]^T\), where \(q_{Z,1}\) and \(q_{Z,2}\) are the last diagonal entries of the PTMs of two Pauli channels as defined in~\eqref{eq:Pauli-channel-parameters}.
Then, the measurement outputs a \(\kett{0}\) with probability \(\mathbb{P}(\kett{0})=\frac{1 + q_{Z,1} q_{Z,2}}{2}\).
By conducting the unicast protocol multiple times and taking the average of the measurement results, one can estimate \(\mathbb{P}(\kett{0})\), the probability that $\kett{0}$ is measured from which one can estimate the value of \(q_{Z,1}q_{Z,2}\).

Three Pauli channels \(\mathcal{P}_1, \mathcal{P}_2,\) and \(\mathcal{P}_3\), connecting at the same node \nodeCharlie, forms a simple star network.
By applying the unicast protocol in Figure~\ref{fig:simple-unicast} to the three pairs \(\{\mathcal{P}_1, \mathcal{P}_2\}\), \(\{\mathcal{P}_2, \mathcal{P}_3\}\), and \(\{\mathcal{P}_1, \mathcal{P}_3\}\),
one can estimate the values of \(q_{Z,1}q_{Z,2}\), \(q_{Z,2}q_{Z,3}\), and \(q_{Z,1}q_{Z,3}\).
From these three values, one obtains two solutions for the channel parameters \(\{q_{Z,1}, q_{Z,2}, q_{Z,3}\}\) and \(\{-q_{Z,1}, - q_{Z,2}, - q_{Z,3}\}\)---notice \(q_{Z,i} = 1 - 2(p_{X,i} + p_{Y,i}) \in [-1, 1]\) can be either negative or positive.
That is, the final solutions have a \emph{sign ambiguity}, and one cannot uniquely identify the channel parameters via unicast, even in a simple star network.
This example illustrates the ambiguity for coefficients \(q_Z\).
For an arbitrary Pauli channel in~\eqref{eq:Pauli-channel-parameters}, this ambiguity also appears for the estimation of the other two coefficients \(q_X\) and \(q_Y\), yielding \(8\) sets of possible solutions \(\{\pm q_X, \pm q_Y, \pm q_Z\}\)  at last.
Similar identifiability issues exist for multicast protocols in star quantum network tomography~\citep{de2024quantum}.


This sign ambiguity arises from the symmetric range \([-1,1]\) and multiplicative relations of the Pauli channel coefficients \(q_{X,i}, q_{Y,i}, q_{Z,i}\). In contrast, classical network tomography benefits from non-negative channel coefficients and its typically additive relation, which inherently avoids such sign ambiguities. Consequently, classical protocols---such as unicast and multicast---are inapplicable in quantum settings, as they do not account for the extra layer of ambiguity present in quantum networks. Addressing this issue, therefore, requires the development of novel protocols tailored to ensure parameter identifiability in QNT.



\subsection{\mergecast: QNT with Identifiability for Any Pauli Channels} \label{subsec:mergecast}


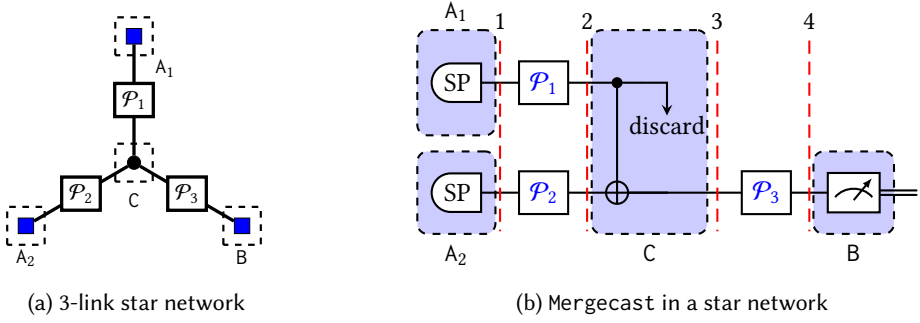
\begin{figure}
    \centering
    \begin{subfigure}[b]{0.25\textwidth}
        \centering
        \resizebox{\textwidth}{!}{%
            \begin{tikzpicture}
                \node[draw, circle, minimum size=6pt, inner sep=1pt, fill] (A) at (0,0) {};
                \node[draw, rectangle, minimum size=7pt, inner sep=1pt, fill=blue]  (B) at (0,2) {};
                \node[draw, rectangle, minimum size=7pt, inner sep=1pt, fill=blue]  (C) at (1.732,-1) {};
                \node[draw, rectangle, minimum size=7pt, inner sep=1pt, fill=blue]  (D) at (-1.732,-1) {};

                \draw[line width=1.5pt] (A) -- node[draw, rectangle, midway, fill=white] {\(\mathcal{P}_{1}\)} (B);
                \draw[line width=1.5pt] (A) -- node[draw, rectangle, midway, fill=white] {\(\mathcal{P}_{3}\)} (C);
                \draw[line width=1.5pt] (A) -- node[draw, rectangle, midway, fill=white] {\(\mathcal{P}_{2}\)} (D);

                \draw[dashed, line width=1pt] (-0.3, 1.7) rectangle (0.3, 2.3);
                \node at (0.5, 1.5) {\(\nodeAlice_1\)};

                \draw[dashed, line width=1pt] (-0.3, -0.3) rectangle (0.3, 0.3);
                \node at (0, -0.6) {\nodeCharlie};

                \draw[dashed, line width=1pt] (1.432, -1.3) rectangle (2.032, -0.7);
                \node at (1.732, -1.5) {\nodeBob};

                \draw[dashed, line width=1pt] (-2.032, -1.3) rectangle (-1.432, -0.7);
                \node at (-1.732, -1.5) {\(\nodeAlice_2\)};
            \end{tikzpicture}
        }
        \caption{\(3\)-link star network}
        \label{subfig:star-mergecast}
    \end{subfigure}
    \hspace{0.1\textwidth}
    \begin{subfigure}[b]{0.55\textwidth}
        \centering
        \resizebox{\textwidth}{!}{%
            \begin{quantikz}
                & \inputD{\text{SP}} \slice{1}\gategroup[1,steps=1,style={dashed,rounded corners,fill=blue!20, inner xsep=2pt},background,label style={label position=above,anchor=north,yshift=0.3cm}]{\(\nodeAlice_1\)}
                & \gate{\textcolor{blue}{\mathcal{P}_1}}\slice{2}
                &  \ctrl{1}\gategroup[2,steps=2,style={dashed,rounded corners,fill=blue!20, inner xsep=2pt},background,label style={label position=below,anchor=north,yshift=-0.2cm}]{\nodeCharlie}
                &[-0.5cm] \trash{\text{discard}}\slice{3}
                & \wireoverride{n}
                & \wireoverride{n}
                &  \wireoverride{n}
                &  \wireoverride{n}
                \\
                &\inputD{\text{SP}}\gategroup[1,steps=1,style={dashed,rounded corners,fill=blue!20, inner xsep=2pt},background,label style={label position=below,anchor=north,yshift=-0.2cm}]{\(\nodeAlice_2\)}
                &\gate{\textcolor{blue}{\mathcal{P}_2}}
                & \targ{}
                & \qw
                & \gate{\textcolor{blue}{\mathcal{P}_3}}\slice{4}
                & \meter{}\gategroup[1,steps=1,style={dashed,rounded corners,fill=blue!20, inner xsep=2pt},background,label style={label position=below,anchor=north,yshift=-0.2cm}]{\nodeBob}
                & \setwiretype{c}
            \end{quantikz}
        }
        \caption{\mergecast in a star network}
        \label{fig:mergecast}
    \end{subfigure}
    \caption{\mergecast in a \(3\)-link star network, where node \nodeCharlie is in the center of the star topology, and nodes \(\nodeAlice_1\), \(\nodeAlice_2\), and \nodeBob are the three peripheral nodes.
        This protocol allows unambiguous identification of all three channels \(\mathcal{P}_1\), \(\mathcal{P}_2\), and \(\mathcal{P}_3\) without sign ambiguity.
    }
    \label{fig:mergecast-for-star}
\end{figure}

To overcome the identifiability challenge, we propose \mergecast, a QNT protocol designed to uniquely identify general Pauli channels in star topologies.
Later, we generalize \mergecast to arbitrary network topologies in Section~\ref{subsec:progressive-etching}.
Figure~\ref{fig:mergecast} illustrates \mergecast for a star network composed of three arbitrary Pauli channels \(\mathcal{P}_1\), \(\mathcal{P}_2\), and \(\mathcal{P}_3\). In this protocol:
\begin{itemize}
    \item Nodes \(\nodeAlice_1\) and \(\nodeAlice_2\) prepare \(\kett{0}\) states and transmit them via \(\mathcal{P}_1\) and \(\mathcal{P}_2\) to central node \nodeCharlie.
    \item Node \nodeCharlie entangles the incoming qubits using a CNOT gate, discards (traces out) the qubit from channel \(\mathcal{P}_1\),\footnote{The discarding operation can be easily implemented via resetting the qubit in practice, e.g., dissipation engineering~\citep{huber2025parametric,watanabe2025nondemolition,kim2025fast}.} and relays the remaining qubit via \(\mathcal{P}_3\) to node \nodeBob.
    \item Node \nodeBob measures the received qubit in the computational (\(Z\)) basis, yielding a \(\ket{0}\) outcome with probability
          \(
          p_{\mergecast} = \frac{1 + q_{Z,1}\,q_{Z,2}\,q_{Z,3}}{2},
          \)
          where \(q_{Z,i}\) are the \(Z\)-basis diagonal entries of the Pauli transfer matrices (PTMs) of \(\mathcal{P}_i\) for \(i=1,2,3\) (see Appendix~\ref{app:mergecast} for derivation).
\end{itemize}

By comparison, applying the simple unicast protocol (Figure~\ref{fig:simple-unicast}) to the channel pair \(\{\mathcal{P}_2, \mathcal{P}_3\}\) yields a \(\ket{0}\) measurement probability of
\(
p_{\text{Unicast}} = \frac{1 + q_{Z,2}\,q_{Z,3}}{2}.
\)
Notice that one can unambiguously isolate
\(
q_{Z,1} = \frac{q_{Z,1}\,q_{Z,2}\,q_{Z,3}}{q_{Z,2}\,q_{Z,3}}.
\)
Consequently, we devise an estimator for \(q_{Z,1}\) as follows,
\begin{align}\label{eq:mergecast-estimator}
    \hat{q}_{Z,1} \coloneqq \frac{2 \, \hat p_{\mergecast} - 1}{2 \, \hat p_{\text{Unicast}} - 1}
\end{align}
attaining unambiguous identification of the first channel,
where \(\hat p_{\mergecast}\) and \(\hat p_{\text{Unicast}}\) are the estimated probabilities from \mergecast and unicast, respectively.
Similarly, applying unicast to pairs \(\{\mathcal{P}_1, \mathcal{P}_2\}\) and \(\{\mathcal{P}_1, \mathcal{P}_3\}\) enables the determination of \(q_{Z,3}\) and \(q_{Z,2}\), respectively.




\noindent\textbf{Remarks.}
\mergecast\ differs fundamentally from traditional unicast and multicast QNT protocols by employing a CNOT-based entanglement step followed by a strategic merge-and-trace operation. This approach effectively consolidates channel effects into a single quantum state prior to measurement, representing an innovation unique to our QNT methodology.

Notably, \mergecast\ eliminates the need for the intermediate node \nodeCharlie to perform state preparation---a task typically inaccessible to end users, which is the primary focus of network tomography. This design choice significantly simplifies both the implementation and practical usability of the protocol. In contrast, quantum multicast protocols, such as those presented in \citep{de2023characterization,de2024quantum}, require intermediate nodes to engage in state preparations.

\mergecast\ does rely on the synchronization of the two root nodes, \(\nodeAlice_1\) and \(\nodeAlice_2\), to ensure the simultaneous arrival of their respective qubits at the intermediate node, \nodeCharlie, for the merging process. However, achieving such synchronization may be challenging in practical scenarios due to relatively high qubit loss rates, such as photon loss in optical fibers.
In Section~\ref{subsec:photon-loss}, we present experimental evaluations of \mergecast's performance in the presence of photon loss. Our findings demonstrate that, when a quantum memory (with decoherence) is employed at node \nodeCharlie to store the first arrived qubit, \mergecast\ maintains its effectiveness even in the presence of moderate photon loss.



\begin{figure}[tb]
    \centering
    \resizebox{0.65\textwidth}{!}{%
        \begin{quantikz}
            &\gate{\mathcal{P}}
            &
            &\setwiretype{n} \rstick{\({\mathcal{P}}=\text{diag}(1,q_X,q_Y,q_Z)\)}
            \\
            \gate{\mathcal{H}}
            & \gate{\mathcal{P}}
            & \gate{\mathcal{H}}
            &\setwiretype{n} \rstick{\(\mathcal{H}{\mathcal{P}}\mathcal{H}=\text{diag}(1,q_Z,q_Y,q_X)\)}
            \\
            \gate{\mathcal{H}\mathcal{S}}
            & \gate{\mathcal{P}}
            & \gate{\mathcal{S}^\dagger\mathcal{H}}
            &\setwiretype{n} \rstick{\(\mathcal{H}\mathcal{S}{\mathcal{P}}\mathcal{S}^\dagger\mathcal{H}=\text{diag}(1,q_Z,q_X,q_Y)\)}
        \end{quantikz}}
    \caption{Channel dressing for Pauli channels: the dressings in second and third rows swap the positions of the Pauli parameters such that \(q_X\) and \(q_Y\) appears at the position of \(q_Z\) in the first row, respectively.}\label{fig:channel-dressing}
\end{figure}
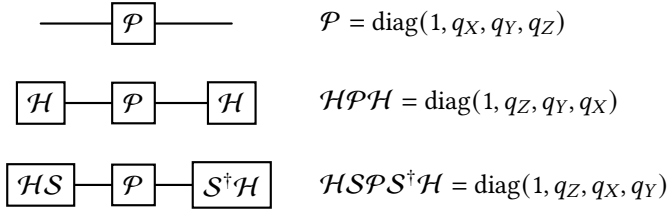

\noindent\textbf{Channel dressing for estimating other Pauli parameters.}
A single qubit Pauli channel can be described by the diagonal matrix \(\diag{1, q_X, q_Y, q_Z}\), where \(q_Z\) can be estimated via \mergecast in Figure~\ref{fig:mergecast}. To estimate the other two parameters \(q_X\) and \(q_Y\), we propose a ``channel dressing'' technique illustrated in Figure~\ref{fig:channel-dressing}.
Specifically, by applying a Hadamard gate \(H\) before and after the channel \(\mathcal{P}\) (i.e., \emph{``dressing''} channel \(\mathcal{P}\) with two Hadamard gates \(H\)), one obtains a new channel \(\mathcal{H}\mathcal{P}\mathcal{H} = \text{diag}(1,q_Z,q_Y,q_X)\),
which swaps \(q_X\) with \(q_Z\) in the diagonal matrix.
Similarly, by replacing the Hadamard gate with
the composition of phase gate and Hadamard gate, i.e., \(\mathcal{H}\mathcal{S}\),
one obtains \((\mathcal{H}\mathcal{S})\mathcal{P}(\mathcal{H}\mathcal{S})^\dagger = \mathcal{H}\mathcal{S}\mathcal{P}\mathcal{S}^\dagger\mathcal{H} = \text{diag}(1,q_Z,q_X, q_Y)\),
which swaps \(q_Y\) with \(q_Z\).
After dressing the channel, one can apply \mergecast to estimate the ``new'' parameter \(q_Z'\) (which is actually \(q_X\) or \(q_Y\)) and then identify the other two parameters \(q_X\) and \(q_Y\) of the Pauli channel.

This {channel dressing} technique can be implemented more economically by performing an appropriate basis transformation.
Take, for instance, the Hadamard dressing used in \mergecast{} to estimate the \(q_Y\) parameters.
Instead of placing Hadamard gates before and after every channel, one may initialize the qubits directly in the \(Y\) basis, \(\{\kett{+},\kett{-}\}\), replace the CNOT gate (originally meant to be sandwiched by Hadamard gates) with a CNOT whose control and target qubits are swapped, and also change the final measurement to the \(Y\) basis.  This modification attains the same dressing effect and can be viewed simply as a change of basis applied at the input state.

\subsection{Progressive Etching: QNT for General Topology}\label{subsec:progressive-etching}

This subsection proposes the \emph{``progressive etching''} procedure for QNT in general topologies.
For ease of presentation, we assume all nodes in the network have either degree one (peripheral) or degree greater than two.
When degree two nodes are present, each one is removed and the two adjacent channels are replaced by a single channel reflecting their composite behavior. This produces a network with a desired topology,
a standard practice in classical network tomography~\citep{he2021network} (details in Appendix~\ref{app:degree-2-nodes}).
Below, we present the details of progressive etching. Figure~\ref{fig:toy-example-general-tomograph},  as an example, illustrates its two key steps:
a) identifying peripheral channels via a generalized \mergecast protocol (Figure~\ref{fig:mergecast-peripheral}), and
b) etching into intermediate channels.

\tikzset{
    noisy/.style={
            starburst,
            fill=yellow,
            draw=red,
            line width=2pt,
            inner xsep=4pt,  
            inner ysep=5pt
        },
    cloudy/.style={
            cloud,
            cloud puffs=10,
            cloud ignores aspect,
            fill=cyan!50!white,
            draw=black,
            line width=1pt,
            inner xsep=-7pt,
            inner ysep=-1pt
        }
}

\begin{figure}[htbp]
    \centering
    \begin{subfigure}[b]{0.575\textwidth}
        \resizebox{\textwidth}{!}{
            \begin{quantikz}
                & \inputD{\text{SP}} \gategroup[1,steps=1,style={dashed,rounded corners,fill=blue!20, inner xsep=2pt},background,label style={label position=above,anchor=north,yshift=0.3cm}]{\(\nodeAlice_1\)}
                &[-0.15cm] \gate{\textcolor{blue}{\mathcal{P}_1}}
                & [-0.15cm] \ctrl{1}\gategroup[2,steps=2,style={dashed,rounded corners,fill=blue!20, inner xsep=2pt},background,label style={label position=below,anchor=north,yshift=-0.2cm}]{\nodeCharlie}
                &[-0.6cm] \trash{\text{discard}}
                &[-0.25cm] \wireoverride{n}
                &[-0.1cm] \wireoverride{n}
                & \wireoverride{n}
                &  \wireoverride{n}
                &  \wireoverride{n}
                \\
                &\inputD{\text{SP}}\gategroup[1,steps=1,style={dashed,rounded corners,fill=blue!20, inner xsep=2pt},background,label style={label position=below,anchor=north,yshift=-0.2cm}]{\(\nodeAlice_2\)}
                &
                \gate[style={cloudy},label style=blue]{\{\mathcal{P}',\dots\}_2}
                & \targ{}
                & \qw
                &
                \gate[style={cloudy},label style=blue]{\{\mathcal{P}'',\dots\}_3}
                & \meter{}\gategroup[1,steps=1,style={dashed,rounded corners,fill=blue!20, inner xsep=2pt},background,label style={label position=below,anchor=north,yshift=-0.2cm}]{\nodeBob}
                & \setwiretype{c}
            \end{quantikz}
        }
        \caption{Generalized \mergecast (for channel \(\mathcal{P}_1\))}
        \label{subfig:mergecast-peripheral}
    \end{subfigure}
    \begin{subfigure}[b]{0.375\textwidth}
        \resizebox{\linewidth}{!}{
            \begin{tikzpicture}
                \node[draw, circle, minimum size=6pt, inner sep=1pt, fill] (A1) at (0,1) {};
                \node[draw, circle, minimum size=6pt, inner sep=1pt, fill] (A2) at (0,-1) {};

                \node[draw, circle, minimum size=6pt, inner sep=1pt, fill] (B1) at (-1.732,2) {};
                \node[draw, circle, minimum size=6pt, inner sep=1pt, fill] (B2) at (-3.464,1) {};
                \node[draw, circle, minimum size=6pt, inner sep=1pt, fill] (B3) at (-3.464,-1) {};

                \node[draw, circle, minimum size=6pt, inner sep=1pt, fill] (C1) at (1.732,2) {};
                \node[draw, circle, minimum size=6pt, inner sep=1pt, fill] (C2) at (3.464,1) {};

                \node[draw, rectangle, minimum size=7pt, inner sep=1pt, fill=blue] (D1) at (-1.732,4) {};
                \node[draw, rectangle, minimum size=7pt, inner sep=1pt, fill=blue] (D2) at (-5.196,2) {};

                \node[draw, rectangle, minimum size=7pt, inner sep=1pt, fill=blue] (E1) at (1.732,4) {};

                \draw[line width=2pt] (B1) -- node[draw, rectangle, midway, fill=white] {\(\mathcal{P}_{12}\)} (D1);
                \draw[line width=2pt] (B2) -- node[draw, rectangle, midway, fill=white] {\(\mathcal{P}_{13}\)} (D2);

                \draw[line width=2pt] (C1) -- node[draw, rectangle, midway, fill=white] {\(\mathcal{P}_{19}\)} (E1);

                \draw[line width=2pt] (A1) -- node[draw, rectangle, midway, fill=white] {\(\mathcal{P}_2\)} (B1);
                \draw[line width=2pt] (B1) -- node[draw, rectangle, midway, fill=white] {\(\mathcal{P}_3\)} (B2);
                \draw[line width=2pt] (B2) -- node[draw, rectangle, midway, fill=white] {\(\mathcal{P}_4\)} (B3);

                \draw[line width=2pt] (A1) -- node[draw, rectangle, midway, fill=white] {\(\mathcal{P}_{11}\)} (C1);
                \draw[line width=2pt] (C1) -- node[draw, rectangle, midway, fill=white] {\(\mathcal{P}_{10}\)} (C2);

                \draw[line width=2pt] (A1) -- node[draw, rectangle, midway, fill=white] {\(\mathcal{P}_1\)} (A2);




                \draw[dashed, line width=1pt] (-2.032, 3.7) rectangle (-1.432, 4.3);
                \node at (-2.2, 3.5) {\LARGE \(\nodeAlice_1\)};

                \draw[dashed, line width=1pt] (-5.496, 1.7) rectangle (-4.896, 2.3);
                \node at (-5.5, 1.5) {\LARGE \(\nodeAlice_2\)};

                \draw[dashed, line width=1pt] (1.432, 3.7) rectangle (2.032, 4.3);
                \node at (1.3, 3.5) {\LARGE \(\nodeBob\)};

                \draw[dashed, line width=1pt] (-2.032, 1.7) rectangle (-1.432, 2.3);
                \node at (-1.732, 1.4) {\LARGE \(\nodeCharlie\)};

                \draw[->, red, line width=4pt, opacity=0.5]
                ($(D1)$) -- ($(B1)$) ;

                \draw[-, blue, line width=4pt, opacity=0.5] ($(D2)$) -- ($(B2)$);
                \draw[->, blue, line width=4pt, opacity=0.5] ($(B2)$) -- ($(B1)$);

                \draw[-, green, line width=4pt, opacity=0.5] ($(B1)$) -- ($(A1)$);
                \draw[-, green, line width=4pt, opacity=0.5] ($(A1)$) -- ($(C1)$);
                \draw[->, green, line width=4pt, opacity=0.5] ($(C1)$) -- ($(E1)$);
            \end{tikzpicture}
        }
        \caption{Segment of Figure~\ref{fig:qnt} (for channel \(\mathcal{P}_{12}\))}\label{subfig:general-mergecast-example}
    \end{subfigure}
    \caption{\mergecast for peripheral channel tomography:
        Figure~\ref{subfig:mergecast-peripheral} shows the generalized \mergecast protocol to estimate the parameter of a peripheral channel \(\mathcal{P}_1\) in a general network topology.
        Figure~\ref{subfig:general-mergecast-example} is an example of applying \mergecast to a general network.
        Node labels \(\nodeAlice_1\), \(\nodeAlice_2\), \(\nodeBob\), and \(\nodeCharlie\) are the same across both subfigures.
    }\label{fig:mergecast-peripheral}
\end{figure}

\noindent\textbf{Initialize from peripheral channels.}
We generalize \mergecast to estimate the parameters of peripheral channels in a general topology.
Recall that in the original \mergecast in Figure~\ref{fig:mergecast}, three peripheral channels connect at the intermediate node \(\nodeCharlie\).
The generalization of \mergecast only needs one peripheral channel connecting to the intermediate node \(\nodeCharlie\), while the other two can be paths (sets of quantum channels).
For example, in Figure~\ref{fig:mergecast-peripheral}, the
internal node \(\nodeCharlie\) connects to
node \(\nodeAlice_1\) through a single target channel \(\mathcal{P}_1\)
and to the other two end nodes \(\nodeAlice_2\) and \(\nodeBob\) through two paths---two sets of channels  \(\{\mathcal{P}',\dots\}_2\) and \(\{\mathcal{P}'',\dots\}_3\).
Applying \mergecast can estimate the parameters of the single channel \(\mathcal{P}_1\).
Take a segment of the topology in Figure~\ref{fig:qnt} as an example, illustrated in Figure~\ref{subfig:general-mergecast-example}.
If the target channel \(\mathcal{P}_1\) here corresponds to the peripheral channel \(\mathcal{P}_{12}\),
then the two sets of channels \(\{\mathcal{P}',\dots\}_2\) and \(\{\mathcal{P}'',\dots\}_3\) correspond to the two branches starting from the non-peripheral end node of channel \(\mathcal{P}_{12}\),
e.g., \(\{\mathcal{P}_{13}, \mathcal{P}_{3}\}\)
and \(\{\mathcal{P}_{2}, \mathcal{P}_{11}, \mathcal{P}_{19}\}\).

Then, we apply \mergecast to estimate the parameter \(q_{Z,1} Q_{Z,2} Q_{Z,3}\), where \(Q_{Z,i} \coloneqq \prod_{\ell: \mathcal{P}_\ell \in \{\mathcal{P}',\dots\}_i} q_{Z,\ell}\)
is the product of all the \(q_Z\) parameters of the channels in the sets \(\{\mathcal{P}',\dots\}_i\) for \(i\in \{2,3\}\).
Last, an application of the unicast protocol from node \(\nodeAlice_2\) to \nodeCharlie node, then to node \nodeBob, yields an estimate of the term \(\widehat{Q_{Z,2}Q_{Z,3}}\).
Combining the unicast estimate \(\widehat{Q_{Z,2}Q_{Z,3}}\)  and the estimate \(\widehat{q_{Z,1} Q_{Z,2} Q_{Z,3}}\)  from the generalized \mergecast together,
and noticing that \(q_{Z,1} = q_{Z,1} Q_{Z,2} Q_{Z,3} / Q_{Z,2}Q_{Z,3}\),
one then uniquely estimates the parameter \(q_{Z,1}\) (details similar to Section~\ref{subsec:mergecast}).
This process is illustrated by the channel highlighted as \textcolor{green!80!black}{green} in Figure~\ref{subfig:initial-peripheral-channels}.
Denote the set of all identified peripheral channels as set \(\mathfrak{I}\).


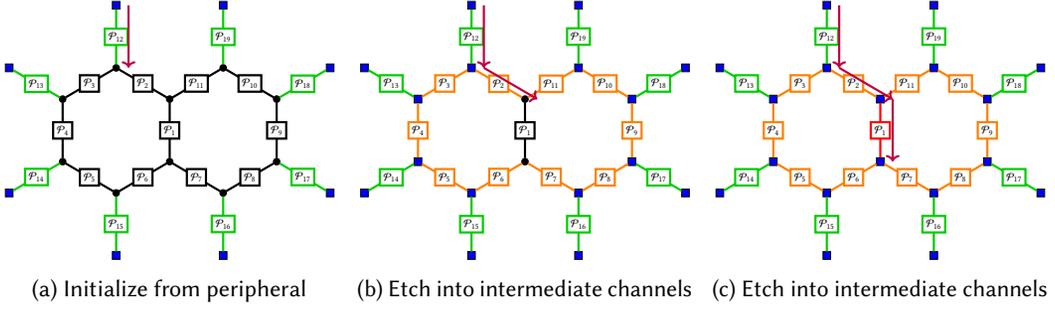
\begin{figure}[tbp]
    \centering
    \begin{subfigure}[b]{0.32\textwidth}
        \resizebox{\linewidth}{!}{
            \begin{tikzpicture}
                \node[draw, circle, minimum size=6pt, inner sep=1pt, fill] (A1) at (0,1) {};
                \node[draw, circle, minimum size=6pt, inner sep=1pt, fill] (A2) at (0,-1) {};

                \node[draw, circle, minimum size=6pt, inner sep=1pt, fill] (B1) at (-1.732,2) {};
                \node[draw, circle, minimum size=6pt, inner sep=1pt, fill] (B2) at (-3.464,1) {};
                \node[draw, circle, minimum size=6pt, inner sep=1pt, fill] (B3) at (-3.464,-1) {};
                \node[draw, circle, minimum size=6pt, inner sep=1pt, fill] (B4) at (-1.732,-2) {};

                \node[draw, circle, minimum size=6pt, inner sep=1pt, fill] (C1) at (1.732,2) {};
                \node[draw, circle, minimum size=6pt, inner sep=1pt, fill] (C2) at (3.464,1) {};
                \node[draw, circle, minimum size=6pt, inner sep=1pt, fill] (C3) at (3.464,-1) {};
                \node[draw, circle, minimum size=6pt, inner sep=1pt, fill] (C4) at (1.732,-2) {};

                \node[draw, rectangle, minimum size=7pt, inner sep=1pt, fill=blue] (D1) at (-1.732,4) {};
                \node[draw, rectangle, minimum size=7pt, inner sep=1pt, fill=blue] (D2) at (-5.196,2) {};
                \node[draw, rectangle, minimum size=7pt, inner sep=1pt, fill=blue] (D3) at (-5.196,-2) {};
                \node[draw, rectangle, minimum size=7pt, inner sep=1pt, fill=blue] (D4) at (-1.732,-4) {};

                \node[draw, rectangle, minimum size=7pt, inner sep=1pt, fill=blue] (E1) at (1.732,4) {};
                \node[draw, rectangle, minimum size=7pt, inner sep=1pt, fill=blue] (E2) at (5.196,2) {};
                \node[draw, rectangle, minimum size=7pt, inner sep=1pt, fill=blue] (E3) at (5.196,-2) {};
                \node[draw, rectangle, minimum size=7pt, inner sep=1pt, fill=blue] (E4) at (1.732,-4) {};

                \draw[line width=2pt, draw=green!80!black] (B1) -- node[draw, rectangle, midway, fill=white] {\(\mathcal{P}_{12}\)} (D1);
                \draw[line width=2pt, draw=green!80!black] (B2) -- node[draw, rectangle, midway, fill=white] {\(\mathcal{P}_{13}\)} (D2);
                \draw[line width=2pt, draw=green!80!black] (B3) -- node[draw, rectangle, midway, fill=white] {\(\mathcal{P}_{14}\)} (D3);
                \draw[line width=2pt, draw=green!80!black] (B4) -- node[draw, rectangle, midway, fill=white] {\(\mathcal{P}_{15}\)} (D4);

                \draw[line width=2pt, draw=green!80!black] (C1) -- node[draw, rectangle, midway, fill=white] {\(\mathcal{P}_{19}\)} (E1);
                \draw[line width=2pt, draw=green!80!black] (C2) -- node[draw, rectangle, midway, fill=white] {\(\mathcal{P}_{18}\)} (E2);
                \draw[line width=2pt, draw=green!80!black] (C3) -- node[draw, rectangle, midway, fill=white] {\(\mathcal{P}_{17}\)} (E3);
                \draw[line width=2pt, draw=green!80!black] (C4) -- node[draw, rectangle, midway, fill=white] {\(\mathcal{P}_{16}\)} (E4);

                \draw[line width=2pt] (A1) -- node[draw, rectangle, midway, fill=white] {\(\mathcal{P}_2\)} (B1);
                \draw[line width=2pt] (B1) -- node[draw, rectangle, midway, fill=white] {\(\mathcal{P}_3\)} (B2);
                \draw[line width=2pt] (B2) -- node[draw, rectangle, midway, fill=white] {\(\mathcal{P}_4\)} (B3);
                \draw[line width=2pt] (B3) -- node[draw, rectangle, midway, fill=white] {\(\mathcal{P}_5\)} (B4);
                \draw[line width=2pt] (B4) -- node[draw, rectangle, midway, fill=white] {\(\mathcal{P}_6\)} (A2);

                \draw[line width=2pt] (A1) -- node[draw, rectangle, midway, fill=white] {\(\mathcal{P}_{11}\)} (C1);
                \draw[line width=2pt] (C1) -- node[draw, rectangle, midway, fill=white] {\(\mathcal{P}_{10}\)} (C2);
                \draw[line width=2pt] (C2) -- node[draw, rectangle, midway, fill=white] {\(\mathcal{P}_9\)} (C3);
                \draw[line width=2pt] (C3) -- node[draw, rectangle, midway, fill=white] {\(\mathcal{P}_{8}\)} (C4);
                \draw[line width=2pt] (C4) -- node[draw, rectangle, midway, fill=white] {\(\mathcal{P}_{7}\)} (A2);

                \draw[line width=2pt] (A1) -- node[draw, rectangle, midway, fill=white] {\(\mathcal{P}_1\)} (A2);

                \draw[->, purple, line width=2pt]
                ($(D1)+(0.4,0)$) -- ($(B1)+(0.4,0)$) ;



            \end{tikzpicture}
        }
        \caption{Initialize from peripheral}\label{subfig:initial-peripheral-channels}
    \end{subfigure}
    \hfill
    \begin{subfigure}[b]{0.32\textwidth}
        \resizebox{\linewidth}{!}{
            \begin{tikzpicture}
                \node[draw, circle, minimum size=6pt, inner sep=1pt, fill] (A1) at (0,1) {};
                \node[draw, circle, minimum size=6pt, inner sep=1pt, fill] (A2) at (0,-1) {};

                \node[draw, rectangle, minimum size=7pt, inner sep=1pt, fill=blue] (B1) at (-1.732,2) {};
                \node[draw, rectangle, minimum size=7pt, inner sep=1pt, fill=blue] (B2) at (-3.464,1) {};
                \node[draw, rectangle, minimum size=7pt, inner sep=1pt, fill=blue] (B3) at (-3.464,-1) {};
                \node[draw, rectangle, minimum size=7pt, inner sep=1pt, fill=blue] (B4) at (-1.732,-2) {};

                \node[draw, rectangle, minimum size=7pt, inner sep=1pt, fill=blue] (C1) at (1.732,2) {};
                \node[draw, rectangle, minimum size=7pt, inner sep=1pt, fill=blue] (C2) at (3.464,1) {};
                \node[draw, rectangle, minimum size=7pt, inner sep=1pt, fill=blue] (C3) at (3.464,-1) {};
                \node[draw, rectangle, minimum size=7pt, inner sep=1pt, fill=blue] (C4) at (1.732,-2) {};

                \node[draw, rectangle, minimum size=7pt, inner sep=1pt, fill=blue] (D1) at (-1.732,4) {};
                \node[draw, rectangle, minimum size=7pt, inner sep=1pt, fill=blue] (D2) at (-5.196,2) {};
                \node[draw, rectangle, minimum size=7pt, inner sep=1pt, fill=blue] (D3) at (-5.196,-2) {};
                \node[draw, rectangle, minimum size=7pt, inner sep=1pt, fill=blue] (D4) at (-1.732,-4) {};

                \node[draw, rectangle, minimum size=7pt, inner sep=1pt, fill=blue] (E1) at (1.732,4) {};
                \node[draw, rectangle, minimum size=7pt, inner sep=1pt, fill=blue] (E2) at (5.196,2) {};
                \node[draw, rectangle, minimum size=7pt, inner sep=1pt, fill=blue] (E3) at (5.196,-2) {};
                \node[draw, rectangle, minimum size=7pt, inner sep=1pt, fill=blue] (E4) at (1.732,-4) {};

                \draw[line width=2pt, draw=green!80!black] (B1) -- node[draw, rectangle, midway, fill=white] {\(\mathcal{P}_{12}\)} (D1);
                \draw[line width=2pt, draw=green!80!black] (B2) -- node[draw, rectangle, midway, fill=white] {\(\mathcal{P}_{13}\)} (D2);
                \draw[line width=2pt, draw=green!80!black] (B3) -- node[draw, rectangle, midway, fill=white] {\(\mathcal{P}_{14}\)} (D3);
                \draw[line width=2pt, draw=green!80!black] (B4) -- node[draw, rectangle, midway, fill=white] {\(\mathcal{P}_{15}\)} (D4);

                \draw[line width=2pt, draw=green!80!black] (C1) -- node[draw, rectangle, midway, fill=white] {\(\mathcal{P}_{19}\)} (E1);
                \draw[line width=2pt, draw=green!80!black] (C2) -- node[draw, rectangle, midway, fill=white] {\(\mathcal{P}_{18}\)} (E2);
                \draw[line width=2pt, draw=green!80!black] (C3) -- node[draw, rectangle, midway, fill=white] {\(\mathcal{P}_{17}\)} (E3);
                \draw[line width=2pt, draw=green!80!black] (C4) -- node[draw, rectangle, midway, fill=white] {\(\mathcal{P}_{16}\)} (E4);

                \draw[line width=2pt, draw=orange] (A1) -- node[draw, rectangle, midway, fill=white] {\(\mathcal{P}_2\)} (B1);
                \draw[line width=2pt, draw=orange] (B1) -- node[draw, rectangle, midway, fill=white] {\(\mathcal{P}_3\)} (B2);
                \draw[line width=2pt, draw=orange] (B2) -- node[draw, rectangle, midway, fill=white] {\(\mathcal{P}_4\)} (B3);
                \draw[line width=2pt, draw=orange] (B3) -- node[draw, rectangle, midway, fill=white] {\(\mathcal{P}_5\)} (B4);
                \draw[line width=2pt, draw=orange] (B4) -- node[draw, rectangle, midway, fill=white] {\(\mathcal{P}_6\)} (A2);

                \draw[line width=2pt, draw=orange] (A1) -- node[draw, rectangle, midway, fill=white] {\(\mathcal{P}_{11}\)} (C1);
                \draw[line width=2pt, draw=orange] (C1) -- node[draw, rectangle, midway, fill=white] {\(\mathcal{P}_{10}\)} (C2);
                \draw[line width=2pt, draw=orange] (C2) -- node[draw, rectangle, midway, fill=white] {\(\mathcal{P}_9\)} (C3);
                \draw[line width=2pt, draw=orange] (C3) -- node[draw, rectangle, midway, fill=white] {\(\mathcal{P}_{8}\)} (C4);
                \draw[line width=2pt, draw=orange] (C4) -- node[draw, rectangle, midway, fill=white] {\(\mathcal{P}_{7}\)} (A2);

                \draw[line width=2pt] (A1) -- node[draw, rectangle, midway, fill=white] {\(\mathcal{P}_1\)} (A2);

                \draw[->, purple, line width=2pt]
                ($(D1)+(0.4,0)$) -- ($(B1)+(0.4,0)$) ;

                \draw[->, purple, line width=2pt]
                ($(B1)+(0.4,0)$) -- ($(A1)+(0.4,0)$);

            \end{tikzpicture}
        }
        \caption{Etch into intermediate channels}\label{subfig:etch-into}
    \end{subfigure}
    \hfill
    \begin{subfigure}[b]{0.32\textwidth}
        \resizebox{\linewidth}{!}{
            \begin{tikzpicture}
                \node[draw, rectangle, minimum size=7pt, inner sep=1pt, fill=blue] (A1) at (0,1) {};
                \node[draw, rectangle, minimum size=7pt, inner sep=1pt, fill=blue] (A2) at (0,-1) {};

                \node[draw, rectangle, minimum size=7pt, inner sep=1pt, fill=blue] (B1) at (-1.732,2) {};
                \node[draw, rectangle, minimum size=7pt, inner sep=1pt, fill=blue] (B2) at (-3.464,1) {};
                \node[draw, rectangle, minimum size=7pt, inner sep=1pt, fill=blue] (B3) at (-3.464,-1) {};
                \node[draw, rectangle, minimum size=7pt, inner sep=1pt, fill=blue] (B4) at (-1.732,-2) {};

                \node[draw, rectangle, minimum size=7pt, inner sep=1pt, fill=blue] (C1) at (1.732,2) {};
                \node[draw, rectangle, minimum size=7pt, inner sep=1pt, fill=blue] (C2) at (3.464,1) {};
                \node[draw, rectangle, minimum size=7pt, inner sep=1pt, fill=blue] (C3) at (3.464,-1) {};
                \node[draw, rectangle, minimum size=7pt, inner sep=1pt, fill=blue] (C4) at (1.732,-2) {};

                \node[draw, rectangle, minimum size=7pt, inner sep=1pt, fill=blue] (D1) at (-1.732,4) {};
                \node[draw, rectangle, minimum size=7pt, inner sep=1pt, fill=blue] (D2) at (-5.196,2) {};
                \node[draw, rectangle, minimum size=7pt, inner sep=1pt, fill=blue] (D3) at (-5.196,-2) {};
                \node[draw, rectangle, minimum size=7pt, inner sep=1pt, fill=blue] (D4) at (-1.732,-4) {};

                \node[draw, rectangle, minimum size=7pt, inner sep=1pt, fill=blue] (E1) at (1.732,4) {};
                \node[draw, rectangle, minimum size=7pt, inner sep=1pt, fill=blue] (E2) at (5.196,2) {};
                \node[draw, rectangle, minimum size=7pt, inner sep=1pt, fill=blue] (E3) at (5.196,-2) {};
                \node[draw, rectangle, minimum size=7pt, inner sep=1pt, fill=blue] (E4) at (1.732,-4) {};

                \draw[line width=2pt, draw=green!80!black] (B1) -- node[draw, rectangle, midway, fill=white] {\(\mathcal{P}_{12}\)} (D1);
                \draw[line width=2pt, draw=green!80!black] (B2) -- node[draw, rectangle, midway, fill=white] {\(\mathcal{P}_{13}\)} (D2);
                \draw[line width=2pt, draw=green!80!black] (B3) -- node[draw, rectangle, midway, fill=white] {\(\mathcal{P}_{14}\)} (D3);
                \draw[line width=2pt, draw=green!80!black] (B4) -- node[draw, rectangle, midway, fill=white] {\(\mathcal{P}_{15}\)} (D4);

                \draw[line width=2pt, draw=green!80!black] (C1) -- node[draw, rectangle, midway, fill=white] {\(\mathcal{P}_{19}\)} (E1);
                \draw[line width=2pt, draw=green!80!black] (C2) -- node[draw, rectangle, midway, fill=white] {\(\mathcal{P}_{18}\)} (E2);
                \draw[line width=2pt, draw=green!80!black] (C3) -- node[draw, rectangle, midway, fill=white] {\(\mathcal{P}_{17}\)} (E3);
                \draw[line width=2pt, draw=green!80!black] (C4) -- node[draw, rectangle, midway, fill=white] {\(\mathcal{P}_{16}\)} (E4);

                \draw[line width=2pt, draw=orange] (A1) -- node[draw, rectangle, midway, fill=white] {\(\mathcal{P}_2\)} (B1);
                \draw[line width=2pt, draw=orange] (B1) -- node[draw, rectangle, midway, fill=white] {\(\mathcal{P}_3\)} (B2);
                \draw[line width=2pt, draw=orange] (B2) -- node[draw, rectangle, midway, fill=white] {\(\mathcal{P}_4\)} (B3);
                \draw[line width=2pt, draw=orange] (B3) -- node[draw, rectangle, midway, fill=white] {\(\mathcal{P}_5\)} (B4);
                \draw[line width=2pt, draw=orange] (B4) -- node[draw, rectangle, midway, fill=white] {\(\mathcal{P}_6\)} (A2);

                \draw[line width=2pt, draw=orange] (A1) -- node[draw, rectangle, midway, fill=white] {\(\mathcal{P}_{11}\)} (C1);
                \draw[line width=2pt, draw=orange] (C1) -- node[draw, rectangle, midway, fill=white] {\(\mathcal{P}_{10}\)} (C2);
                \draw[line width=2pt, draw=orange] (C2) -- node[draw, rectangle, midway, fill=white] {\(\mathcal{P}_9\)} (C3);
                \draw[line width=2pt, draw=orange] (C3) -- node[draw, rectangle, midway, fill=white] {\(\mathcal{P}_{8}\)} (C4);
                \draw[line width=2pt, draw=orange] (C4) -- node[draw, rectangle, midway, fill=white] {\(\mathcal{P}_{7}\)} (A2);

                \draw[line width=2pt, draw=red] (A1) -- node[draw, rectangle, midway, fill=white] {\(\mathcal{P}_1\)} (A2);

                \draw[->, purple, line width=2pt]
                ($(D1)+(0.4,0)$) -- ($(B1)+(0.4,0)$) ;

                \draw[->, purple, line width=2pt]
                ($(B1)+(0.4,0)$) -- ($(A1)+(0.4,0)$);

                \draw[->, purple, line width=2pt]
                ($(A1)+(0.4,0)$) -- ($(A2)+(0.4,0)$);
            \end{tikzpicture}
        }
        \caption{Etch into intermediate channels}\label{subfig:etch-end}\label{fig:etch-experiment-topology}
    \end{subfigure}
    \caption{Example for progressive etching in general topology: The purple arrow sequence indicates the order of channel etching from peripheral to internal channels.
        Figure~\ref{subfig:initial-peripheral-channels} shows the first step of identifying peripheral channels \(\mathcal{P}_{12}\), \(\mathcal{P}_{13}\), \(\mathcal{P}_{14}\), \(\mathcal{P}_{15}\), \(\mathcal{P}_{16}\), \(\mathcal{P}_{17}\), \(\mathcal{P}_{18}\), and \(\mathcal{P}_{19}\) via the generalized \mergecast in Figure~\ref{fig:mergecast-peripheral}.
        Figure~\ref{subfig:etch-into} shows the second step of identifying internal channels \(\mathcal{P}_{2}\), \(\mathcal{P}_{3}\), \(\mathcal{P}_{4}\), \(\mathcal{P}_{5}\), \(\mathcal{P}_{6}\), \(\mathcal{P}_{7}\), \(\mathcal{P}_{8}\), \(\mathcal{P}_{9}\), \(\mathcal{P}_{10}\), and \(\mathcal{P}_{11}\), all of which connect to the identified peripheral channels in the first step.
        Figure~\ref{subfig:etch-end} shows the last step of etching into the channel \(\mathcal{P}_1\).}
    \label{fig:toy-example-general-tomograph}
\end{figure}

\noindent\textbf{Etch into intermediate channels.}
After identifying all peripheral channels, one \emph{``etches into''} the intermediate channels.
For any identified peripheral channel (with one monitor node in the peripheral and another internal node),
with the state preparation and measurement ability of the monitor and the identified channel parameters, one can then ``control'' the states in the other internal node of the peripheral channels.
This implies that one can ``push'' the monitor from the peripheral, through the identified channel, to the internal node, which then becomes an \emph{equivalent} monitor node.
The change from Figure~\ref{subfig:initial-peripheral-channels} to Figure~\ref{subfig:etch-into} illustrates that: after the channel \(\mathcal P_{12}\) has been identified, the peripheral monitor (blue square) on the top is then ``pushed'' to its internal end node, which is represented as a blue square in Figure~\ref{subfig:etch-into}; similar updates for other peripheral channels.
Then, taking these equivalent ``monitor'' nodes into account, one can find more ``peripheral'' channels and use generalized \mergecast (in Figure~\ref{fig:mergecast-peripheral}) to identify them.
After that, including these identified channels into set \(\mathfrak{I}\),
and repeat the ``etching'' step until all channels are estimated.

Figure~\ref{fig:toy-example-general-tomograph} illustrates this process.
Starting from Figure~\ref{subfig:initial-peripheral-channels}, all real peripheral channels---\(\mathcal{P}_{12}, \mathcal{P}_{13}, \mathcal{P}_{14},\) \(\mathcal{P}_{15}, \mathcal{P}_{16}, \mathcal{P}_{17}, \mathcal{P}_{18}, \mathcal{P}_{19}\)---are first identified.
Next, the monitors are advanced into their corresponding internal nodes (shown as the enlarged blue square nodes in Figure~\ref{subfig:etch-into}, compared to Figure~\ref{subfig:initial-peripheral-channels}), enabling the identification of the internal channels
\(
\mathcal{P}_{2}, \mathcal{P}_{3}, \mathcal{P}_{4}, \mathcal{P}_{5}, \mathcal{P}_{6}, \mathcal{P}_{7}, \mathcal{P}_{8}, \mathcal{P}_{9}, \mathcal{P}_{10}, \mathcal{P}_{11}
\) via general \mergecast,
which are highlighted in \textcolor{orange}{orange} in Figure~\ref{subfig:etch-into}.
After these channels are identified, the monitor can be further propagated deeper into the network.
As shown in Figure~\ref{subfig:etch-end}, the top and bottom end nodes of channel \(\mathcal{P}_{1}\) each effectively become monitors.
Consequently, since both end nodes of \(\mathcal{P}_{1}\) now serve as monitors, the channel itself can be identified (highlighted in \textcolor{red}{red} in Figure~\ref{subfig:etch-end}).
This progressive etching protocol, summarized in Algorithm~\ref{alg:progressive-etching}, enables the identification of all Pauli channels in a general network topology.

\begin{algorithm}[H]
    \caption{Progressive Etching for General Topology}\label{alg:progressive-etching}
    \begin{algorithmic}[1]
        \Initial \(\mathfrak{L}\gets \emptyset\),
        \(\mathfrak{N} \gets \{\text{all peripheral nodes}\}\)

        \State \(\mathfrak{I} \gets \{\text{channels with peripheral end nodes in }\mathfrak{N}\}\)

        \While{exists a channel \(\ell\) in \(\mathfrak{I}\setminus \mathfrak{L}\)} \Comment{exists channels to be identified}
        \State Apply \mergecast (Figure~\ref{fig:mergecast-peripheral}) to identify channel \(\ell\)
        \State \(\mathfrak{L} \gets \mathfrak{L}\cup \{\ell\}\) \Comment{record the identified channel}
        \State \(\mathfrak{N} \gets \mathfrak{N} \cup \{\text{the non-peripheral node of channel \(\ell\)}\}\)
        \RightComment{treat the non-peripheral node of an identifiable channel as peripheral}
        \State \(\mathfrak{I} \gets \{\text{channels with peripheral nodes in }\mathfrak{N}\}\)
        \Comment{update peripheral channels}
        \EndWhile
    \end{algorithmic}
\end{algorithm}

\subsection{\bypassunicast: QNT with Identifiability for Bypassable Pauli Channels}\label{subsec:bypassable-channel-protocol}



\begin{figure}
    \centering
    \begin{subfigure}[b]{0.25\textwidth}
        \centering
        \resizebox{\textwidth}{!}{%
            \begin{tikzpicture}
                \node[draw, circle, minimum size=6pt, inner sep=1pt, fill] (A) at (0,0) {};
                \node[draw, rectangle, minimum size=7pt, inner sep=1pt, fill=blue]  (B) at (0,2) {};
                \node[draw, rectangle, minimum size=7pt, inner sep=1pt, fill=blue]  (C) at (1.732,-1) {};
                \node[draw, rectangle, minimum size=7pt, inner sep=1pt, fill=blue]  (D) at (-1.732,-1) {};

                \draw[line width=1.5pt] (A) -- node[draw, rectangle, midway, fill=white] {\(\mathcal{B}_{1}\)} (B);
                \draw[line width=1.5pt] (A) -- node[draw, rectangle, midway, fill=white] {\(\mathcal{B}_{3}\)} (C);
                \draw[line width=1.5pt] (A) -- node[draw, rectangle, midway, fill=white] {\(\mathcal{B}_{2}\)} (D);

                \draw[dashed, line width=1pt] (-0.3, 1.7) rectangle (0.3, 2.3);
                \node at (0.5, 1.5) {\(\nodeAlice\)};

                \draw[dashed, line width=1pt] (-0.3, -0.3) rectangle (0.3, 0.3);
                \node at (0, -0.6) {\nodeCharlie};


                \draw[dashed, line width=1pt] (-2.032, -1.3) rectangle (-1.432, -0.7);
                \node at (-1.732, -1.5) {\(\nodeBob\)};
            \end{tikzpicture}
        }
        \caption{\(3\)-link star network}
        \label{subfig:star-bypassable-unicast}
    \end{subfigure}
    \hspace{0.06\textwidth}
    \begin{subfigure}[b]{0.6\textwidth}
        \centering
        \resizebox{0.9\textwidth}{!}{%
            \begin{quantikz}
                \inputD{\text{SP}}\gategroup[1,steps=2,style={dashed,rounded corners,fill=blue!20, inner xsep=2pt},background,label style={label position=below,anchor=north,yshift=-0.2cm}]{\nodeAlice}
                & \gate{H}
                \slice{1}
                & \gate{{\mathcal{B}_1}}
                & \gate{H}\gategroup[1,steps=1,style={dashed,rounded corners,fill=blue!20, inner xsep=2pt},background,label style={label position=below,anchor=north,yshift=-0.2cm}]{\nodeCharlie}
                \slice{2}
                & \gate{\textcolor{blue}{\mathcal{B}_2}}
                \slice{3}
                & \meter{}\gategroup[1,steps=1,style={dashed,rounded corners,fill=blue!20, inner xsep=2pt},background,label style={label position=below,anchor=north,yshift=-0.2cm}]{\nodeBob}
                & \setwiretype{c}
            \end{quantikz}
        }
        \caption{\bypassunicast for bypassing channel \(\mathcal{B}_1\) and identifying \(\mathcal{B}_2\)}
        \label{subfig:bypassable-unicast}
    \end{subfigure}
    \caption{\bypassunicast for bypassable channels in a star network
    }
    \label{fig:bypassable-unicast}
\end{figure}
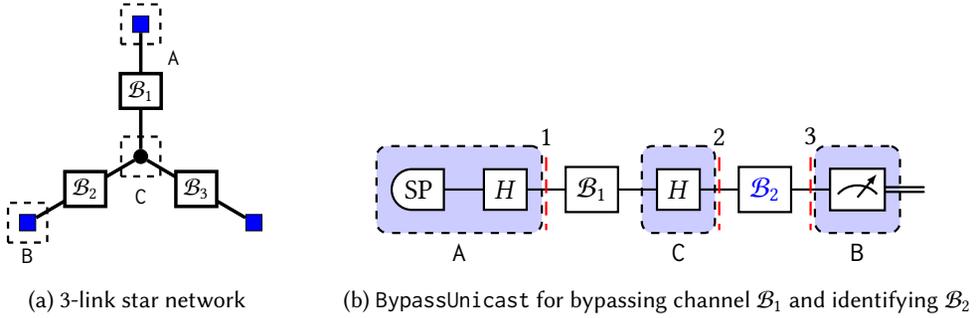

Last, we consider a special class of quantum networks where all channels are bypassable (Definition~\ref{def:bypassable-channel}).
As the name suggests, a bypassable channel can be ``bypassed'' by certain quantum states.
In this section, we propose \bypassunicast,
a QNT protocol that utilizes this property to resolve the identifiability challenge for general topologies.

Without loss of generality, we illustrate the behavior of \bypassunicast on a \(3\)-link star network consisting of three bit-flip channels \(\mathcal{B}_1\), \(\mathcal{B}_2\), and \(\mathcal{B}_3\) in Figure~\ref{subfig:star-bypassable-unicast}.
Figure~\ref{subfig:bypassable-unicast} depicts the unicast protocol for identifying channel \(\mathcal{B}_2\).
First, node \nodeAlice\ prepares the state \(\kett{0}\), applies a Hadamard gate to prepare \(\kett{+}\) (whose Pauli-Liouville representation is \([1,1,0,0]^T\) at slice 1), and sends it through the bit-flip channel \(\mathcal{B}_1 = \mathrm{diag}(1,1,q_{Y,1},q_{Z,1})\) to node \nodeCharlie. As shown in~\eqref{eq:bypass-bit-flip-channel}, the \(\kett{+}\) state remains unchanged by \(\mathcal{B}_1\), effectively \emph{bypassing} it.

Upon receiving the state, node \nodeCharlie\ applies another Hadamard gate, transforming \(\kett{+}\) back to \(\kett{0}\) (slice 2), then routes the resulting state through the second bit-flip channel \(\mathcal{B}_2 = \mathrm{diag}(1,1,q_{Y,2},q_{Z,2})\) to node \nodeBob. Because \(\kett{0}\) cannot bypass bit-flip channels, the state emerging from \(\mathcal{B}_2\) becomes \([1,0,0,q_{Z,2}]^T\) at slice 3. Measuring the final received state at node \nodeBob in the computational basis yields \(\kett{0}\) with probability \(p_{\text{bypass}} = \tfrac{1 + q_{Z,2}}{2}\). By executing the protocol multiple times, one can estimate the probability \(\hat p_{\text{bypass}}\) using a mean estimator, from which one can estimate the channel parameter \(q_{Z,2}\) using estimator \(\hat q_{Z,2} \coloneqq 2\hat p_{\text{bypass}} - 1\).

By reversing the procedure---starting with \(\mathcal{B}_2\), bypassing it via Hadamards, and then measuring the final received qubit after going through \(\mathcal{B}_1\)---one can similarly isolate and estimate \(q_{Z,1}\).
Consequently, each bit-flip channel parameter can be estimated individually, and no identifiability issue remains.
Then, we can pick another path of the three-link star network, say from \(\mathcal{B}_1\) to \(\mathcal{B}_3\), and apply the same protocol to estimate \(q_{Z,3}\) uniquely (i.e., addressing the identifiability issue illustrated at the beginning of Section~\ref{sec:qnt-identifiability}).

Together with the channel dressing technique in Figure~\ref{fig:channel-dressing}, one can extend the protocol in Figure~\ref{fig:bypassable-unicast} to any other bypassable channel.
For any general network topology consisting of bypassable channels, say the example in Figure~\ref{fig:toy-example-general-tomograph}, one can replace \mergecast in Figure~\ref{fig:mergecast-peripheral} with the unicast protocol in Figure~\ref{fig:bypassable-unicast} to identify peripheral channels.
Then, one can apply the progressive etching protocol in Algorithm~\ref{alg:progressive-etching} to identify all channels in the network.
A prior work by~\citet{de2023characterization} implicitly utilizes the bypassable idea in a multicast protocol designed for a bit-flip star network.
Here, we formalize the idea with the rigorously defined bypassable Pauli channels (Definition~\ref{def:bypassable-channel}),
demonstrate how to utilize the bypassable property in unicast protocols (Figure~\ref{fig:bypassable-unicast}),
and illustrate how it can be extended to general network topologies.


\section{QNT \lowercase{with} SPAM Errors}
\label{sec:qnt-spam}

In Section~\ref{sec:qnt-identifiability}, we assumed perfect state preparation and measurement (SPAM) operations.
We relax this assumption in this section and introduce SPAM errors in QNT.
We first introduce a parametric model for SPAM errors in Section~\ref{subsec:spam-formulation}. We then propose protocols to estimate the SPAM model parameters in Section~\ref{subsec:spam-estimation},
and in Section~\ref{subsec:qnt-protocol-spam-extension}, we extend the QNT protocols of Section~\ref{sec:qnt-identifiability} to account for SPAM errors.
Lastly, we present an integrated workflow for QNT channel estimation under SPAM errors in Section~\ref{subsec:qnt-spam-workflow}.

\subsection{SPAM Error Model}\label{subsec:spam-formulation}


We follow the SPAM error formulation of~\citet{lin2021independent}.
Most other formulations, e.g.,~\citet{jayakumar2024universal,cai2023quantum},
map to this model.
Specifically, we only consider preparation of the basis state \(\kett{0}\) and measurement under the computation Z-basis \(\{\kett{0}, \kett{1}\}\).
Preparation and measurement of other states/bases can be formulated by applying corresponding gates before and after the above considered state preparation and measurement, transferring them to corresponding states/bases.

In general, one can denote \(\kett{\tilde \rho_0} = [1, s_X, s_Y, s_Z]^T\) and \(\kett{\tilde{\mathcal{M}}_0} = [m_I, m_X, m_Y, m_Z]^T\)
as the state preparation  of state \(\kett{0}\) and measurement of state \(\kett{0}\) with \emph{errors}.
Using a phase cycling technique introduced by~\citep{lin2021independent,levitt2008spin} (detailed in Appendix~\ref{rmk:spam-simplification}), SPAM errors can be represented as \begin{align}\label{eq:spam-formulation}
    \kett{\tilde \rho} = [1, 0, 0, s]^T
    \quad \text{and} \quad
    \kett{\tilde{\mathcal{M}}_0} = [1, 0, 0, m]^T,
\end{align} with only two parameters \(s, m\in [0,1]\) needed to describe the SPAM errors.
In particular, \(s=m=1\) refers to the perfect state preparation and measurement (no SPAM error), and \(s=m=0\) refers to the worst SPAM errors (completely random outputs from maximal mixed states).
Recall that the probability of measuring outcome \(\kett{0}\) given the prepared state \(\tilde \rho_0\) and the measurement operator \(\tilde{\mathcal{M}}_0\) is \(\frac 1 2 \bbra{\tilde \rho}\kett{\tilde{\mathcal M}_0} = \frac{1 + ms}{2}\).

\subsection{SPAM Error Estimation}\label{subsec:spam-estimation}

In this section, we propose protocols to estimate the SPAM error parameters \(s\) and \(m\) in the QNT scenario.
In particular, we devise protocols that do not rely on knowledge of the channel parameters and can be applied to any scenario.
While certain methods in the literature---such as those by~\citet{jayakumar2024universal,cai2023quantum}---enable monitors to locally estimate their SPAM error parameters,
these local approaches require that the monitors conduct both state preparation and measurement operations, which may not be feasible in some network settings.
For example, in satellite-based quantum networks, the monitors (optical ground stations) may be only capable of either state preparation (for the upload link from ground to satellite) or measurement operations (for the download link from satellite to group), since these two are typically performed by different devices (e.g., single-photon sources and single-photon detectors)~\citep{chen2021integrated}.
In these scenarios, our protocols play a crucial role in estimating the SPAM errors in the network without requiring monitors to perform both operations.

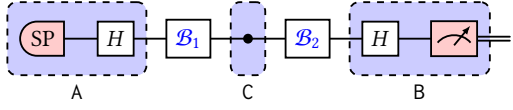
\begin{figure}[tb]
    \centering
    \resizebox{0.5\textwidth}{!}{
        \begin{quantikz}
            \inputD[style={fill=red!20}]{{\text{SP}}}\gategroup[1,steps=2,style={dashed,rounded corners,fill=blue!20, inner xsep=2pt},background,label style={label position=below,anchor=north,yshift=-0.2cm}]{\nodeAlice}
            & \gate{H}
            & \gate{\textcolor{blue}{\mathcal{B}_1}}
            & \ctrl{0}\gategroup[1,steps=1,style={dashed,rounded corners,fill=blue!20, inner xsep=2pt},background,label style={label position=below,anchor=north,yshift=-0.2cm}]{\nodeCharlie}
            & \gate{\textcolor{blue}{\mathcal{B}_2}}
            & \gate{H}\gategroup[1,steps=2,style={dashed,rounded corners,fill=blue!20, inner xsep=2pt},background,label style={label position=below,anchor=north,yshift=-0.2cm}]{\nodeBob}
            & \meter[style={fill=red!20}]{}
            & \wireoverride{c}
        \end{quantikz}
    }
    \caption{Estimate \(ms\) via bypassing channels, where the light red colored ``SP'' box and measurement meter represents the existence of SPAM errors.}
    \label{fig:spam-estimation-bypassable}
\end{figure}

\noindent\textbf{Challenges of estimating \(m\) and \(s\) separately.}
For QNT with bypassable channels, we can estimate the product \(ms\) by bypassing all channels in the network.
For example, for two consecutive bit-flip channels, one can bypass both channels, as shown in Figure~\ref{fig:spam-estimation-bypassable},
and then estimates the probability of the measurement outcome \(\kett{0}\) (which is \(\frac{1 + ms}{2}\))
yielding an estimate of \(ms\).

However, in the case of non-bypassable channels, e.g., the depolarization channel, it is impossible to estimate \(ms\) directly as in the bypassable case.
For example, in the unicast protocol (Figure~\ref{fig:simple-unicast}) with one state preparation (SP) and one measurement (AM) with errors, the probability of measurement outcome \(\kett{0}\) becomes  \(
\frac{1 + msq_{Z_1}q_{Z_2}}{2},
\)
where the two parameters \(m\) and \(s\) appear together and are always coupled.
Furthermore, without knowledge of the channel parameters \(q_{Z,1}, q_{Z,2}\), it is also difficult to estimate \(ms\) as one term.
Instead, we aim to propose protocols to separate \(m\) and \(s\) and estimate them individually.


\begin{figure}[tb]
    \centering
    \resizebox{0.5\textwidth}{!}{
        \begin{quantikz}
            &\inputD[style={fill=red!20}]{\text{SP}}\gategroup[2,steps=3,style={dashed,rounded corners,fill=blue!20, inner xsep=2pt},background,label style={label position=below,anchor=north,yshift=-0.2cm}]{\nodeAlice}
            & \ctrl{1}
            & [-0.55cm] \trash{\text{discard}}
            & \setwiretype{n}
            & & & &
            \\
            & \inputD[style={fill=red!20}]{\text{SP}}
            \slice{1}
            & \targ{}
            &
            \slice{2}
            & \gate{\textcolor{blue}{\mathcal{P}_1}}
            &\ctrl{0} \gategroup[1,steps=1,style={dashed,rounded corners,fill=blue!20, inner xsep=2pt},background,label style={label position=below,anchor=north,yshift=-0.2cm}]{\nodeCharlie}
            & \gate{\textcolor{blue}{\mathcal{P}_2}}
            & \meter[style={fill=red!20}]{}\gategroup[1,steps=1,style={dashed,rounded corners,fill=blue!20, inner xsep=2pt},background,label style={label position=below,anchor=north,yshift=-0.2cm}]{\nodeBob}
            & \wireoverride{c}
        \end{quantikz}
    }
    \caption{Estimate the \(s\) parameter of SPAM errors in QNT with any Pauli channels via unicast}
    \label{fig:unicast-spam-estimate-s}
\end{figure}
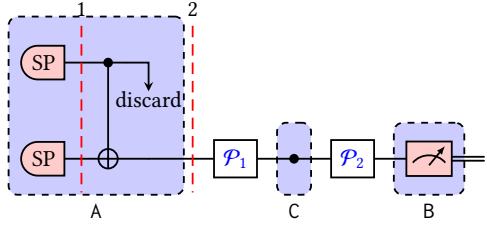

\noindent\textbf{Estimate preparation error parameter \(s\).}
Figure~\ref{fig:unicast-spam-estimate-s} illustrates a protocol for estimating the state preparation error parameter \(s\) in the presence of arbitrary Pauli channels. A CNOT gate at node \nodeAlice generates a state whose Pauli-Liouville representation contains an \(s^2\) term. Specifically, after tracing out the second qubit in slice 1, the remaining single-qubit state becomes \([1,0,0,s^2]^T\). Passing through the two Pauli channels and proceeding to measurement at the “End” node, the probability of obtaining the outcome \(|0\rangle\) is
\(
p_{\mathrm{SPAM},1}  = \frac{1 + ms^2q_{Z,1}q_{Z,2}}{2}.
\)
By contrast, in the unicast protocol (prepare \(\kett{\tilde \rho_0}\) at “Root” only), the corresponding success probability is
\begin{align}\label{eq:spam-estimate-s-p0}
    p_{\mathrm{SPAM},0} = \frac{1 + msq_{Z,1}q_{Z,2}}{2}.
\end{align}
Thus, one can express the parameter \(s\) as
\(
s = \frac{ms^2q_{Z,1}q_{Z,2}}{msq_{Z,1}q_{Z,2}}.
\)
By estimating both probabilities \(p_{\mathrm{SPAM},0}\) and \(p_{\mathrm{SPAM},1}\) denoted by \(\hat{p}_{\mathrm{SPAM},0}\) and \(\hat{p}_{\mathrm{SPAM},1}\), a practical estimator for \(s\) is given by
\begin{align}
    \label{eq:spam-estimate-s}
    \hat{s} \coloneqq \frac{2\hat{p}_{\mathrm{SPAM},1} - 1}{2\hat{p}_{\mathrm{SPAM},0} - 1}.
\end{align}

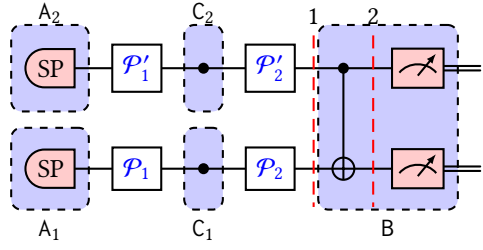
\begin{figure}[tb]
    \centering
    \resizebox{0.5\textwidth}{!}{
        \begin{quantikz}
            & \inputD[style={fill=red!20}]{\text{SP}}
            \gategroup[1,steps=1,style={dashed,rounded corners,fill=blue!20, inner xsep=2pt},background,label style={label position=below,anchor=north,yshift=1.4cm}]{\(\nodeAlice_2\)}
            & \gate{\textcolor{blue}{\mathcal{P}_1'}}
            &\ctrl{0}
            \gategroup[1,steps=1,style={dashed,rounded corners,fill=blue!20, inner xsep=2pt},background,label style={label position=below,anchor=north,yshift=1.4cm}]{\(\nodeCharlie_2\)}
            & \gate{\textcolor{blue}{\mathcal{P}_2'}}\slice{1}
            & \ctrl{1}\gategroup[2,steps=2,style={dashed,rounded corners,fill=blue!20, inner xsep=2pt},background,label style={label position=below,anchor=north,yshift=-0.2cm}]{\nodeBob}
            & \meter[style={fill=red!20}]{}
            & \setwiretype{c}
            \\[0.25cm]
            & \inputD[style={fill=red!20}]{\text{SP}}
            \gategroup[1,steps=1,style={dashed,rounded corners,fill=blue!20, inner xsep=2pt},background,label style={label position=below,anchor=north,yshift=-0.2cm}]{\(\nodeAlice_1\)}
            & \gate{\textcolor{blue}{\mathcal{P}_1}}
            &\ctrl{0}\gategroup[1,steps=1,style={dashed,rounded corners,fill=blue!20, inner xsep=2pt},background,label style={label position=below,anchor=north,yshift=-0.2cm}]{\(\nodeCharlie_1\)}
            & \gate{\textcolor{blue}{\mathcal{P}_2}}
            & \targ{}
            \slice{2}
            & \meter[style={fill=red!20}]{}
            & \setwiretype{c}
        \end{quantikz}
    }
    \caption{Estimate the \(m\) parameter of SPAM errors in QNT with any Pauli channels via unicast}
    \label{fig:unicast-spam-estimate-m}
\end{figure}

\noindent\textbf{Estimate measurement error parameter \(m\).}
Figure~\ref{fig:unicast-spam-estimate-m} outlines a scheme to estimate the measurement error parameter \(m\) under arbitrary Pauli channel noise.
The figure shows two branches of channels (i.e., one across nodes \(\nodeAlice_1\) and \(\nodeCharlie_1\), and another across nodes \(\nodeAlice_2\) and \(\nodeCharlie_2\)) connecting the node \nodeBob.
This two-branch structure is mainly for illustration; one can use a single branch two times to achieve the same effect.
Specifically, each branch sends a qubit to node \nodeBob,
the two received qubits are entangled via a CNOT gate at node \nodeBob,
and then node \nodeBob performs a joint measurement on both qubits. As detailed in Appendix~\ref{app:spam-m-calculation}, the joint outcome probabilities are as follows,
\begin{align}
    \mathbb{P}(\kett{00})
     & =
    \frac{1 + ms^2q_{Z,1}q_{Z,2}q_{Z,1}'q_{Z,2}' + msq_{Z,1}'q_{Z,2}' + m^2 s q_{Z,1}q_{Z,2}}{4}, \\
    \mathbb{P}(\kett{11})
     & =
    \frac{1 - ms^2q_{Z,1}q_{Z,2}q_{Z,1}'q_{Z,2}' - msq_{Z,1}'q_{Z,2}' + m^2 s q_{Z,1}q_{Z,2}}{4},
\end{align}
and we omit the detailed expressions for \(\mathbb{P}(\kett{10})\) and \(\mathbb{P}(\kett{01})\).
Then, we denote the sum of the above two expressions as
\begin{align}
    p_{\text{SPAM}, 2} \coloneqq \mathbb{P}(\kett{00}) + \mathbb{P}(\kett{11}) = \frac{1 + m^2 s  q_{Z,1}q_{Z,2}}{2}
\end{align}
Notice that one can express the parameter \(m\) as \(
m = \frac{m^2s q_{Z,1}q_{Z,2}}{ms q_{Z,1} q_{Z,2}}.
\)
By estimating the above probability \(p_{\text{SPAM}, 2}\) (denote as \(\hat{p}_{\text{SPAM}, 2}\)) as well as the \(p_{\text{SPAM}, 0}\) from unicast in~\eqref{eq:spam-estimate-s-p0}, the measurement error parameter \(m\) can be estimated as follows,
\begin{align}\label{eq:spam-estimate-m}
    \hat m \coloneqq \frac{2 \hat{p}_{\text{SPAM}, 2} - 1}{2 \hat{p}_{\text{SPAM}, 0} - 1}.
\end{align}


\begin{figure}[tbp]
    \centering
    \begin{subfigure}[b]{0.46\textwidth}
        \centering
        \resizebox{0.9\linewidth}{!}{
            \begin{quantikz}
                & \inputD{\text{SP}}\gategroup[2,steps=4,style={dashed,rounded corners,fill=blue!20, inner xsep=2pt},background,label style={label position=below,anchor=north,yshift=-0.2cm}]{\nodeAlice}
                & \gate[style={fill=red!20}]{\text{Error}_{\text{SP}}}
                & \ctrl{1}
                & [-0.55cm]\trash{\text{discard}}
                & \wireoverride{n}
                \\
                &\inputD{\text{SP}}
                & \gate[style={fill=red!20}]{\text{Error}_{\text{SP}}}
                & \targ{}
                & \qw
                & \qw
            \end{quantikz}
        }
        \caption{\mergecast interpretation for node \nodeAlice of Figure~\ref{fig:unicast-spam-estimate-s}} \label{subfig:mergecast-root}
    \end{subfigure}
    \hspace{0.05\textwidth}
    \begin{subfigure}[b]{0.46\textwidth}
        \centering
        \resizebox{0.9\linewidth}{!}{
            \begin{quantikz}
                & \qw \gategroup[2,steps=4,style={dashed,rounded corners,fill=blue!20, inner xsep=2pt},background,label style={label position=below,anchor=north,yshift=-0.2cm}]{\nodeBob}
                & \ctrl{1}
                & \gate[style={fill=red!20}]{\text{Error}_{\text{M}}}
                & \meter{}
                & \setwiretype{c}
                \\
                &\qw
                & \targ{}
                & \gate[style={fill=red!20}]{\text{Error}_{\text{M}}}
                & \meter{}
                & \setwiretype{c}
            \end{quantikz}
        }
        \caption{Multicast interpretation for node \nodeBob of Figure~\ref{fig:unicast-spam-estimate-m}}
        \label{subfig:multicast-end}
    \end{subfigure}
    \caption{\mergecast and multicast interpretations of the SPAM error estimation protocols}
    \label{fig:mergecast-multicast-spam-interpretation}
\end{figure}
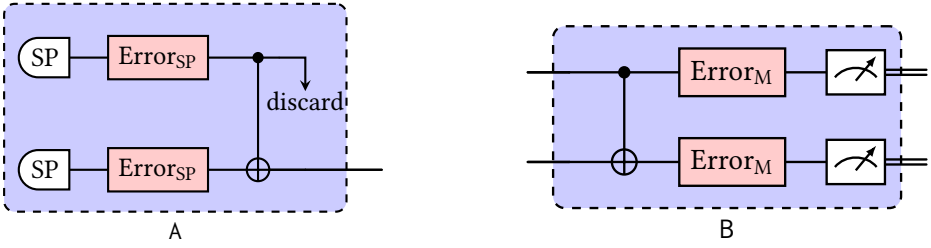

\begin{remark}[Alternate interpretations of Figures~\ref{fig:unicast-spam-estimate-s} and~\ref{fig:unicast-spam-estimate-m}]
    \label{rmk:mergecast-multicast-interpretation}
    One can represent the state preparation (SP) with an error channel by a perfect SP followed by a bit-flip channel with error parameter \(q_Z = s\), and similarly, represent the measurement (Meter) with error to a bit-flip channel with error parameter \(q_Z = m\) followed by a perfect measurement.
    From this point of view,
    the procedure in node \nodeAlice of Figure~\ref{fig:unicast-spam-estimate-s} can be interpreted as a ``\emph{\mergecast}'' of two channels in Figure~\ref{subfig:mergecast-root},
    and the procedure in node \nodeBob of Figure~\ref{fig:unicast-spam-estimate-s} can be interpreted as a ``multicast'' of two channels in Figure~\ref{subfig:multicast-end}.
\end{remark}

\subsection{QNT Protocols with SPAM Errors}\label{subsec:qnt-protocol-spam-extension}

After estimating the SPAM error parameters \(m\) and \(s\), one can easily extend the protocols of Section~\ref{sec:qnt-identifiability} to address QNT with SPAM errors.
The main difference is that the probability of the final measurement outcome \(\kett{0}\) in node \nodeBob nodes of these protocols in Section~\ref{sec:qnt-identifiability} becomes \(\frac{1 + f(m,s)q_{Z_1}q_{Z_2}}{2}\), for some known polynomial function \(f(m,s)\) of parameters \(m\) and \(s\), and its specific form depends on the protocol.
Although this prefactor \(f(m,s)\) will introduce additional errors to the estimates of the QNT protocol (illustrated via later experiments in Section~\ref{sec:qnt-experiment}), the identifiability of the channel parameters are guaranteed as in the no-SPAM error case.



\subsection{Integrated Workflow for QNT Channel Estimation}\label{subsec:qnt-spam-workflow}

This subsection outlines a systematic, actionable workflow for combining the QNT protocols introduced in this paper to estimate channel parameters in quantum networks. The procedure comprises two primary stages:
\begin{enumerate}
    \item \textbf{SPAM Error Characterization.}
          First, estimate the SPAM error parameters \(s\) and \(m\) by:
          \begin{itemize}
              \item If the monitors (i.e., peripheral end nodes) can perform \emph{both} state preparation and measurement, then employ the local protocols of prior works~\citep{liu2022quantum,jayakumar2024universal}; \emph{otherwise}
              \item Employing the network-based protocols proposed in Figures~\ref{fig:unicast-spam-estimate-s} (for \(s\)) and~\ref{fig:unicast-spam-estimate-m} (for \(m\)).
          \end{itemize}
    \item \textbf{Channel Parameter Estimation.}
          Next, determine the channel parameters \(q_{P,i}\) for each Pauli basis \(P \in \{X,Y,Z\}\) and any network channel \(i\), following a two-step process:
          \begin{enumerate}
              \item Simplify nodes with degree two (i..e, line patterns) as in Appendix~\ref{app:degree-2-nodes}, and then
              \item Apply \mergecast (Section~\ref{subsec:mergecast}) together with progressive etching (Section~\ref{subsec:progressive-etching}), processing channels sequentially from the network periphery toward its interior.
          \end{enumerate}
\end{enumerate}

This integrated workflow applies to general quantum networks with arbitrary Pauli channels. In the special case where all channels are bypassable, the procedure simplifies further: first deploy the unicast protocol of Figure~\ref{fig:spam-estimation-bypassable} to estimate the combined error parameter \(ms\); then individually estimate each channel parameter using \bypassunicast presented in Figures~\ref{fig:bypassable-unicast}.

\section{Experiments and Evaluations}\label{sec:qnt-experiment}

This section reports the experimental evaluation of \mergecast with and without SPAM errors.
Specifically, we investigate the performance of \mergecast with SPAM errors in Section~\ref{subsec:mergecast-experiment}, the performance of the SPAM error estimation protocols in Section~\ref{subsec:spam-estimation-exp}, and the performance of the progressive etching protocol in Section~\ref{subsec:etching-experiment}.
Last, we consider the impact of the quantum memory decoherence and photon loss on \mergecast in Section~\ref{subsec:photon-loss}.

\noindent\textbf{NetSquid simulation.} We use \texttt{NetSquid}~\citep{coopmans2021netsquid}, a Python-based simulator for quantum networks for our study.
For simplicity, we assume that all channels are depolarizing channels, with equal coefficients \(q_{X}\), \(q_{Y}\), and \(q_{Z}\).
We focus on estimating the Pauli coefficients \(q_{Z}\), as the other two coefficients \(q_{X}\) and \(q_{Y}\) can be estimated similarly via channel dressing.
SPAM errors are simulated via bit-flip channels with parameters \(s\) and \(m\), after perfect state preparation and before perfect measurement, respectively, as illustrated in Remark~\ref{rmk:mergecast-multicast-interpretation}.

We focus on the mean squared error (MSE) of the estimates.
All experiments are conducted for \(1000\) trials with the MSE taken as an average over all of the trials. The MSE is plotted as a line curve and its standard deviation as a shaded region.

\subsection{Performance of \mergecast on a Star Network}\label{subsec:mergecast-experiment}

In this first subsection, we conduct experiments to evaluate the performance of \mergecast on the most basic star network topology.

\begin{figure}[b]
    \centering
    \begin{subfigure}{0.245\textwidth}
        \centering
        \includegraphics[width=\textwidth]{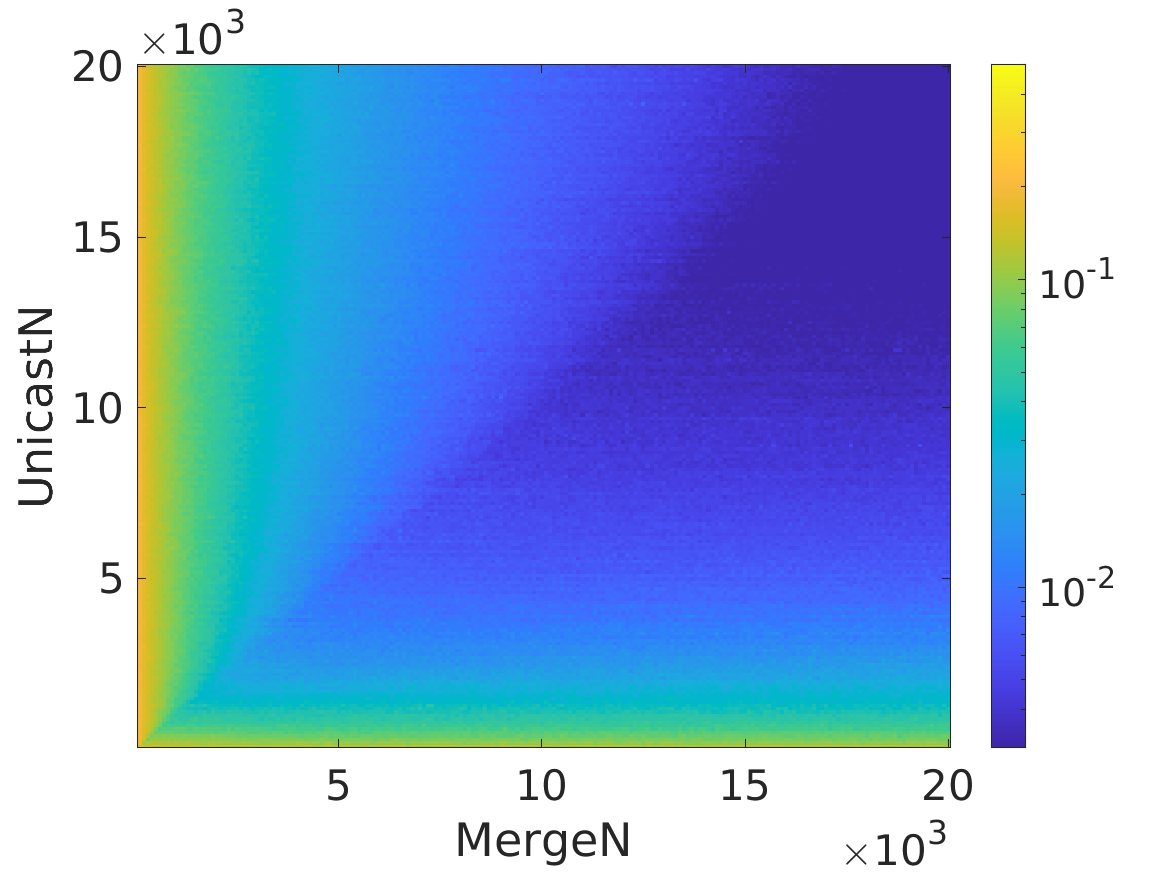}
        \caption{\(s,m=1\)}
        \label{subfig:mergecast-3d}
        \label{fig:mergecast-spam-heatmap-perfect}
    \end{subfigure}
    \hfill
    \begin{subfigure}{0.245\textwidth}
        \centering
        \includegraphics[width=\textwidth]{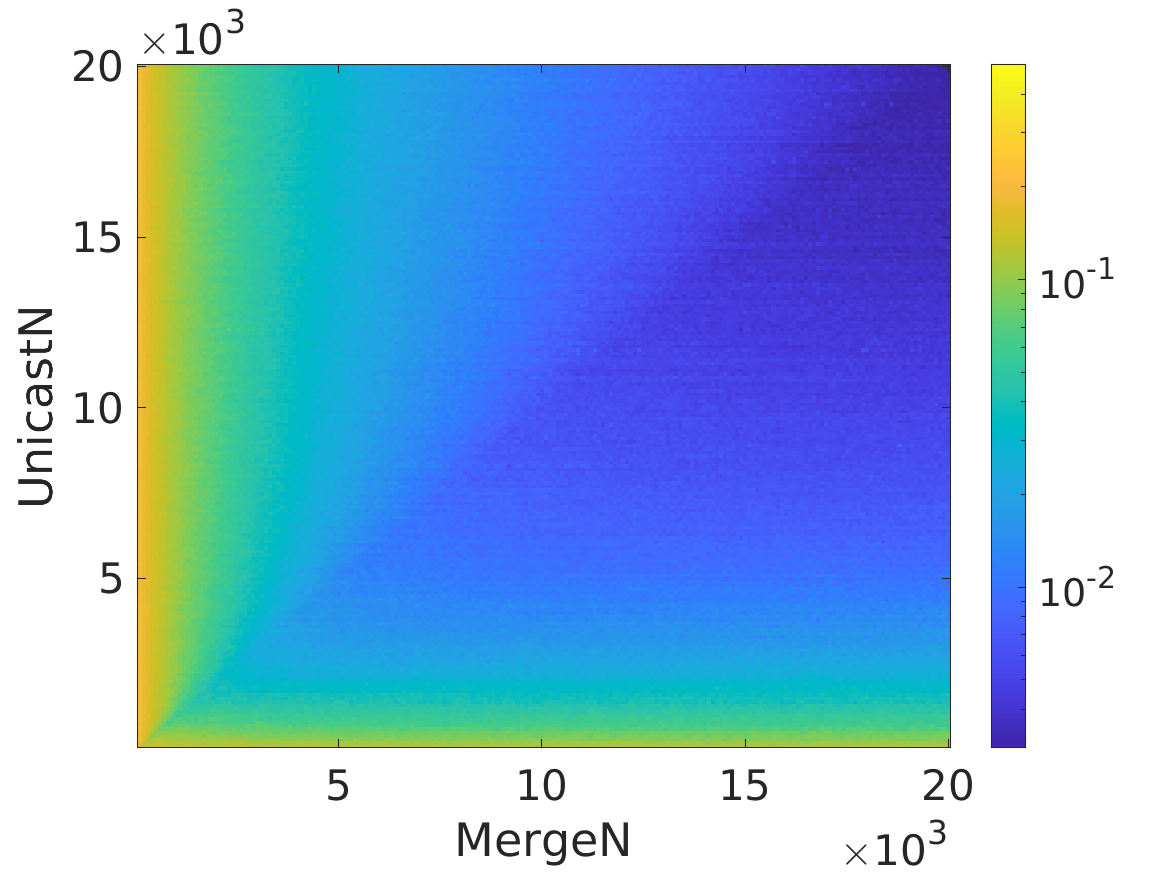}
        \caption{\(s,m=0.95\)}
        \label{subfig:mergecast-spam-95-3d}
    \end{subfigure}
    \hfill
    \begin{subfigure}{0.245\textwidth}
        \centering
        \includegraphics[width=\textwidth]{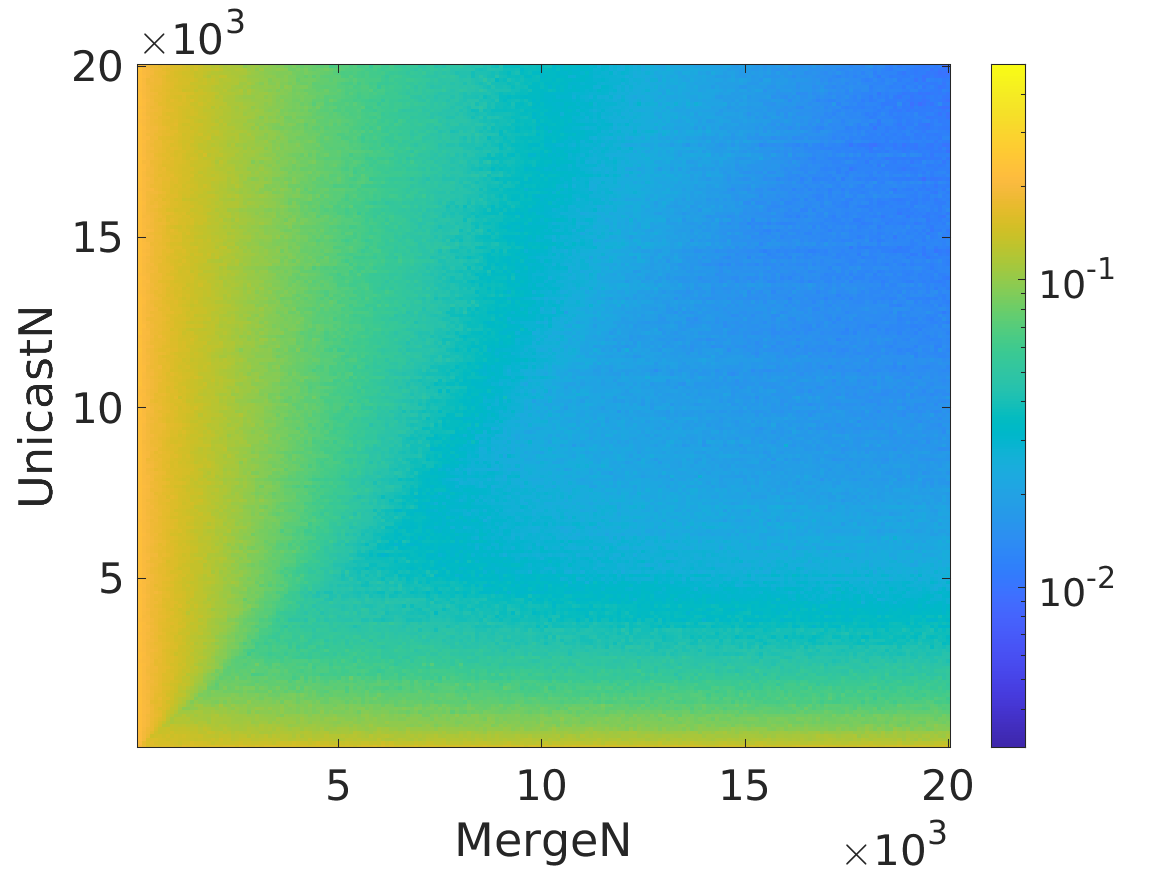}
        \caption{\(s,m=0.80\)}
        \label{subfig:mergecast-spam-80-3d}
    \end{subfigure}
    \hfill
    \begin{subfigure}{0.245\textwidth}
        \centering
        \includegraphics[width=\textwidth]{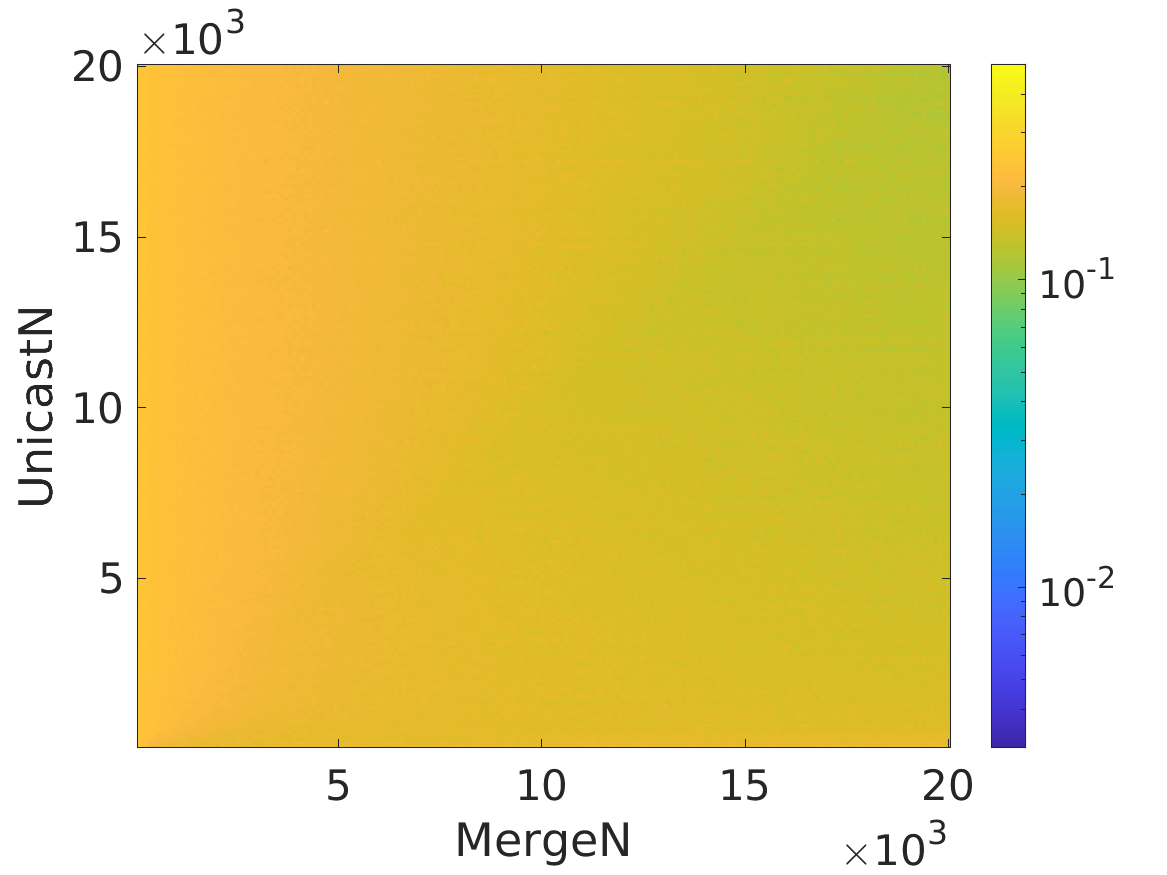}
        \caption{\(s,m=0.50\)}
        \label{subfig:mergecast-spam-50-3d}
    \end{subfigure}
    \\
    \begin{subfigure}{0.245\textwidth}
        \centering
        \includegraphics[width=\textwidth]{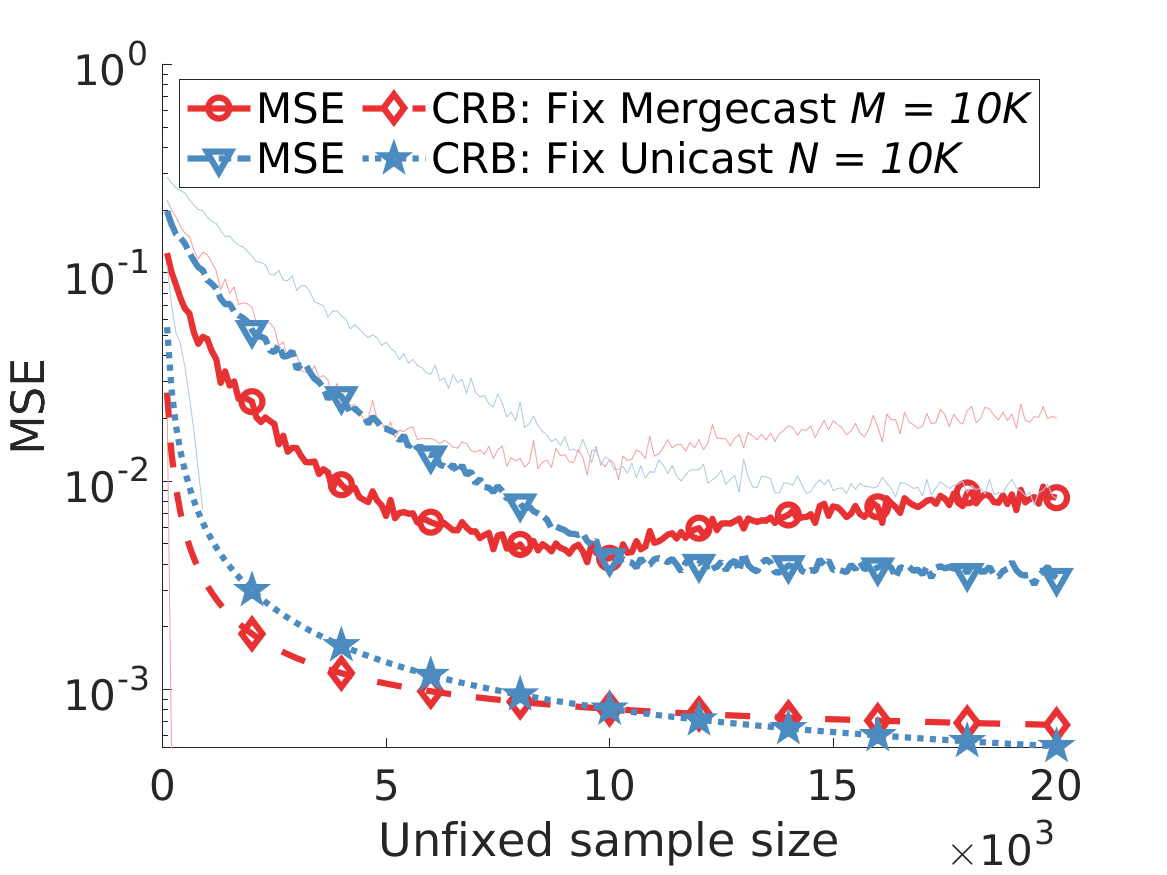}
        \caption{\(s,m=1\)}
        \label{subfig:mergecast}
    \end{subfigure}
    \hfill
    \begin{subfigure}{0.245\textwidth}
        \centering
        \includegraphics[width=\textwidth]{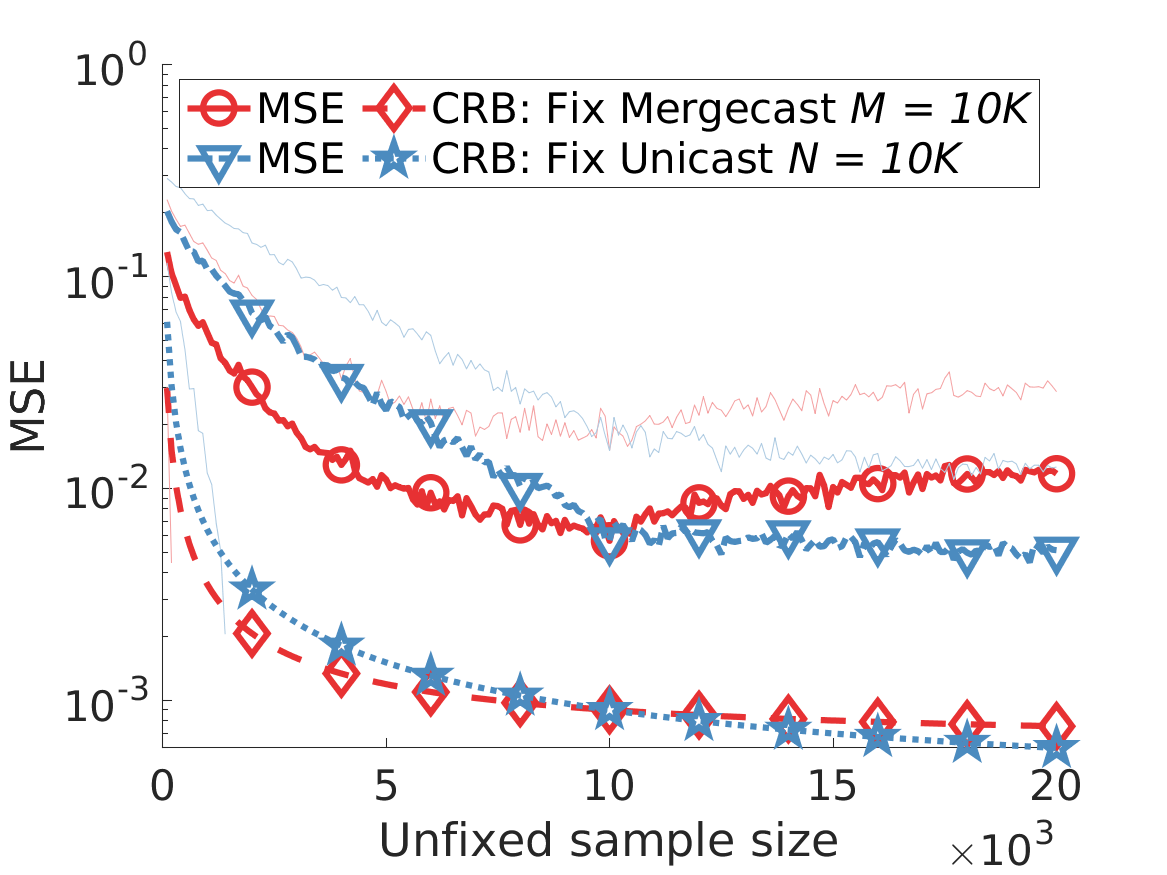}
        \caption{\(s,m=0.95\)}
        \label{subfig:mergecast-spam-95}
    \end{subfigure}
    \hfill
    \begin{subfigure}{0.245\textwidth}
        \centering
        \includegraphics[width=\textwidth]{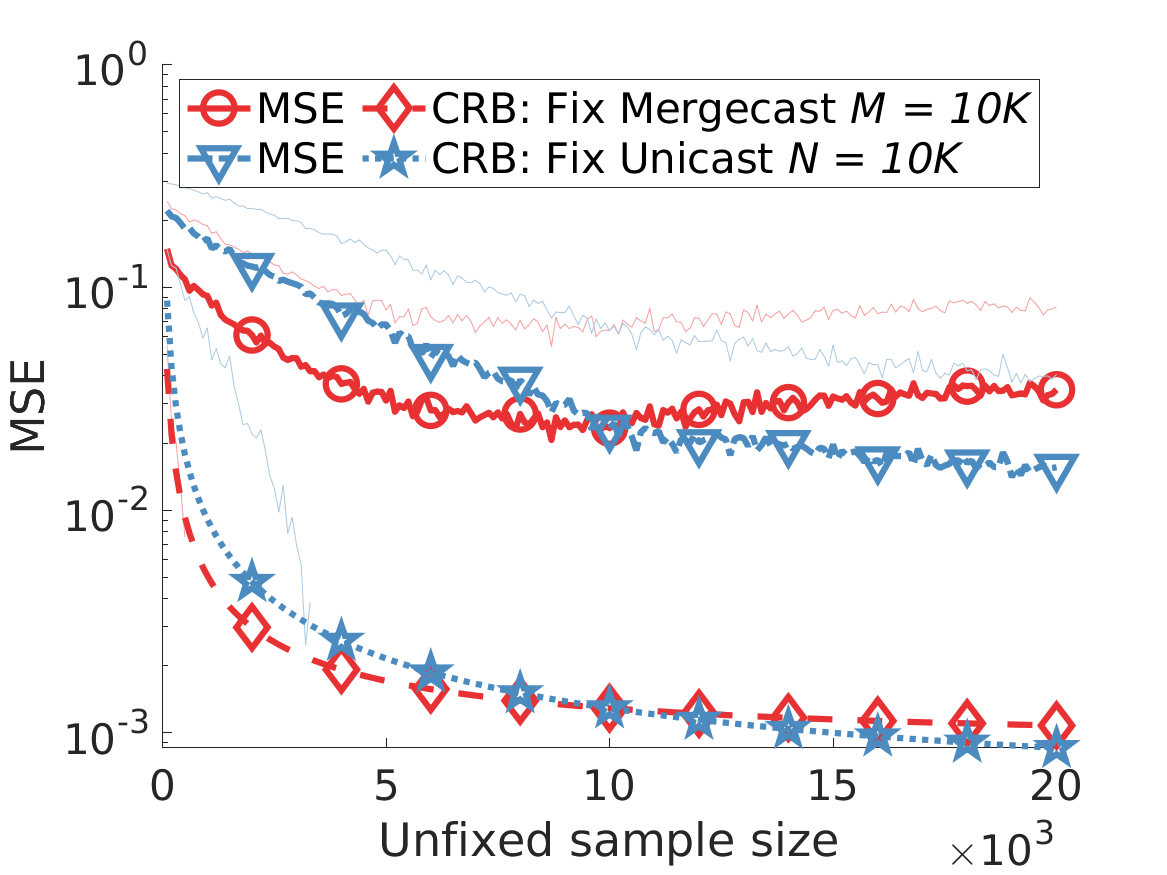}
        \caption{\(s,m=0.80\)}
        \label{subfig:mergecast-spam-80}
    \end{subfigure}
    \hfill
    \begin{subfigure}{0.245\textwidth}
        \centering
        \includegraphics[width=\textwidth]{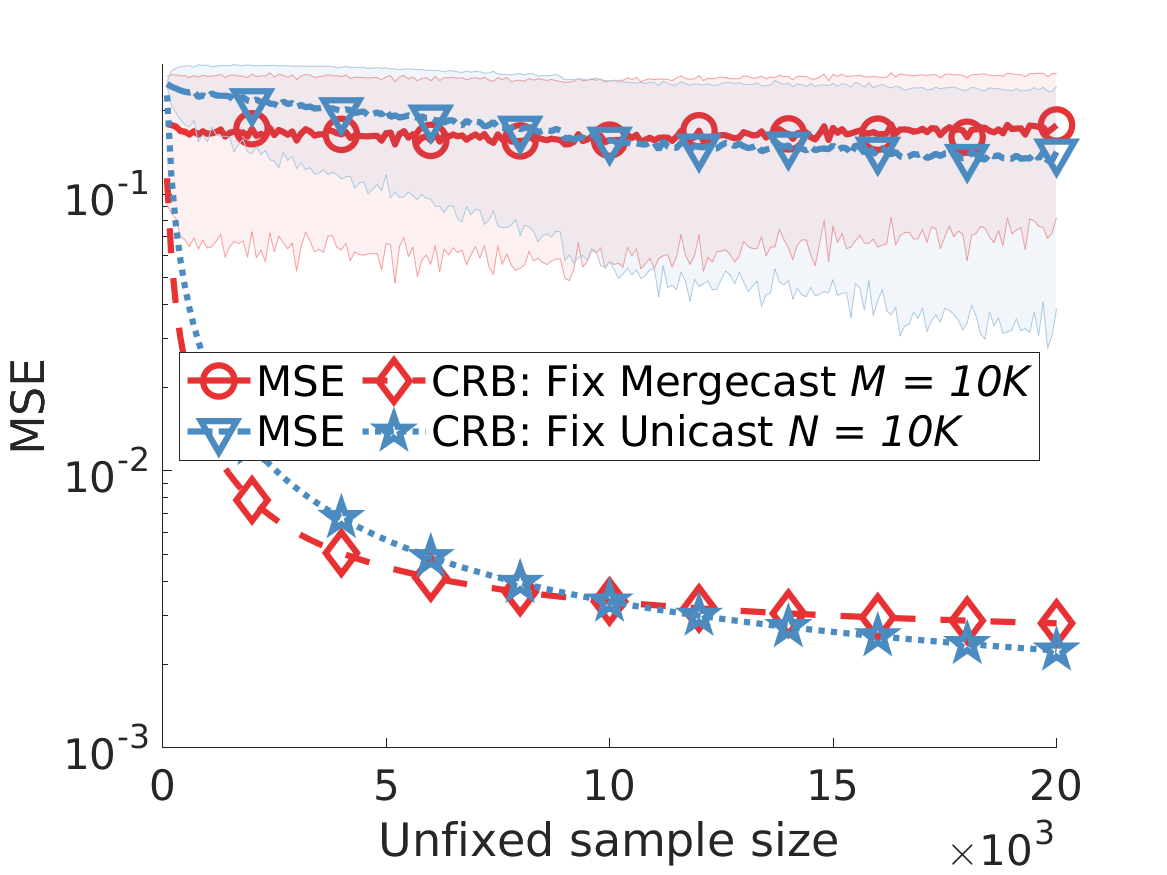}
        \caption{\(s,m=0.50\)}
        \label{subfig:mergecast-spam-50}
    \end{subfigure}
    \caption{MSE of \mergecast with different SPAM errors: (a)--(d) vary both sample sizes \(N\) and \(M\),
        and (e)--(h) fix one, vary the other}
    \label{fig:mergecast-spam-heatmap}
    \label{fig:mergecast-spam-fix}
\end{figure}

\noindent\textbf{Setup.} We consider a three-channel star network with \(q_{Z,1}=0.5, q_{Z,2}=0.25, q_{Z,3}=0.35\) (corresponding to depolarization probabilities \(p_1 = 0.5, p_2 = 0.75\) and \(p_3 = 0.65\)) and focus on estimating \(q_{Z,1}\).
We consider four different SPAM error scenarios, \(s=m=1, 0.95, 0.80, 0.50\) (known).
Recall \(s=m=1\) is the perfect case with no SPAM errors.
Recall that the estimator for \(q_{Z,1}\) in~\eqref{eq:mergecast-estimator} needs two sets of samples, one from \mergecast itself and the other from unicast,
whose sample sizes are denoted as \(M\) and \(N\), respectively.
In Figure~\ref{fig:mergecast-spam-heatmap},
we very \(M\) and \(N\) from \(100\) to \(20000\) in steps of size \(100\) to generate the mesh heatmaps.
In Figure~\ref{fig:mergecast-spam-fix},
we fix one of the sample sizes as \(10000\) and vary the other from \(100\) to \(20000\) in a steps of  size \(100\).

\noindent\textbf{Observations.}
Figures~\ref{fig:mergecast-spam-heatmap-perfect},~\ref{subfig:mergecast-spam-95-3d},~\ref{subfig:mergecast-spam-80-3d},~\ref{subfig:mergecast-spam-50-3d} show the MSE of \mergecast with SPAM errors in a heatmap.
When there are no SPAM errors (\(s,m=1\)), Figure~\ref{fig:mergecast-spam-heatmap-perfect},
the MSE decreases as sample sizes \(M\) and \(N\) increase.
Furthermore, the heatmap suggests that
MSE is minimized when \(M\) and \(N\) are close in value.
Similar observations can be made for the other three cases in Figure~\ref{fig:mergecast-spam-heatmap}.
Second, across these four heatmaps, for fixed \(M\) and \(N\), the MSE increases (as the color becomes lighter) with the value of the SPAM errors. This aligns with the intuition that the SPAM errors degrade the performance of \mergecast.
Figures~\ref{subfig:mergecast},~\ref{subfig:mergecast-spam-95},~\ref{subfig:mergecast-spam-80},~\ref{subfig:mergecast-spam-50} show the MSE of \mergecast when fixing one of two sample sizes and varying the other.
We add the Cram\'er-Rao bound (CRB)  to compare with the MSE.
The CRB is a theoretical lower bound of the MSE for \emph{unbiased} estimators~\citep{lehmann2006theory}, and we defer its detailed expression to Appendix~\ref{app:spam-crb}.
Although the estimator for \(q_{Z,1}\) of \mergecast produces a biased estimate, and the CRB is not a strict lower bound for the MSE of a biased estimate, it provides a reference for the performance of \mergecast.
Similar observations to Figure~\ref{fig:mergecast-spam-heatmap} can be made for Figure~\ref{fig:mergecast-spam-fix}, where the V shapes of MSE curves (especially for the red lines) correspond to the darker diagonal lines in the heatmap.

\subsection{Performance of SPAM Error Estimations}\label{subsec:spam-estimation-exp}

The above subsection investigates the performance of \mergecast under the \emph{known} SPAM errors.
In practice, SPAM errors may be unknown, and one need to first apply the error estimation protocols proposed in Section~\ref{sec:qnt-spam}.
In this subsection, we evaluate the performance of these protocols.

\begin{figure}[b]
    \centering
    \begin{subfigure}{0.245\textwidth}
        \centering
        \includegraphics[width=\textwidth]{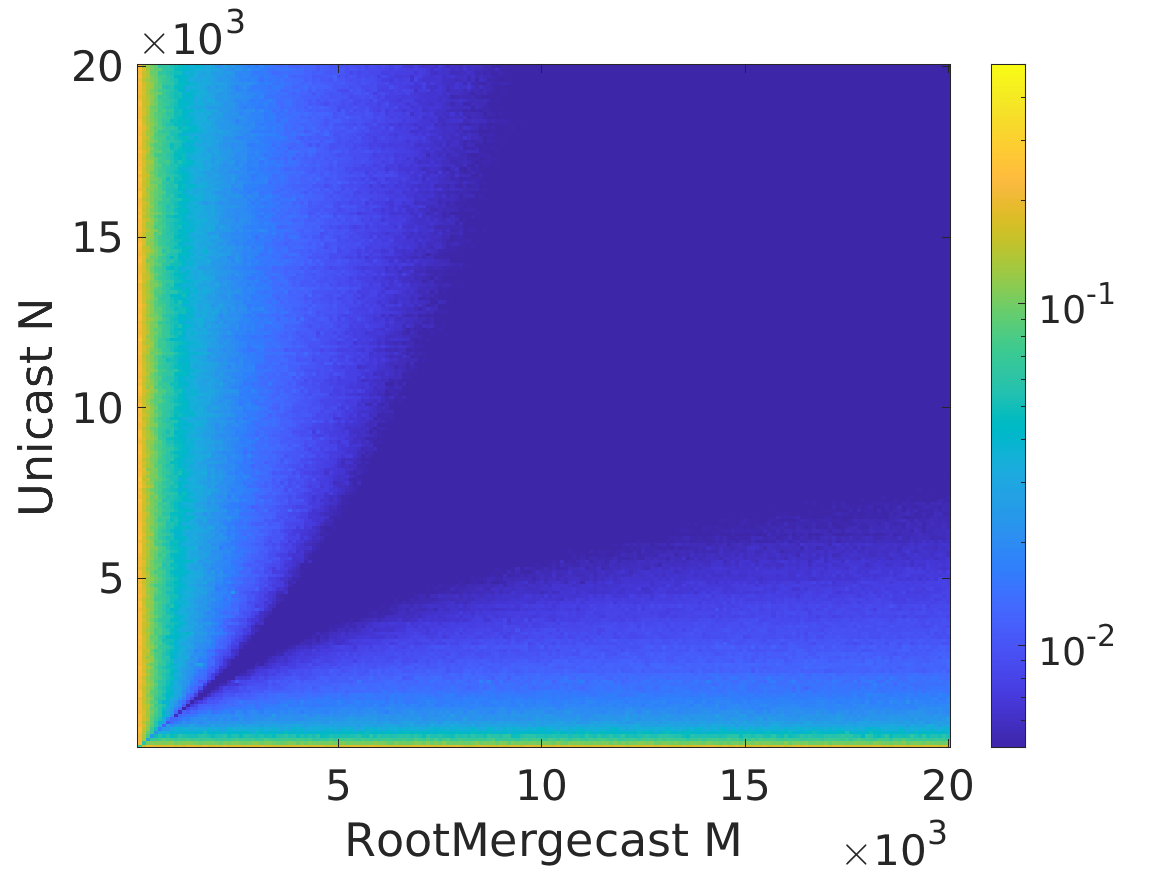}
        \caption{Estimate \(s\) (\(s,m=0.90\))}
        \label{subfig:spam-s-3d-90}
    \end{subfigure}
    \hfill
    \begin{subfigure}{0.245\textwidth}
        \centering
        \includegraphics[width=\textwidth]{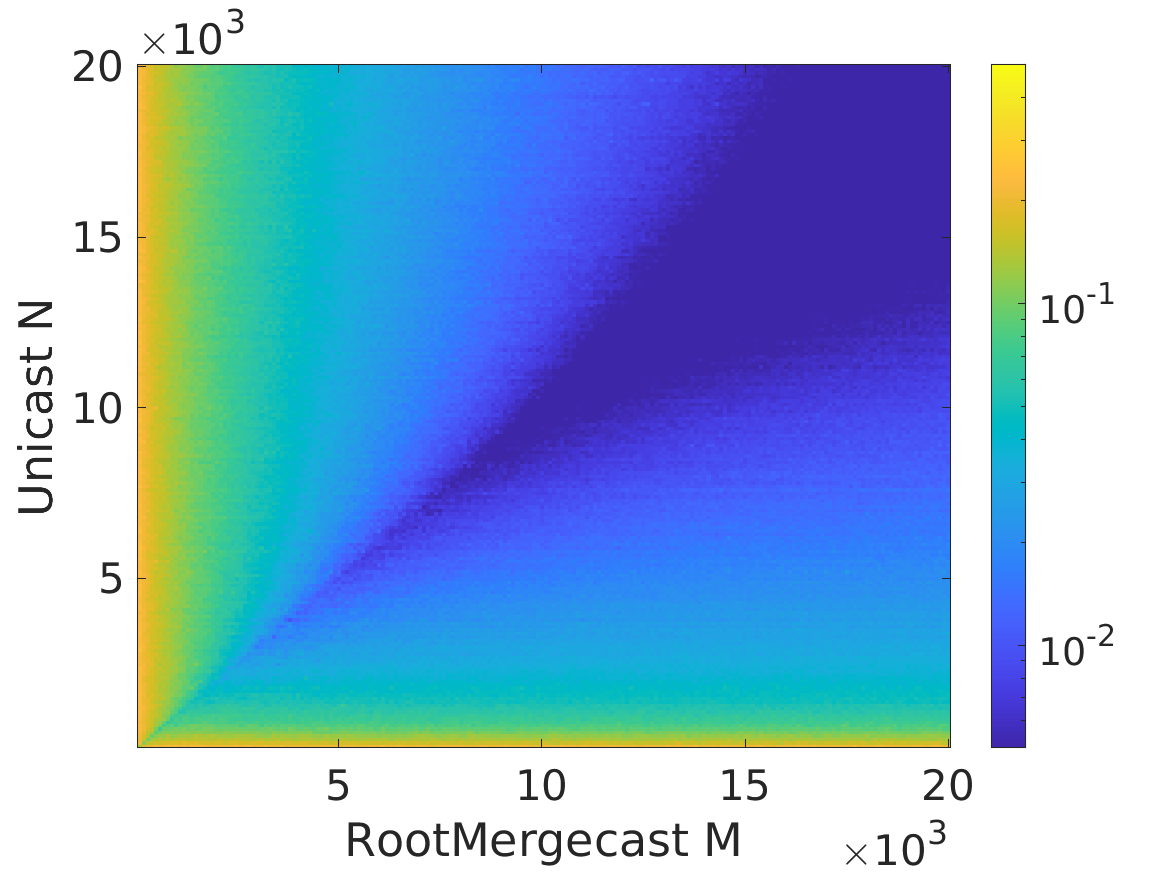}
        \caption{Estimate \(s\) (\(s,m=0.70\))}
        \label{subfig:spam-s-3d-70}
    \end{subfigure}
    \hfill
    \begin{subfigure}{0.245\textwidth}
        \centering
        \includegraphics[width=\textwidth]{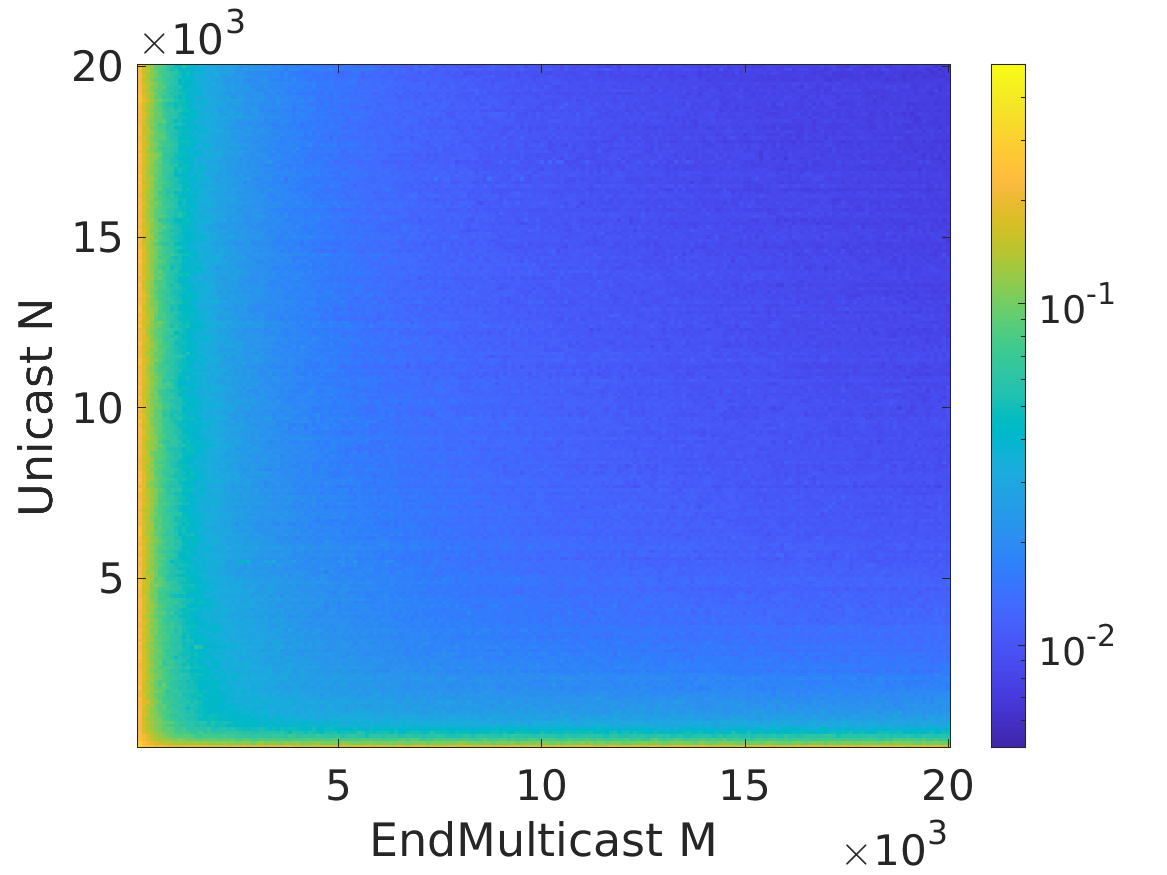}
        \caption{Estimate \(m\) (\(s,m=0.90\))}
        \label{subfig:spam-m-3d-90}
    \end{subfigure}
    \hfill
    \begin{subfigure}{0.245\textwidth}
        \centering
        \includegraphics[width=\textwidth]{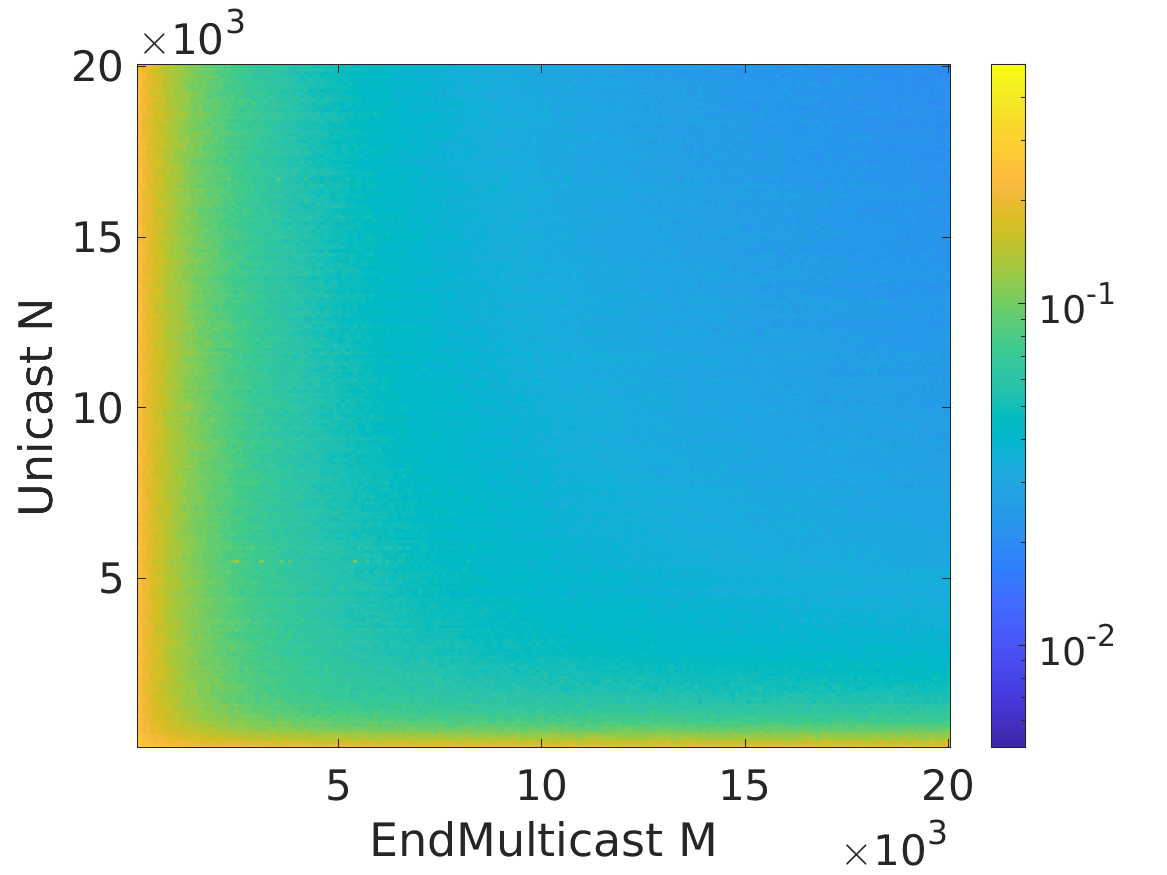}
        \caption{Estimate \(m\) (\(s,m=0.70\))}
        \label{subfig:spam-m-3d-70}
    \end{subfigure}
    \\
    \begin{subfigure}{0.245\textwidth}
        \centering
        \includegraphics[width=\textwidth]{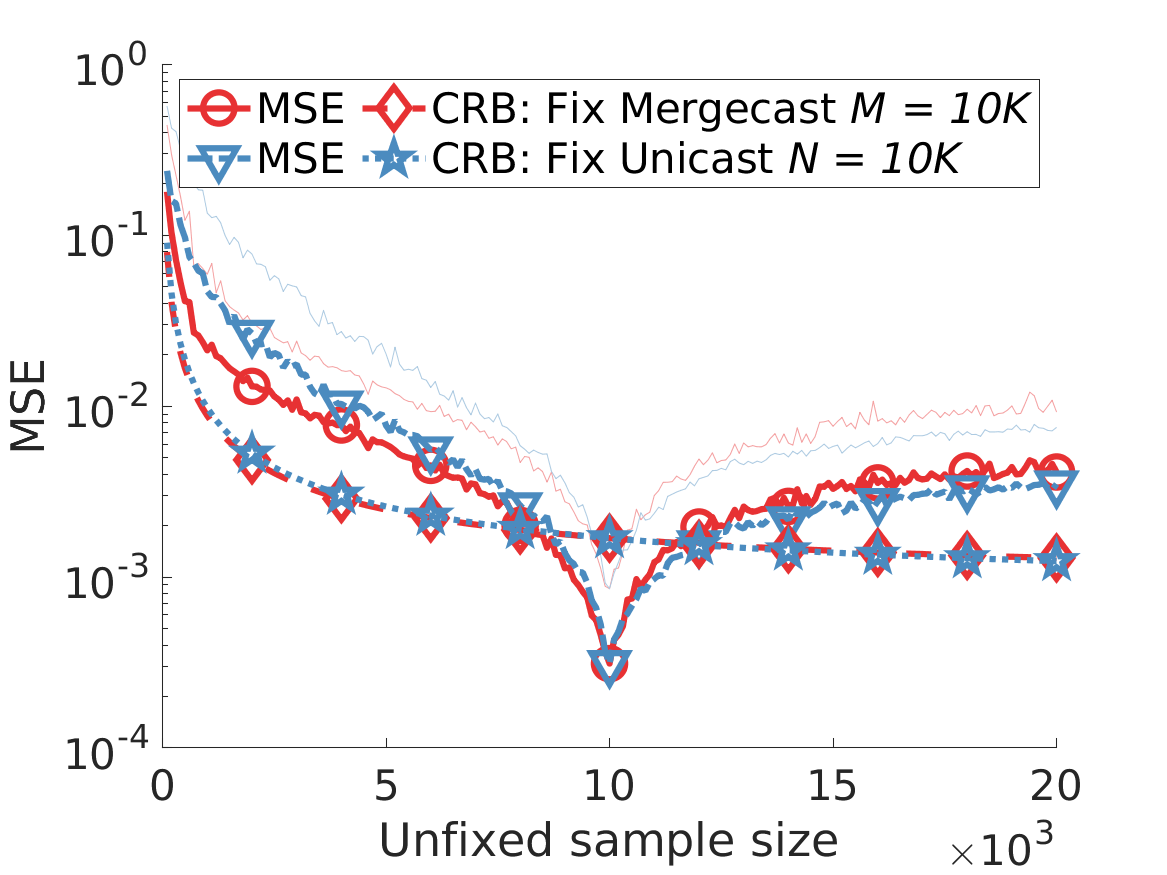}
        \caption{Estimate \(s\) (\(s,m=0.90\))}
        \label{subfig:spam-s-90}
    \end{subfigure}
    \hfill
    \begin{subfigure}{0.245\textwidth}
        \centering
        \includegraphics[width=\textwidth]{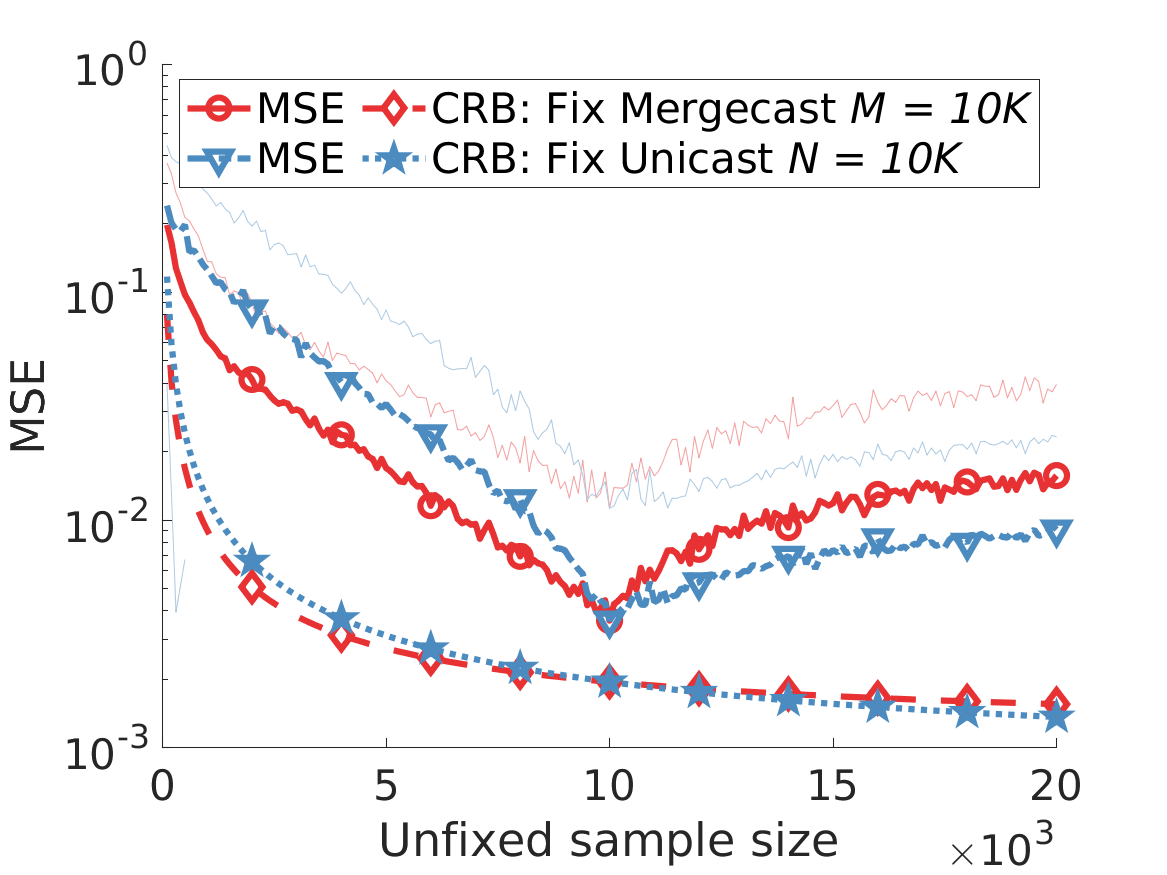}
        \caption{Estimate \(s\) (\(s,m=0.70\))}
        \label{subfig:spam-s-70}
    \end{subfigure}
    \hfill
    \begin{subfigure}{0.245\textwidth}
        \centering
        \includegraphics[width=\textwidth]{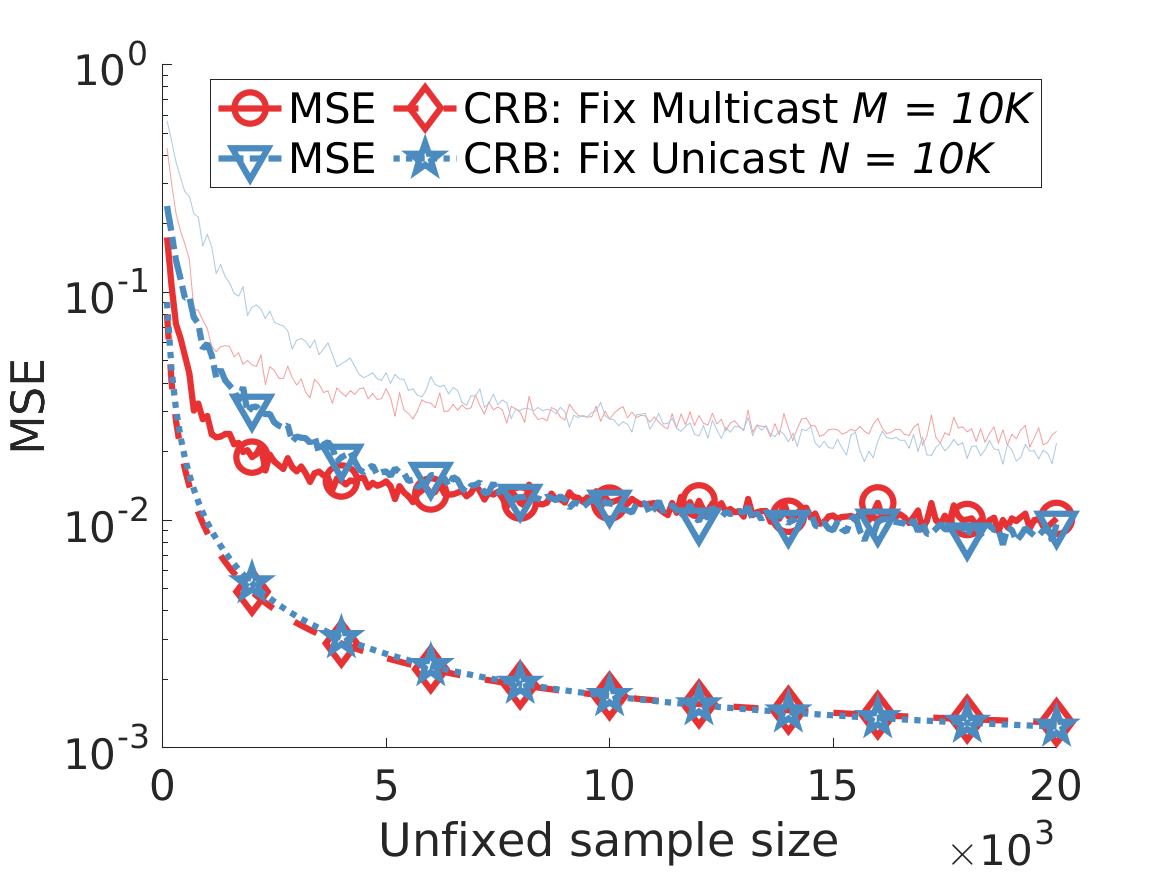}
        \caption{Estimate \(m\) (\(s,m=0.90\))}
        \label{subfig:spam-m-90}
    \end{subfigure}
    \hfill
    \begin{subfigure}{0.245\textwidth}
        \centering
        \includegraphics[width=\textwidth]{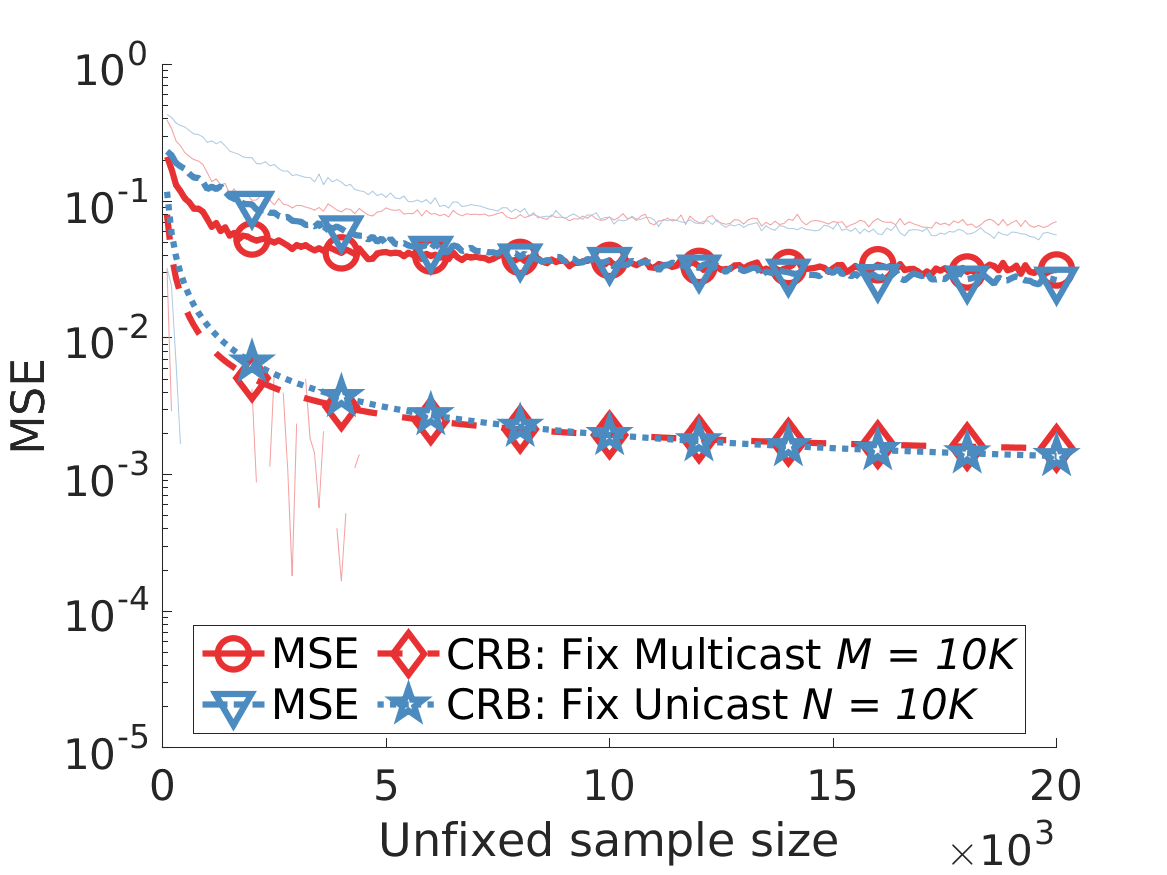}
        \caption{Estimate \(m\) (\(s,m=0.70\))}
        \label{subfig:spam-m-70}
    \end{subfigure}
    \caption{MSE of estimating SPAM error parameters \(s\) and \(m\): (a)--(d) vary both sample sizes  \(N\) and \(M\), and (e)--(h) fix one, vary the other}
    \label{fig:spam-fix-one}
    \label{fig:spam-heatmap}
\end{figure}

\noindent\textbf{Setup.} We consider two consecutive channels with parameters \(q_{Z,1}=0.5\) and \(q_{Z,2}=0.25\) (as Figure~\ref{fig:unicast-spam-estimate-s} shows) with two cases of SPAM errors: \(s,m=0.90\) and \(s,m=0.70\).
Similar to \mergecast,
the protocol for estimating the preparation error \(s\) (resp., measurement error \(m\)) needs two sets of samples, one from unicast and the other from the ``root \mergecast'' in Figure~\ref{fig:unicast-spam-estimate-s}
(resp., ``end multicast'' in Figure~\ref{fig:unicast-spam-estimate-m}),
and we slightly abuse the notation by using \(N\) and \(M\) to denote the sample sizes of the unicast and the RootMergecast (resp., EndMulticast), respectively.
Other settings are the same as Section~\ref{subsec:mergecast-experiment}.

\noindent\textbf{Observations.}
Figures~\ref{subfig:spam-s-3d-90} and~\ref{subfig:spam-s-3d-70} report the mean squared error (MSE) of estimating the preparation error parameter \(s\) under two SPAM error scenarios (\(s,m=0.90\) and \(s,m=0.70\)), visualized as heatmaps. Both heatmaps exhibit patterns similar to the \mergecast results in Figure~\ref{fig:mergecast-spam-heatmap}, indicating that the MSE is minimized when the unicast sample size \(N\) and the root-mergecast sample size \(M\) are comparable. Figures~\ref{subfig:spam-s-90} and~\ref{subfig:spam-s-70} further illustrate the MSE when one of the sample sizes is fixed at \(10{,}000\) while the other varies from \(100\) to \(20{,}000\) in increments of \(100\). In Figure~\ref{subfig:spam-s-90}, the resulting MSE curves exhibit a characteristic V-shape, in some cases even dipping below the Cramér-Rao bound (CRB).
Figures~\ref{subfig:spam-m-3d-90} and~\ref{subfig:spam-m-3d-70} show the MSE of estimating the measurement error parameter \(m\) under the same SPAM scenarios, again as heatmaps over unicast sample size \(N\) and end-multicast sample size \(M\). Figures~\ref{subfig:spam-m-90} and~\ref{subfig:spam-m-70} display the corresponding MSE curves alongside the CRB when one sample size is fixed at \(10{,}000\). In this case, the observed MSEs closely follow the CRB. Although the general trends resemble those of the \(s\)-estimator, no V-shaped behavior appears—this is due to the different coefficients in the estimator for \(\hat{m}\), which prevent the same type of error cancellation.

\noindent\textbf{Explanation of the V-shape.}
This V-shaped behavior arises from bias in the estimators---specifically, from their ratio form in the \mergecast and the preparation error parameter \(s\) cases. As an example, rewrite the estimator for \(s\) in~\eqref{eq:spam-estimate-s} as
\(
\hat{s} = \frac{\hat{X}}{\hat{Y}},
\)
where \(\hat{X}\) and \(\hat{Y}\) denote the empirical means of the \(\kett{0}\) outcome frequencies from root-mergecast and unicast, respectively. Applying the delta method and Taylor approximations yields
\begin{align}
    \hat{s}
    = \frac{\E[\hat{X}] + \delta_X}{\E[\hat{Y}] + \delta_Y}
    \approx \frac{\E[\hat{X}]}{\E[\hat{Y}]}
    + \frac{1}{\E[\hat{Y}]} \delta_X
    - \frac{\E[\hat{X}]}{\E[\hat{Y}]^2} \delta_Y
    \overset{(\text{a})}{=} s
    + \frac{1}{\E[\hat{Y}]} \delta_X(M)
    - \frac{\E[\hat{X}]}{\E[\hat{Y}]^2},
\end{align}
where \(\delta_X = \hat{X}-\E[\hat{X}]\) and \(\delta_Y = \hat{Y}-\E[\hat{Y}]\), and the dependence of \(\delta_X\) and \(\delta_Y\) on sample sizes \(M\) and \(N\) is made explicit in equation (a).
From the above expression, the approximate MSE is
\begin{align}\label{eq:mse-approx}
    \mathrm{MSE}(M,N) \approx \E\!\left[\! \left( \frac{1}{\E[\hat{Y}]} \delta_X(M)
        - \frac{\E[\hat{X}]}{\E[\hat{Y}]^2} \delta_Y(N) \right)^{\!2} \right],
\end{align}
with expectation taken over repeated experiments. Since both \(\hat{X}\) and \(\hat{Y}\) are averages of Bernoulli random variables, their fluctuations \(\delta_X(M)\) and \(\delta_Y(N)\) converge at similar rates when \(M\) and \(N\) are comparable. In particular, when
\(
\frac{1}{\E[\hat{Y}]} \approx \frac{\E[\hat{X}]}{\E[\hat{Y}]^2},
\)
the error terms in~\eqref{eq:mse-approx} partially cancel, minimizing the MSE when \(M \approx N\). As \(M\) and \(N\) diverge, this cancellation diminishes, leading to the V-shaped curve observed in Figure~\ref{subfig:spam-s-90}.

\subsection{Performance of Progressive Etching Procedure}\label{subsec:etching-experiment}

While Section~\ref{subsec:mergecast-experiment} reports the performance of \mergecast in a star network, when the network topology becomes complex (e.g., Figure~\ref{fig:qnt}),
one needs to conduct the progressive etching procedure (Section~\ref{subsec:progressive-etching}) to apply \mergecast several times (first peripheral channels, then internal ones).
In this subsection, we study the performance of \mergecast for estimating peripheral and internal channels in progressive etching in  Figure~\ref{fig:qnt}'s topology.

\begin{wrapfigure}{r}{0.4\textwidth}
    \centering
    \includegraphics[width=0.35\textwidth]{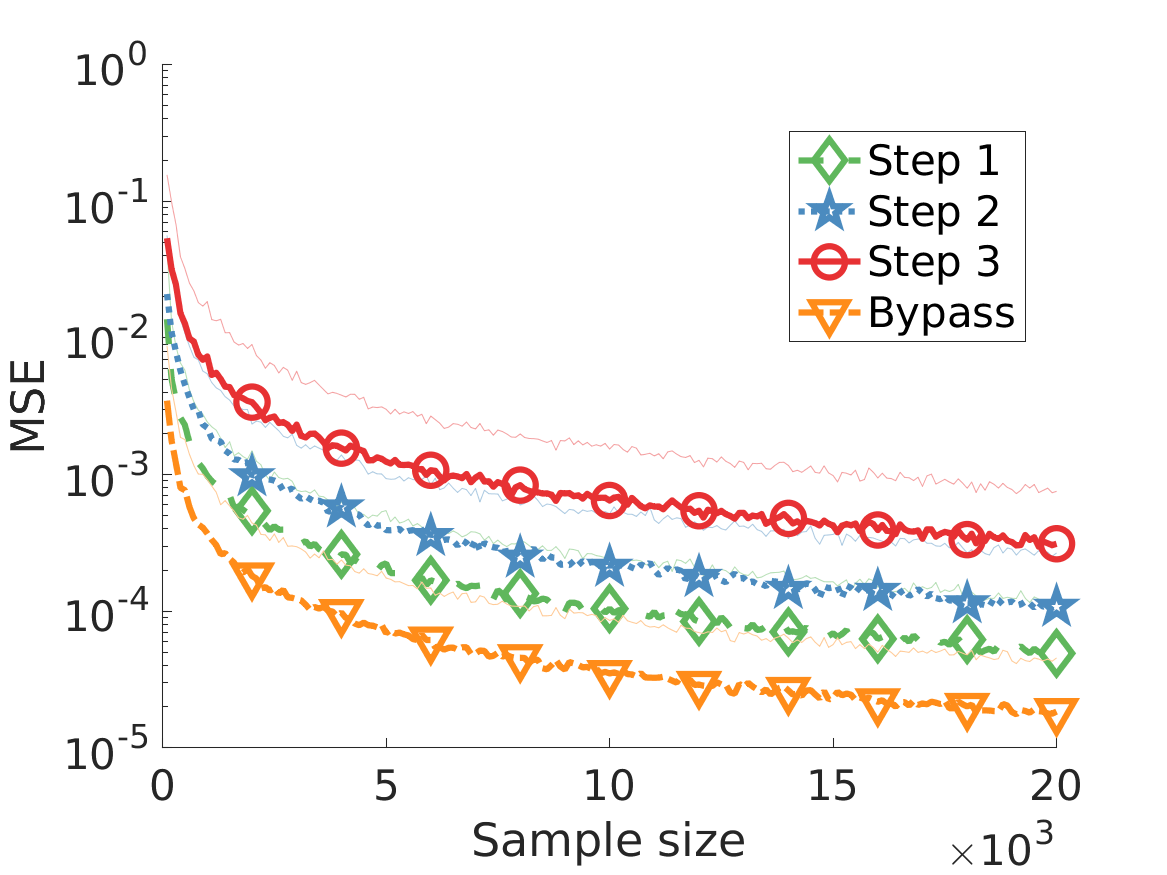}
    \caption{MSE of progressive etching in the three steps (Step 1, Step 2, and Step 3) of Figure~\ref{fig:toy-example-general-tomograph} and comparing to the \bypassunicast (Bypass in the legend)}\label{fig:ethcing-experiment}\label{subfig:ethcing-with-crb}
\end{wrapfigure}

\noindent\textbf{Setup.}
We follow the progressive etching procedure in Figure~\ref{fig:toy-example-general-tomograph}, consisting of three steps:
Step 1. {peripheral} channels \(\mathcal{P}_{12}, \mathcal{P}_{13},\dots, \mathcal{P}_{19}\);
Step 2. {intermediate} channels \(\mathcal{P}_2, \mathcal{P}_3, \dots, \mathcal{P}_{11}\);
and
Step 3. the channel \(\mathcal{P}_1\) at the center.
To mitigate the impact of different channel parameters, we pick the same \(q_{Z,\ell} = 0.8\) for all channels \(\ell\) in the network.
As \mergecast relies on two sets of samples from the unicast and the mergecast, we vary the sample sizes of mergecast from \(M=100\) to \(20000\) in a step size of \(100\) and set the sample sizes of unicast \(N\) the same as mergecast for all cases (as the case \(M = N\) attains low MSEs as shown above).
When proceeding to estimate an inner channel in the next etching step, we use the estimated channel parameters from the previous steps at the sample sizes \(N=M=10000\).
To compare with the performance of \bypassunicast for the case with bypassable channels (Figure~\ref{fig:bypassable-unicast}), we also plot its MSE when varying its sample sizes from \(100\) to \(20000\) in a step size of \(100\) and assuming the identified channels are bit-flip (bypassable) channels with parameter \(q_{Z,\ell} = 0.8\).

\noindent\textbf{Observations.}
Figure~\ref{subfig:ethcing-with-crb} reports the MSEs of the progressive etching protocol for estimating three channels \(\mathcal{P}_{12}\) in the first step, \(\mathcal{P}_3\) in the second step, and \(\mathcal{P}_1\) in the third step.
As expected, the MSE of the first step is the lowest, and the MSE of the second step is higher than that of the first step, and the MSE of the third step is the highest, which illustrates the error propagation in the progressive etching process.
On the other hand, the MSEs of \bypassunicast are the lowest among all three steps, as it bypasses all the peripheral and intermediate channels and directly estimates the target channels. This shows the advantage of applying \bypassunicast when the channels in a network are bypassable, as suggested in the workflow of Section~\ref{subsec:qnt-spam-workflow}.

\subsection{Realistic Performance of \mergecast with Photon Loss and Memory Decoherence}
\label{subsec:photon-loss}

In this subsection, we evaluate the performance of the \mergecast protocol under realistic photon loss conditions in photonic quantum networks using the \texttt{NetSquid} simulator. Photon loss---due to absorption or scattering in optical fibers---causes the two states directly transmitted from peripheral node \nodeAlice to arrive at intermediate node \nodeCharlie at different times.
The first-arriving qubit must be held in a \emph{quantum memory} until its partner arrives, during which it experiences decoherence. We study how both photon loss and memory decoherence affect the accuracy of the final estimate.
The experiments demonstrate the robust performance of \mergecast.

\noindent\textbf{Network and loss model.}
We adopt the same three-channel star network as in Section~\ref{subsec:mergecast-experiment}, with true parameters
\(
q_{Z,1}=0.50,\; q_{Z,2}=0.25,\; q_{Z,3}=0.35
\),
and focus on estimating \(q_{Z,1}\).  Each fiber has length \(L=10\,\mathrm{km}\) and propagation speed \(v=2\times10^{5}\,\mathrm{km/s}\).  Photon loss is modeled by an initial loss probability \(p_{0}=0.50\) and an additional loss rate \(\alpha=0.05\) per kilometer (in total, around \(69.7\%\) loss)~\citep{agrawal2012fiber}.  Both Roots transmit their qubits simultaneously in periodic intervals chosen from
\(
T_{\mathrm{send}}\in\{0.1,\,0.3,\,0.5,\,0.7,\,0.9\}\,\mathrm{s}.
\)

\noindent\textbf{Memory decoherence and cutoff.}
We model the quantum memory with two standard decoherence times: relaxation \(T_{1}=10\,\mathrm{s}\) and dephasing \(T_{2}=1\,\mathrm{s}\), typical of ion trap platforms~\citep{ionq-aria-practical-performance}.  Over a storage interval \(\Delta t\), the density matrix evolves under both amplitude-damping and phase-damping channels; when \(\Delta t\) reaches \(T_{1}\) or \(T_{2}\), the state becomes effectively ground or maximally mixed.
To prevent excessive noise accumulation, we introduce a \emph{cutoff time} \(T_{c}\), beyond which the stored qubit is discarded.  We consider
\(
T_{c}\in\{0.05,\,0.35,\,0.75,\,5,\,10\}\,\mathrm{s},
\)
where \(T_{c}=0.05\,(< 0.1)\) enforces perfect synchrony (no memory noise), and larger \(T_{c}\) allows increasingly noisy storage.

\noindent\textbf{Simulation details.}
For each pair \((T_{\mathrm{send}},T_{c})\), we run \mergecast continuously for \(3600\,\mathrm{s}\) (one hour) and repeat each setting over \(1000\) trials.  We record the total number of successful qubits that are merged at node \nodeCharlie and are received at node \nodeBob, as well as the resulting mean square error (MSE) in the estimate of \(q_{Z,1}\), averaging over \(1000\) trials.

\begin{figure}[t]
    \centering
    \begin{subfigure}{0.35\linewidth}
        \centering
        \includegraphics[width=\linewidth]{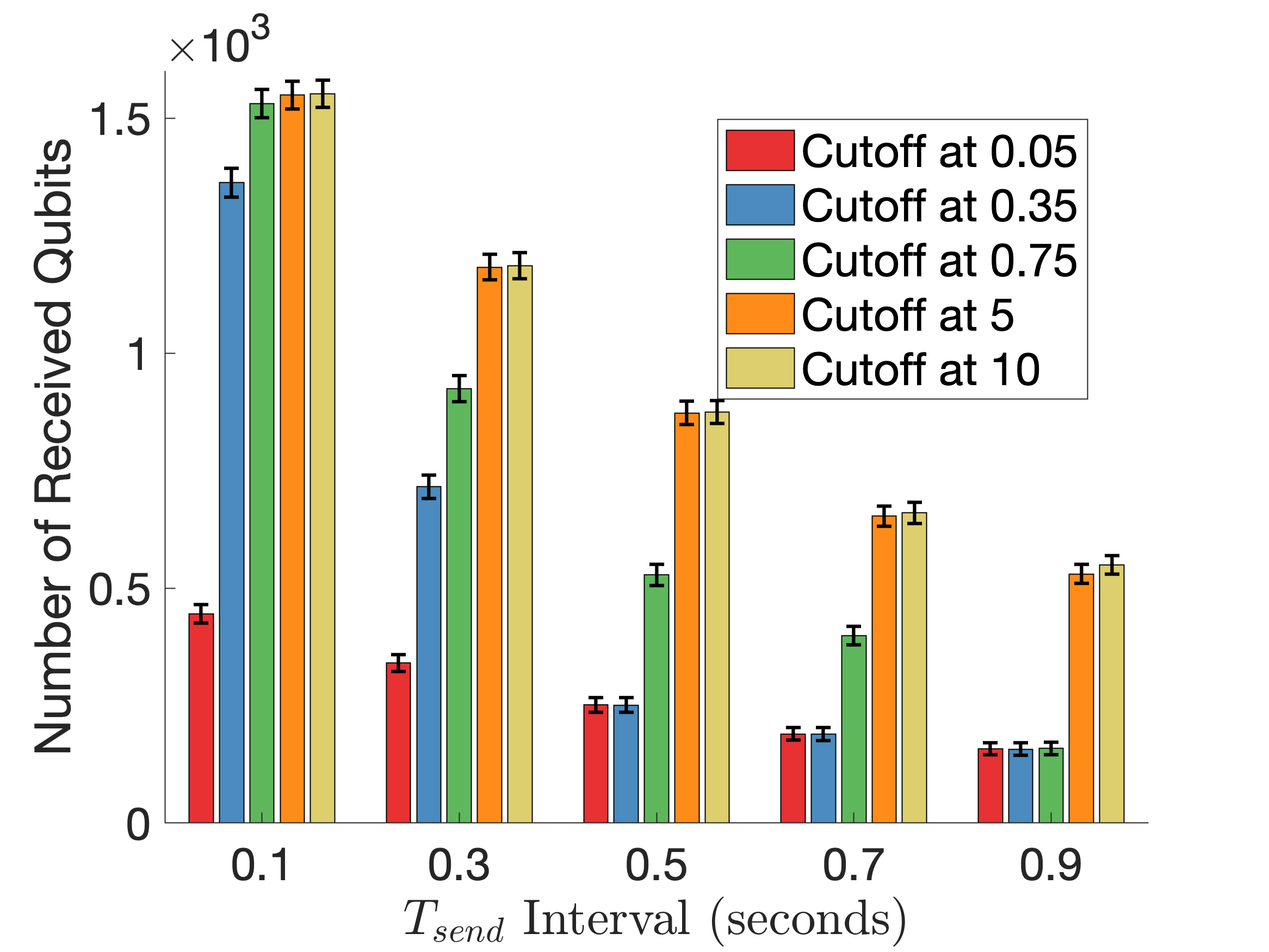}
        \caption{Number of Received Qubits}
        \label{subfig:photon-loss-bar}
    \end{subfigure}
    \hspace{0.05\linewidth}
    \begin{subfigure}{0.35\linewidth}
        \centering
        \includegraphics[width=\linewidth]{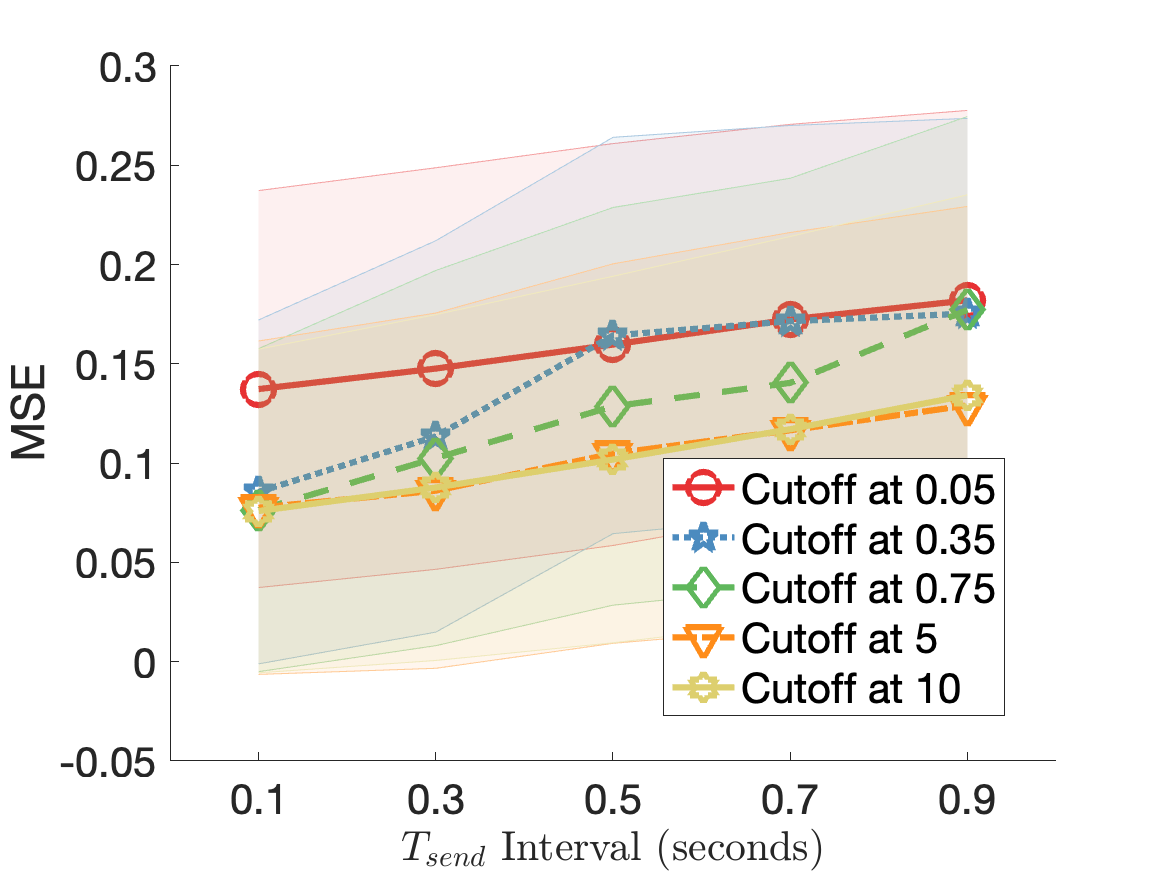}
        \caption{Mean-Square Error}
        \label{subfig:photon-loss-mse}
    \end{subfigure}
    \caption{\texttt{NetSquid} simulation of \mergecast with photon loss and memory decoherence: (a) Total merged qubits and (b) estimation MSE under various send intervals \(T_{\mathrm{send}}\) and cutoff times \(T_{c}\) for one hour
    }
    \label{fig:photon-loss}
\end{figure}

\noindent\textbf{Observations.}
Figure~\ref{subfig:photon-loss-bar} illustrates two clear trends.  First, for a fixed send interval \(T_{\mathrm{send}}\), the total number of merged qubits increases as the cutoff time \(T_{c}\) increases—longer cutoffs maintain more early arrivals.
Moreover, for a fixed \(T_{\text{send}}\), the number of received qubits for different \(T_{c} < T_{\text{send}}\) coincide because all arrival-time mismatches fall within the cutoff window.
Second, for a fixed \(T_{c}\), increasing \(T_{\mathrm{send}}\) yields more merge opportunities and hence more qubits.
Figure~\ref{subfig:photon-loss-mse} shows the MSE versus \(T_{\mathrm{send}}\).  Holding \(T_{c}\) constant, the MSE grows with \(T_{\mathrm{send}}\), because longer intervals reduce the merge rate (i.e., the sample size).
However, for fixed \(T_{\mathrm{send}}\), raising \(T_{c}\)---though increasing the sample size---\emph{does not} monotonically improve accuracy.
Because larger \(T_{c}\) introduces additional decoherence noise.
For example, at \(T_{\mathrm{send}}=0.5\,\mathrm{s}\), the \(T_{c}=0.05\,\mathrm{s}\) case achieves lower MSE than \(T_{c}=0.35\,\mathrm{s}\) despite merging fewer qubits.

Overall, the mildly increasing rate of MSE in Figure~\ref{subfig:photon-loss-mse} indicates that \mergecast performs robustly, even when the two input quantum states (for merging in the intermediate node) may arrive asynchronously due to photon loss and memory decoherence.

\section{Conclusion and Outlook}
\label{sec:conclusion}

In this paper, we introduced a new type of tomography mechanism in quantum networks (QNs), termed {\mergecast}.
Together with the progressive etching procedure, it enables the unique identification of internal quantum channels in networks characterized by arbitrary topologies and Pauli channels.
This procedure pushes the boundary of quantum network tomography (QNT) beyond the previously works~\citep{de2022quantum,de2023characterization,de2024quantum}
in three key aspects:
(i) It works for QNs with arbitrary topologies (only star topology were previously considered);
(ii) It can identify any Pauli channels, rather than being limited to specific bit-flip and depolarizing channels as in prior works.
(iii) It removes the assumption of state preparation ability of the intermediate nodes.
Besides the \mergecast for general Pauli channels, we also identify a special case of QNs with \emph{bypassable} channels and propose a more efficient \bypassunicast protocol for such networks.
We also extended our investigation to a more realistic QNT framework that incorporates state preparation and measurement (SPAM) errors. We rigorously defined SPAM errors, proposed estimation protocols for these errors within QNT, and adapted our \mergecast approach to handle networks affected by SPAM errors.
The \mergecast and progressive etching are scalable and efficient methods for identifying internal quantum channels in quantum networks. In experiments, we have shown that our protocol is robust against SPAM errors, photon losses, memory decoherence, and can be implemented in a variety of network topologies.

Future works should focus on further relaxing the requirements on the operations and intermediate nodes for the quantum network tomography task.
For example,
while \mergecast removed the assumption of state preparation and measurement ability of the intermediate nodes~\citep{de2022quantum,de2023characterization,de2024quantum}, it still require the intermediate node to be able to perform a CNOT operation.
Further relaxation of the requirements on the intermediate nodes would allow for a more general class of quantum networks to be considered.
Another direction is to further extend the \mergecast protocol beyond Pauli channels.
While the Pauli twirling can transfer any channel to a Pauli, this transformation only provides an approximation of the actual channel.
Directly estimating the internal channel in a quantum network can give more fine-grained channel estimates than the approximation from Pauli twirling.
Lastly, while this paper proposes the novel \mergecast protocol and devises its corresponding static estimators, further investigating the convergence rate and estimation accuracy in QNT in the theoretical perspective is an interesting direction.


\bibliographystyle{ACM-Reference-Format}
\bibliography{bibliography}


\newpage
\appendix

\section{Proof of Lemma~\ref{lem:bypassable-channel}} \label{app:bypassable-lemma-proof}

Let $\mathcal{P}$ be a single-qubit Pauli channel, acting on a density matrix $\rho$ as
\[
    \mathcal{P}(\rho) = \sum_{P \in \{I, X, Y, Z\}} p_P\, P \rho P,
    \quad \text{with } \sum p_P = 1, \ p_P \ge 0.
\]
In the Pauli transfer matrix (PTM) representation, the channel acts linearly on the Pauli basis
$\{I, X, Y, Z\}$ as
\[
    \mathcal{P}(P_j) = \lambda_j P_j, \quad P_j \in \{I, X, Y, Z\},
\]
so that the PTM is diagonal:
\[
    T_{\mathcal{P}} = \operatorname{diag}(1, \lambda_X, \lambda_Y, \lambda_Z),
\]
where $\lambda_I = 1$ always holds because $\mathcal{P}$ is trace-preserving and unital.

\smallskip
\noindent\textbf{($\Rightarrow$)}~
If $\mathcal{P}$ is bypassable, then by definition there exists a non-identity Pauli operator
$P_k \in \{X, Y, Z\}$ such that $\mathcal{P}(P_k) = P_k$.
Hence, $\lambda_k = 1$.
Since $\lambda_I = 1$ for all Pauli channels,
the PTM diagonal must contain at least two entries equal to one: $\lambda_I = 1$ and $\lambda_k = 1$.

\smallskip
\noindent\textbf{($\Leftarrow$)}~
Conversely, if at least two entries on the PTM diagonal are equal to one,
then because $\lambda_I = 1$ always, at least one of
$\lambda_X, \lambda_Y, \lambda_Z$ must also equal one.
This implies the existence of a non-identity Pauli operator $P_k$ such that
$\mathcal{P}(P_k) = P_k$, i.e., the channel leaves $P_k$ invariant.
Hence, $\mathcal{P}$ is bypassable.

\smallskip
\noindent
Therefore, a single-qubit Pauli channel is bypassable if and only if
the diagonal of its Pauli transfer matrix contains at least two entries equal to one.
\qed

\section{Address Degree-2 Nodes in General Topology}\label{app:degree-2-nodes}

Due to the intrinsic constraint of the degree-2 node (with a line pattern), it is impossible to individually identify the two channels connected by the node (unless additional assumption is made, e.g., the channels are bypassable).
This impossibility also exists in classical network tomography~\citep{he2021network}.

This appendix provides a detailed explanation for how to modify the topology of a network to ``remove'' all degree-2 nodes by transforming the channels connected by the degree-2 nodes into an equivalent single ``channel.''
We start from giving the definition of identifiable channels.

\begin{figure}[H]
    \centering
    \begin{subfigure}{0.45\columnwidth}
        \centering
        \resizebox{\linewidth}{!}{\begin{tikzpicture}
                \node[draw, circle, minimum size=4pt, inner sep=1pt, fill] (center) at (0,0) {};

                \node[draw, circle, minimum size=4pt, inner sep=1pt, fill] (A1) at (0:2) {};
                \node[draw, circle, minimum size=4pt, inner sep=1pt, fill] (A2) at (180:2) {};
                \node[draw, circle, minimum size=4pt, inner sep=1pt, fill] (A3) at (270:2) {};

                \node[draw, rectangle, minimum size=5pt, inner sep=1pt, fill=blue] (B2) at (180:4) {};

                \node[draw, rectangle, minimum size=5pt, inner sep=1pt, fill=blue] (C2) at (0:4)   {};

                \node[draw, circle, minimum size=4pt, inner sep=1pt, fill=black] (D1) at (-2,-2) {};
                \node[draw, rectangle, minimum size=5pt, inner sep=1pt, fill=blue] (D2) at (-4,-2) {};

                \node[draw, circle, minimum size=4pt, inner sep=1pt, fill=black] (E1) at (2,-2) {};
                \node[draw, rectangle, minimum size=5pt, inner sep=1pt, fill=blue] (E2) at (4,-2) {};

                \draw[line width=1pt] (center) -- node[draw, rectangle, midway, fill=white] {\(\mathcal{P}_3\)} (A1);
                \draw[line width=1pt] (center) -- node[draw, rectangle, midway, fill=white] {\(\mathcal{P}_2\)} (A2);
                \draw[line width=1pt] (center) -- node[draw, rectangle, midway, fill=white] {\(\mathcal{P}_5\)} (A3);

                \draw[line width=1pt] (A2) -- node[draw, rectangle, midway, fill=white] {\(\mathcal{P}_1\)} (B2);

                \draw[line width=1pt] (A1) -- node[draw, rectangle, midway, fill=white] {\(\mathcal{P}_4\)} (C2);

                \draw[line width=1pt] (A3) -- node[draw, rectangle, midway, fill=white] {\(\mathcal{P}_7\)} (D1);
                \draw[line width=1pt] (D1) -- node[draw, rectangle, midway, fill=white] {\(\mathcal{P}_6\)} (D2);

                \draw[line width=1pt] (A3) -- node[draw, rectangle, midway, fill=white] {\(\mathcal{P}_8\)} (E1);
                \draw[line width=1pt] (E1) -- node[draw, rectangle, midway, fill=white] {\(\mathcal{P}_9\)} (E2);
            \end{tikzpicture}}
        \caption{Original topology}\label{subfig:identifiable-channel-original}
    \end{subfigure}
    \hspace{0.05\textwidth}
    \begin{subfigure}{0.45\columnwidth}
        \centering
        \resizebox{\linewidth}{!}{\begin{tikzpicture}
                \node[draw, circle, minimum size=4pt, inner sep=1pt, fill] (center) at (0,0) {};

                \node[draw, circle, minimum size=4pt, inner sep=1pt, fill] (A1) at (0:2) {};
                \node[draw, circle, minimum size=4pt, inner sep=1pt, fill] (A3) at (270:2) {};

                \node[draw, rectangle, minimum size=5pt, inner sep=1pt, fill=blue] (B2) at (180:4) {};

                \node[draw, rectangle, minimum size=5pt, inner sep=1pt, fill=blue] (C2) at (0:4)   {};

                \node[draw, rectangle, minimum size=5pt, inner sep=1pt, fill=blue] (D2) at (-4,-2) {};

                \node[draw, rectangle, minimum size=5pt, inner sep=1pt, fill=blue] (E2) at (4,-2) {};

                \draw[line width=1pt] (center) -- node[draw, rectangle, midway, fill=white] {\(\mathcal{P}_5\)} (A3);

                \draw[line width=1pt] (center) -- node[draw, rectangle, midway, fill=white] {\(\mathcal{P}_1\circ\mathcal{P}_2\)} (B2);

                \draw[line width=1pt] (center) -- node[draw, rectangle, midway, fill=white] {\(\mathcal{P}_3\circ \mathcal{P}_4\)} (C2);

                \draw[line width=1pt] (A3) -- node[draw, rectangle, midway, fill=white] {\(\mathcal{P}_6\circ \mathcal{P}_7\)} (D2);

                \draw[line width=1pt] (A3) -- node[draw, rectangle, midway, fill=white] {\(\mathcal{P}_8\circ \mathcal{P}_9\)} (E2);
            \end{tikzpicture}}
        \caption{Equivalent identifiable topology}
    \end{subfigure}
    \caption{Examples of identifiable channels (\(\mathcal{P}_5\)) and equivalent identifiable channels \(\{\mathcal{P}_1\circ \mathcal{P}_2\), \(\mathcal{P}_3\circ \mathcal{P}_4\), \(\mathcal{P}_6\circ \mathcal{P}_7\), \(\mathcal{P}_8\circ \mathcal{P}_9\}\).}
    \label{fig:identifiable-channels}
\end{figure}
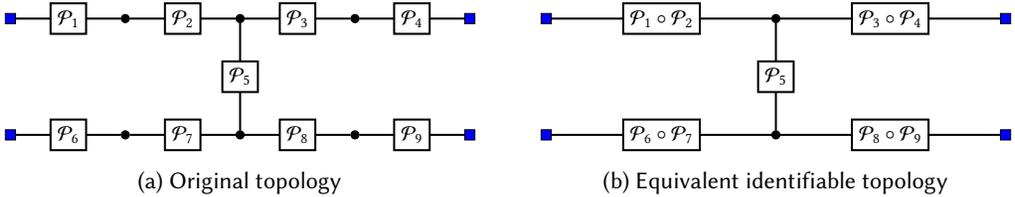

\begin{definition}[Identifiable channel]
    Channels whose two end nodes are either with a star pattern (i.e., connected with three or more channels) or peripheral are called \emph{identifiable channels}.
    \label{def:identifiable-channels}
\end{definition}

All channels in a topology without nodes with degree-2 are identifiable channels.
That is, non-identifiable channels all have at least an end node with degree-2.
This further implies that for each non-identifiable channel, at least one of its two end nodes connects to another non-identifiable channel.
Hence, non-identifiable channels connected together form paths whose two end nodes are either with a star pattern (degree-3) or peripheral (degree-1).
This observation suggests that the path composed of non-identifiable channels can be regarded as one identifiable ``channel'' (see Definition~\ref{def:equivalent-identifiable-channels} as follows and an example in Figure~\ref{fig:identifiable-channels}).

\begin{definition}[Equivalent identifiable channel]
    A path composed of non-identifiable channels whose two end nodes are either with a star pattern or peripheral is called an \emph{equivalent identifiable channel}.
    \label{def:equivalent-identifiable-channels}
\end{definition}

\begin{remark}
    In general, it is impossible to estimate the non-identifiable channel.
    However, if some additional conditions or capacities are assumed, it is possible to identify them.
    Taking the channels \(\mathcal{P}_1\) and \(\mathcal{P}_2\) in Figure~\ref{subfig:identifiable-channel-original} as an instance, if one of the two channels is bypassable (Definition~\ref{def:bypassable-channel}),
    then one can apply the protocol proposed in Section~\ref{subsec:bypassable-channel-protocol} to tomography both channels.
\end{remark}


To sum up, any general topology in a QNT task can be decomposed into identifiable channels and equivalent identifiable channels.
This is a graph with nodes with either degree-1 or degree greater than \(2\).
Then, one can apply the mergecast protocol to identifiable channels as well as equivalent identifiable channels, as Section~\ref{subsec:progressive-etching} described.

\section{Cram\'er-Rao Lower Bounds in Experiments} \label{app:spam-crb}

This appendix presents the detailed expression of the Cram\'er-Rao lower bounds (CRB) plotted in the figures of Section~\ref{sec:qnt-experiment}.

The CRB in Figure~\ref{fig:mergecast-spam-fix} for the mergecast protocol is derived from the quantum Fisher information matrix (FIM) of the quantum state \(\kett{\rho_1}\) and is given by
\begin{align}
        \text{CRB}_{\text{merge}}(M, N)
         & = \frac{q_{Z,1} (1 - ms q_{Z,1} q_{Z,2} q_{Z,3})}{M \cdot ms q_{Z,2} q_{Z,3}}  + \frac{q_{Z,1}^2(1 - ms q_{Z,2} q_{Z,3})}{N \cdot ms q_{Z,2} q_{Z,3}},
\end{align}
where \(M\) and \(N\) are the number of times of the mergecast and unicast protocols, respectively,
\(q_{Z,1}\) is the channel parameter to be estimated, \(q_{Z,2}\) and \(q_{Z,3}\) are the parameters of the other two channels,
and \(m, s\) are the parameters of the SPAM error.

The CRB for the estimation of the SPAM error parameter \(s\) in Figures~\ref{subfig:spam-s-90} and~\ref{subfig:spam-s-70} is as follows,
\begin{align}
        \text{CRB}_{\text{SPAM}, s}(M, N)
         & = \frac{s(1 - msq_{Z,1} q_{Z,2})}{N(1 {-} ms^2 q_{Z,1} q_{Z,2})mq_{Z,1} q_{Z,2}} + \frac{1 -  ms^2 q_{Z,1} q_{Z,2}}{M(1 {-} msq_{Z,1} q_{Z,2})mq_{Z,1} q_{Z,2}},
\end{align}
where \(M\) and \(N\) are the number of times of the root mergecast protocol in Figure~\ref{fig:unicast-spam-estimate-s} and the unicast protocol, respectively,

The CRB for the estimation of the SPAM error parameter \(m\) in Figures~\ref{subfig:spam-m-90} and~\ref{subfig:spam-m-70} is as follows,
\begin{align}
        \text{CRB}_{\text{SPAM}, m}(M, N)
         & = \frac{m(1 - msq_{Z,1} q_{Z,2})}{N(1 {-} m^2s q_{Z,1} q_{Z,2})sq_{Z,1} q_{Z,2}}  + \frac{1 -  m^2 sq_{Z,1} q_{Z,2}}{M(1 {-} msq_{Z,1} q_{Z,2})sq_{Z,1} q_{Z,2}},
\end{align}
where \(M\) and \(N\) are the number of times of the end multicast protocol in Figure~\ref{fig:unicast-spam-estimate-m} and the unicast protocol, respectively,

\section{Details for Mergecast in Figure~\ref{fig:mergecast}}\label{app:mergecast}

Below, we present the quantum states at slices 1, 2, 3, and 4 of Figure~\ref{fig:mergecast}. To simplify the notations of the two-qubit states (\(16\)-entry vectors in the Pauli-Liouville representation), we only write the non-zero entries together with their corresponding Pauli bases as follows:
\begin{align}
    \kett{\rho_1{\otimes} \rho_2}_1
                    & =  I{\otimes} I +   Z{\otimes} I +  I {\otimes} Z +  Z{\otimes} Z,
    \\
    \kett{\rho_1{\otimes} \rho_2}_2
                    & = I{{\otimes}} I +  q_{Z,1} Z{{\otimes}} I + q_{Z,2} I {{\otimes}} Z +  q_{Z,1}q_{Z,2} Z{{\otimes}} Z,
    \\
    \kett{\rho_1{\otimes} \rho_2}_3
                    & = I{\otimes} I +  q_{Z,1} Z{\otimes} I + q_{Z,2} I {\otimes} Z + q_{Z,1}q_{Z,2} Z{\otimes} Z,
    \\
    \kett{\rho_2}_3 & = \Tr_1[\kett{\rho_1{\otimes} \rho_2}_3] =  I + q_{Z,1}q_{Z,2} Z,
    \\
    \kett{\rho_2}_4
                    & = I + q_{Z,1}q_{Z,2}q_{Z,3} Z,
\end{align}
where between the slices 3 and 4, we trace out (discard) the first qubit and have \(
\kett{\rho_2}_3  = \Tr_1[\kett{\rho_1{\otimes} \rho_2}_3].
\)
Hence, the probability of the measurement outcome \(\kett{00}\) at slice 4 is given by \[
    \mathbb{P}(\kett{00}) = \frac{1}{2} \bbra{00} \kett{\rho_2}_4  =\frac{1 + q_{Z,1}q_{Z,2}q_{Z,3}}{2}.
\]


\section{Details for Estimating SPAM Error Parameter \emph{m} in Figure~\ref{fig:unicast-spam-estimate-m}} \label{app:spam-m-calculation}

The quantum states at slices 1 and 2 of Figure~\ref{fig:unicast-spam-estimate-m} are
\begin{align}
     & \kett{\tilde \rho_1 {\otimes} \tilde \rho_2}_1
    = I {\otimes} I + s q_{Z,1} q_{Z,2} I {\otimes} Z + s q_{Z,1}' q_{Z,2}' Z {\otimes} I +  s^2 q_{Z,1} q_{Z,2} q_{Z,1}' q_{Z,2}' Z {\otimes} Z,
    \\
     & \kett{\tilde \rho_1 {\otimes} \tilde \rho_2}_2= I {\otimes} I + s^2 q_{Z,1} q_{Z,2} q_{Z,1}' q_{Z,2}' I {\otimes} Z + s q_{Z,1}' q_{Z,2}' Z {\otimes} I + sq_{Z,1} q_{Z,2} Z {\otimes} Z .
\end{align}
Notice the joint-measurement with SPAM error at the end of this protocol is \begin{align}
    \kett{\tilde M_0 {\otimes} \tilde M_0} =  I {\otimes} I + m I {\otimes} Z + m Z {\otimes} I + m^2 Z {\otimes} Z ,
    \\
    \kett{\tilde M_0 {\otimes} \tilde M_1} =  I {\otimes} I - m I {\otimes} Z + m Z {\otimes} I - m^2 Z {\otimes} Z ,
    \\
    \kett{\tilde M_1 {\otimes} \tilde M_0} =  I {\otimes} I + m I {\otimes} Z - m Z {\otimes} I - m^2 Z {\otimes} Z ,
    \\
    \kett{\tilde M_1 {\otimes} \tilde M_1} =  I {\otimes} I - m I {\otimes} Z - m Z {\otimes} I + m^2 Z {\otimes} Z ,
\end{align}
and the probabilities of observing measurement outcomes  is \(\mathbb{P}(\kett{00}) = \frac{1}{4}\bbra{\tilde M_0 {\otimes} \tilde M_0} \kett{\tilde \rho_1 {\otimes} \tilde \rho_2}_2\),
\(\mathbb{P}(\kett{11}) = \frac{1}{4}\bbra{\tilde M_1 {\otimes} \tilde M_1} \kett{\tilde \rho_1 {\otimes} \tilde \rho_2}_2\)
and similarly for \(\kett{01}, \kett{10}\). This calculation yields the probabilities presented in the main paper.


\section{Simplification of SPAM Error Models}\label{sec:appendix-spam-simplification}\label{rmk:spam-simplification}

The simplification of the parametric formulation of SPAM errors is achieved by averaging the outputs obtained from circuits in which Pauli channels are inserted immediately after state preparation and before measurement---a method conceptually akin to the phase-cycling technique used in nuclear magnetic resonance~\citep{levitt2008spin,lin2021independent}. Below is an illustrative example for a unicast protocol (similar methods apply to other protocols employing Pauli channels):
\begin{center}
        \begin{quantikz}
                \inputD[style={fill=red!20}]{\emph{SP}}
                & \gate[style={fill=blue!50}]{\textcolor{white}{I/Z}}
                & \gate{\mathcal{D}}
                & \gate[style={fill=blue!50}]{\textcolor{white}{I/X^{\circ}}}
                & \gate[style={fill=blue!50}]{\textcolor{white}{I/Z}}
                & \meter[style={fill=red!20}]{}
        \end{quantikz}
\end{center}
Here, each of the blue-coloured gates is randomly selected from its two options. In particular, the superscripted circle in \(X^{\circ}\) indicates that if the \(X\) gate is chosen, the measurement outcome is flipped. By averaging the results across many such randomly compiled circuits, the aggregate effect transforms the prepared state and measurement operator into the vectors \([1,0,0,s_Z]^T\) and \([1,0,0,m_Z]^T\), respectively. With minor modifications, the same technique can similarly map SPAM errors to \([1,0,0,s_Y]^T\) and \([1,0,0,m_Y]^T\), as well as \([1,0,0,s_X]^T\) and \([1,0,0,m_X]^T\).

\end{document}